\documentclass[twocolumn,twocolappendix]{aastex631}
\usepackage{amsmath}
\usepackage{array}
\usepackage{float}

\begin{document}

\title{Detection Prospects of Fast-merging Gravitational Wave Sources in M31}

\author[0000-0003-3862-0726]{Jian-Guo He}
\affiliation{Department of Astronomy, Nanjing University, Nanjing 210023, People's Republic of China}
\affiliation{Key Laboratory of Modern Astronomy and Astrophysics, Nanjing University, Ministry of Education, Nanjing 210023, People's Republic of China}

\author[0000-0003-2506-6906]{Yong Shao}
\email{shaoyong@nju.edu.cn}
\affiliation{Department of Astronomy, Nanjing University, Nanjing 210023, People's Republic of China}
\affiliation{Key Laboratory of Modern Astronomy and Astrophysics, Nanjing University, Ministry of Education, Nanjing 210023, People's Republic of China}

\author[0000-0002-0822-0337]{Shi-Jie Gao}
\affiliation{Department of Astronomy, Nanjing University, Nanjing 210023, People's Republic of China}
\affiliation{Key Laboratory of Modern Astronomy and Astrophysics, Nanjing University, Ministry of Education, Nanjing 210023, People's Republic of China}

\author[0000-0002-0584-8145]{Xiang-Dong Li}
\affiliation{Department of Astronomy, Nanjing University, Nanjing 210023, People's Republic of China}
\affiliation{Key Laboratory of Modern Astronomy and Astrophysics, Nanjing University, Ministry of Education, Nanjing 210023, People's Republic of China}



\begin{abstract}

It is widely accepted that quite a number of double compact objects (DCOs) in the Milky Way can be identified by future space-based  gravitational wave (GW) detectors,  while systematic investigations on the detection of the GW sources in nearby galaxies are still lacking. In this paper, we present  calculations of potential populations of GW sources for all types of DCOs in the Local Group galaxy M31. For M31, we use an age-dependent  model for the evolution of the metallicity and the star-formation rate. By varying assumptions of common-envelope ejection efficiencies and supernova-explosion mechanisms during binary evolution, we make predictions on the properties of DCOs that can be detected by the Laser Interferometer Space Antenna (LISA).  Our calculations indicate that a few (a  dozen) DCOs are likely to be observed by LISA during its 4 (10) yr mission.   We expect that  the sources with black-hole components  are more likely to be firstly identified during a 4-yr mission since these binaries have relatively large chirp masses, while the systems with white-dwarf components dominate the overall population of detectable GW sources during a 10-yr mission. LISA can only detect very tight fast-merging systems in M31, corresponding to the peak  of orbital period distribution from $\sim 2$ min for double white dwarfs to $\sim 20$~min for double black holes. 
%The mass distribution of these GW sources may provide clues to constrain the mechanism of supernova explosions that form neutron stars and black holes. 

\end{abstract}


\section{Introduction} \label{sec:intro}

The discovery of the first double black hole (BHBH) merger GW150914 \citep{Abbott2016} opened the era of gravitational wave (GW) astrophysics. To date, there are nearly 100 double compact object (DCO) mergers detected by the ground-based GW detectors LIGO and Virgo \citep{LVK2021,Nitz2021,LVK2023}. The vast majority of them are identified as BHBH mergers with component masses $\sim 6-95 M_{\odot}$, two are double neutron star (NSNS) mergers, four are BHNS mergers, and one event, GW190814 with a $ \sim2.6M_\odot $ component, is either a BHNS or BHBH merger \citep{Abbott2020}. All these events involve DCO mergers with GW signals emitting at high frequencies of $\mathrm{Hz}-\mathrm{kHz}$. 

Before merger, tight DCO systems may emit GW signals in the $\rm mHz$ band that are likely to be observed by future space-based detectors e.g., LISA \citep{AS2017}, TianQin \citep{Luo2016}, and Taiji \citep{Ruan2020}. For inspiral DCOs in the Milky Way, it is expected that they can spend a typical duration of  $\sim 10^{6} \rm\, yr$ in the LISA band. Previous studies of binary population synthesis (BPS) have shown that LISA can identify a large number of resolved DCOs in the Milky Way \citep[see][for a review]{AS2023} including thousands of double white dwarf (WDWD) systems \citep[e.g.,][]{Nelemans2001,Ruiter2010,Liu2010,Korol2017,Lamberts2019,Breivik2020}, hundreds of NSWD systems \citep[e.g.,][]{Tauris2018,Chen2020,Chen2021,Wang2021}, and tens of other types of DCO systems with BH and/or NS components \citep[e.g.,][]{Belczynski2010a,Liu2014,Lamberts2018,Lau2020,Andrews2020,Sesana2020,Shao2021,Wagg2022a,Gao2022,Qin2023,Feng2023}.

Outside the Milky Way, it is predicted that a number of DCO systems in the Local Group galaxies are detectable by LISA \citep{Korol2018,Seto2019,Andrews2020,Lau2020}.
Since these binaries have a typical distance of Mpc, they are expected to spend a relatively short duration of  $\sim 10^{3} \rm\, yr$ in the LISA band \citep{AS2023}. Regarding M31 (the Andromeda galaxy), the largest galaxy in the Local Group, \citet{Seto2019} estimated that LISA may detect $ \sim 5 $ NSNS systems with a 10-yr observation duration \citep[see also][]{Andrews2020}. In addition, \citet{Korol2018} predicted that 17$-$60 WDWD binaries in M31 can be identified by LISA for its mission duration of 4$-$10 yr.

According to the theory of binary star evolution, tight DCO systems originate from the primordial binaries with two zero-age main-sequence stars when both components have evolved to be compact stars \citep{Tauris2023}. During the evolution, common envelope \citep[CE,][]{Ivanova2013} phases play a vital role to shrink binary orbits probably leading to the formation of close DCO systems \citep[e.g.,][]{Webbink1984}. Recently, stable Roche lobe overflow has also been considered as a potentially important channel for the formation of close DCO binaries \citep[e.g.,][]{Pavlovskii2017,vdH2017,Neijssel2019,Marchant2021,Shao2021,Olejak2021,Gallegos-Garcia2022}.
At the end of stellar evolution, massive stars are expected to leave behind NSs or BHs via the process of supernova explosions. It is still unclear whether there exists a mass gap between NSs and BHs \citep[see][for a review]{Shao2022}. It is believed that the coalescence of DCO binaries can result in various kinds of astrophysical phenomena associating with intense electromagnetic emissions, e.g., type Ia supernovae for WDWD systems \citep{Iben1984,Webbink1984}, long-duration gamma-ray bursts for NSWD/BHWD systems \citep{Fryer1999,Yang2022}, short-duration gamma-ray bursts \citep{Paczynski1986,Abbott2017,Gompertz2020} and kilonovae \citep{Li1998,Zhu2021} for NSNS/BHNS systems.  It is expected that the detection of tight DCO binaries emitting low-frequency GW signals can shed light on the physics of close binary interactions and the origin of DCO coalescence events. 

In this paper, we aim to use a BPS method to assess the properties and the numbers of all types of LISA DCO sources in M31. Our calculations include different treatments of the input parameters related to CE phases and supernova-explosion processes. Furthermore, we adopt a recently reported star-formation history of the M31 disk \citep{Williams2017}, for which the metallicity and the star-formation rate are age-dependent. 

The structure of this paper is organized as follows.
We introduce the method in Section \ref{sec:method}. Then we show the results based on our calculations in Section \ref{sec:result} and make some discussions in Section \ref{sec:discussion}. Finally, we conclude in Section \ref{sec:conclusion}.

\section{Method} \label{sec:method}

\subsection{BPS Code}

Except GW transients, no DCO systems  have been reported outside the Milky Way. Here we present estimations on the potential population of LISA sources in M31 by adopting a BPS approach. Only the systems formed via isolated binary evolution are included in our calculations. We employ the \textit{BSE} code originally developed by \citet{hurley2002} and significantly modified by \citet{Shao2014} that has been used to predict the detection of Galactic LISA sources \citep{Shao2021}. In this code, some important modifications are summarized as follows. During the evolution of primordial binaries, it is assumed that the accretion rate of the secondary stars is dependent on their rotational velocities. Under this assumption, small amount of transferred mass is accreted by rapidly rotating secondary stars. Mass accretion onto the secondary stars can cause them to expand and spin up. On the one hand, the significant expansion of the secondary stars due to rapid mass accretion may lead to the formation of contact binaries as the secondary stars fill their Roche lobes \citep{Nelson2001}. On the other hand, a small amount of accreted mass can spin up the secondary stars to near critical rotation \citep{Packet1981}. As a consequence, the secondary stars are likely to reach critical rotation at the early stage of mass transfer and then accrete at relatively low rates. In this situation, the maximal mass ratios between the primary stars and the secondary stars for avoiding the formation of contact binaries can reach as high as about 6 \citep{Shao2014}.  For stellar wind mass-loss rates, the prescription of \citet{Belczynski2010} has been applied to various types of stars, except that for helium stars we use half of the \citet{Hamann1995} rates \citep{Kiel2006}. When dealing with mass-transfer stability in the binaries with a BH accretor and a massive donor, the criteria of \citet{Shao2021} have been adopted to determinate whether such binaries experience CE evolution or stable mass transfer. These criteria are related to the masses of both components and the radii of donor stars, and show that the binaries with heavier BHs are more likely to undergo stable mass transfer. It's worth noting that these criteria are not strongly dependent on metallicities, except for the systems with donors initially more massive than $ \sim 40 M_\odot $\footnote{The stability of mass transfer is sensitive to the structure of the donor stars \citep{Soberman1997}. At solar metallicity, the donor stars with masses above $ \sim 40-60 M_\odot $ never develop a fully convective envelope which can increase the stability of mass transfer \citep{Pavlovskii2017,Klencki2021}.}. Since the massive donors with increasing metallicities are more likely to experience extensive wind mass loss before mass transfer, the mass transfer tends to be more stable. The use of these criteria  allows the formation of tight DCO binaries via stable mass transfer rather than CE evolution \citep[see also e.g.,][]{Pavlovskii2017,vdH2017,Marchant2021}.

During a CE phase, we use the standard energy conservation equation \citep{Webbink1984} to calculate the orbital evolution of a binary system, i.e.,  
\begin{equation}
\label{energy balance formula}
\alpha_{\mathrm{CE}}\left(\frac{G M_{\mathrm{d,f}} M_\mathrm{a}}{2 a_{\mathrm{f}}}-\frac{G M_\mathrm{d,i} M_\mathrm{a}}{2 a_{\mathrm{i}}}\right)=-E_{\mathrm{bind}},
\end{equation}
where
\begin{equation}
E_{\mathrm{bind}}=-\frac{G M_{ \mathrm{d,i}} M_{\mathrm{d,env}}}{\lambda R_{\rm d, L }}
\end{equation}
is the binding energy of the donor's envelope at the onset of CE evolution. Here $M_{\rm d}$ and $M_{\rm a}$ respectively denote the donor's and accretor's masses, $a$ is the orbital separation of the binary system, $M_{\rm d,env}$ is the mass of donor's envelope that is ejected out of the binary, $R_{\rm d,L}$ is the Roche lobe radius of the donor at the beginning of mass transfer, the subscripts $\mathrm{i}$ and $\mathrm{f}$ correspond to the initial and final stages of a CE phase, respectively. 
$\lambda$ is the binding energy parameter that is determined by the mass distribution of the envelope \citep{Xu2010} and  $\alpha_{\mathrm{CE}}$ is the CE ejection efficiency with the orbital energy of the binary unbinding donor's envelope. When CE evolution is triggered in binaries with Hertzsprung gap donors, we allow them to survive CE phases \citep[see e.g.,][]{Dominik2012}.

Equation (\ref{energy balance formula}) indicates that all of binary orbital energy is used to eject the envelop of engulfing star when $\alpha_{\mathrm{CE}} = 1$. Generally, the larger the $\alpha_{\mathrm{CE}}$, the binaries more likely to survive CE evolution and the longer the separations of post-CE binaries. Recently, \citet{Scherbak2023} suggested that $\alpha_{\mathrm{CE}} \sim 1/3$ according to the analyses of Galactic WDWD binaries. This is consistent with the case of $\alpha_{\mathrm{CE}} \sim 0.2-0.3$ for the post-CE binaries with a WD and a main-sequence companion \citep{Zorotovic2010}. Note that $\alpha_{\mathrm{CE}} \geq 3$ seems to be required to account for the formation of some post-CE binaries such as IK Peg with a $\sim 1.2M_{\odot}$ WD and a $\sim 1.7M_{\odot}$ main-sequence star \citep{Davis2010}. Furthermore, investigations on NS binaries indicate higher CE ejection efficiencies varying from $\alpha_{\mathrm{CE}} \sim 1$ \citep{Zuo2014} to $\alpha_{\mathrm{CE}} \sim 5$ \citep{Fragos2019,Yarza2022}. In our calculations, we set three values for CE ejection efficiencies i.e., $\alpha_{\mathrm{CE}}=0.3$, $\alpha_{\mathrm{CE}}=1.0$ and $\alpha_{\mathrm{CE}}=3.0$.

Following \citet{Shao2021}, we apply three different mechanisms of supernova explosions. These mechanisms involve the rapid \citep{Fryer2012}, the delayed \citep{Fryer2012} and the stochastic \citep{Mandel2020} prescriptions. The rapid recipe allows the formation of a $2-5 \mathrm{~M_{\odot}}$ mass gap among which no compact remnants are formed, while the delayed and stochastic mechanisms do not. So whether existing such a mass gap can provide a clue to constrain the process of supernova explosions \citep[see e.g.,][]{Shao2022}. Both the rapid and the delayed mechanisms treat the remnant masses as a function of CO core masses prior to supernova and fallback-material masses during explosion. In these two mechanisms, it is assumed that the natal kicks of NSs and BHs obey a Maxwellian distribution. We adopt a  dispersion velocity of $\sigma=265 \mathrm{~km} \mathrm{~s}^{-1}$ \citep{Hobbs2005} for NSs formed via  core-collapse supernovae. For BHs, we use reduced kick velocities due to the fallback of some materials compared to the case of NS kicks. In the stochastic mechanism, the remnant masses and the natal kicks are assumed to be probabilistic rather than deterministic. It is assumed that the natal kicks of both NSs and BHs follow a Gaussian distribution and a part of BHs form without any kick \citep{Mandel2020}. In this case, we further apply the scaling pre-factor calibrated by \citet{Kapil2022} to deal with the velocity distribution of NS kicks. In our calculations, we also include the formation of NSs through electron-capture supernovae. It is assumed that these NSs  have the mass of $1.3M_\odot$ and their natal kicks follow a Maxwellian distribution with a relatively low dispersion velocity of $\sigma=30 \mathrm{~km} \mathrm{~s}^{-1}$ \citep[see also][]{Shao2014,Shao2018}.

For primordial binaries we set a grid of initial parameters: the primary masses $M_{1}$ and the secondary masses $M_{2}$ are both limited in the range of $0.5M_{\odot}-100 M_{\odot}$, and the orbital separations $a$ in the range of $3 R_{\odot}- 10000 R_{\odot}$. The fraction of binaries among all stars is taken to be unity and all initial binaries are assumed to have circular orbits. Following \citet{Shao2021}, we calculate the birth rate of a specific DCO system $j$ as
\begin{equation}
R_{j}=\mathrm{SFR} \cdot \mathrm{W_{b}},
\end{equation}
where $\rm SFR$ is the star formation rate of the M31 disk,  $W_\mathrm{{b}}=\Phi(\ln M_{1}) \varphi(\ln M_{2}) \Psi(\ln a) \delta \ln M_{1} \delta \ln M_{2} \delta \ln a$ is a combination parameter to weight the contribution of the system evolved from a specific primordial binary. Then $\Phi (\ln M_{1}) = M_{1} \xi (M_{1})$ represents the distribution of primary masses in logarithmic space and $\xi (M_{1})$ describes the initial mass function \citep[IMF,][]{Kroupa1993}. Here we apply a modified IMF for M31 \citep{Weisz2015}, i.e.,
\begin{equation}
\xi (M_{1})=c_{1}M_{1}^{-(\Gamma+1)}
\begin{cases}
\Gamma = 0.3 & 0.08 M_{\odot}<M_{1} \leqslant 0.5 M_{\odot} \\
\Gamma = 1.3 & 0.5 M_{\odot}<M_{1} \leqslant 1.0 M_{\odot} \\
\Gamma = 1.45 & 1.0 M_{\odot}<M_{1}<100M_{\odot}
\end{cases},
\end{equation}
where $c_{1}=0.2074$ is a normalized parameter. This IMF can slightly increase the fraction of stars with masses above $0.5 M_{\odot}$ compared to the Kroupa's IMF. The secondary masses are assumed to have a uniform distribution \citep{Kobulnicky2007}, then giving $\varphi\left(\ln M_{2}\right)=M_{2}/M_{1}$.
It is assumed that the orbital separations follow a flat distribution in logarithmic space with $\Psi(\ln a) =c_{2}$, where $c_{2}=0.12328$ is a normalized parameter. In addition, the size of each interval $\delta\ln\chi$ in logarithmic space is determined as $\delta \ln \chi=\left(\ln \chi_{\rm max }-\ln \chi_{\rm min }\right)/(n_{\chi}-1)$. Here, $n_{\chi}$ represents the number of grid points for a specific parameter $\chi$ and is set to $n^{1/3}$ (where $n$ is the total number of primordial binary samples in the 3-dimensional parameter space mentioned above). Our calculations include 9 models with different supernova mechanisms and CE ejection efficiencies, and run $\sim 10^7$ primordial binaries for each evolutionary model. 
%The number of a specific DCO binary detectable by LISA can be estimated by multiplying its birthrate $R_{j}$ by the timespan spent in the LISA band.

\subsection{Star Formation History of M31}

The evolutionary history of SFR and stellar metallicity $\rm Z$ is an important factor that can make a considerable impact on the population of observable LISA sources in M31. We follow the work of \citet{Williams2017} which gives wide-area maps of the full star-formation history of the M31 disk by fitting resolved stellar photometry coming from the Panchromatic Hubble Andromeda Treasury survey. According to the result of PARSEC model \citep[see Table 3 of][]{Williams2017}, we obtain the total SFR of M31 during different history stages. For the metallicity evolution we use the age-dependent model \citep{Kemp2022} which linearly interpolates between (age (Gyr), $\rm Z$) coordinates of (0, 0.0002), (2, 0.008), (10, 0.02), (14, 0.022) approximately. The history distribution of $\rm SFR$ and $\rm Z$ of M31 has been mapped by \citet[][see their Figure 4]{Kemp2022}. 

\begin{figure}[htbp]
\centering
\includegraphics[width=8.5cm]{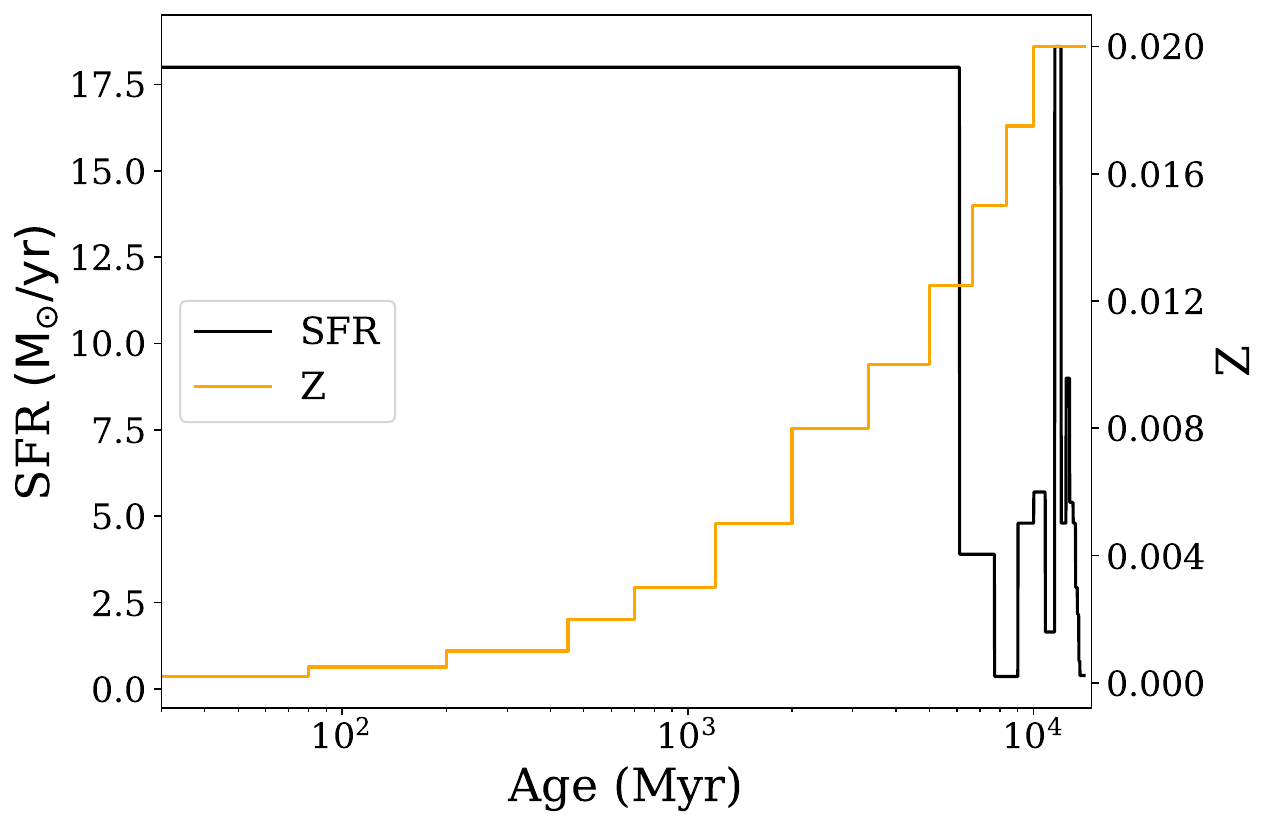}
\caption{The SFR-metallicity-age relation of M31. The black and orange curves denote the star-formation rate and the metallicity, respectively. 
}
\label{fig:SFH of M31}
\end{figure}

Since the metallicity changes over the age, we need to divide all binary stars into different sample spaces according to the metallicity or age. For convenience, we set 12 grids of metallicities as shown in Figure \ref{fig:SFH of M31}. The orange ladder-like curve gives the sample spaces separated by different metallicities. The metallicity of every primordial binary is randomly taken from the 12 individual values. For each metallicity grid, we know the age and the SFR. When calculating $W_\mathrm{{b}}$, the corresponding number of the primordial binaries in each grid is then set to be $ n/12 $. In our calculations, we select the binary samples if the primordial binaries can evolve to be observable LISA sources at the present age of 14 Gyr. For one selected binary sample, its number as LISA sources can be calculated by multiplying the birthrate $R_{j}$ by the timespan spent in the LISA band. As a consequence, we can estimate the total number of detectable LISA DCOs in M31 by accumulating the contribution from all selected binary samples.
 
\subsection{GW Signal}

For GW signal detected over a mission duration (4 or 10 yr), the characteristic strain $h_{c,n}$  at the $n \mathrm{th}$ harmonic can be expressed as \citep{Barack2004},
\begin{equation}
    h_{c, n}^{2}=\frac{1}{\left(\pi d_{\mathrm{L}}\right)^{2}}\left(\frac{2 G}{c^{3}} \frac{\dot{E}_{n}}{\dot{f}_{n}}\right).
\end{equation}
Here $\dot{E}_{n}$ is the power of the $n \mathrm{th}$ harmonic radiation \citep{Peters1963},
\begin{equation}
    \dot{E}_{n}=\frac{32}{5} \frac{G^{7 / 3}}{c^{5}}\left(2 \pi f_{\mathrm {orb }} \mathcal{M}_{c}\right)^{10 / 3} g(n, e),
\end{equation}
where $\mathcal{M}_{c}$
%$\mathcal{M}_{c}=\left(M_{1} M_{2}\right)^{3 / 5}\left(M_{1}+M_{2}\right)^{-1 / 5}$ 
is the chirp mass of a source and $g(n, e)$ (depending on the harmonic order $n$ and the eccentricity $e$) is the Fourier decomposition of GW signal and gives the relative power of gravitational radiation at the nth harmonic. If binaries have eccentric orbits, they would emit most of their energy at higher harmonics and be detected in a higher frequency band \citep{Nelemans2001}.

The change rate of GW frequency $\dot{f}_{n}$ can be written as
\begin{equation}
    \dot{f}_{n}=\frac{48 n}{5 \pi} \frac{\left(G \mathcal{M}_{c}\right)^{5 / 3}}{c^{5}}\left(2 \pi f_{\mathrm {orb }}\right)^{11 / 3} F(e),
\end{equation}
where $F(e)=\left[1+(73 / 24) e^{2}+(37 / 96) e^{4}\right] /\left(1-e^{2}\right)^{7 / 2}$ gives the integrated enhancement factor of gravitational radiation from an eccentric source compared with an equivalent circular source. The luminosity distance $d_{\mathrm{L}}$ of M31 is taken to be $780 \mathrm{~kpc}$ \citep{Ribas2005}, which reduces the characteristic strain amplitudes of tight DCO systems in M31 by over 30 times compared with those in the Milky Way.

\subsection{Signal to Noise Ratio}

In this work, we use the python package LEGWORK developed by \citet{Wagg2022b} to calculate signal-to-noise ratio (S/N) and identify the DCO systems as LISA sources with the condition of $\mathrm{S/N > 5}$ \citep[e.g.,][]{Lamberts2018,Shao2021}. In the following, we briefly describe the method used in the LEGWORK.

The S/N calculation includes the response of detector to signal and the detector's noise. For quasi-stationary sources we consider here, they evolve slowly during detector’s observation mission and their sky locations in the detector-based coordinate system are always changing. The averaged spectral power of GW signal is used to determine which sources can be detected and has the expression of
%Considering the orbital motion of a detector, usually the averaged spectral power of GW signal is used to determine which sources can be detected. Actually, the averaged spectral power is suitable for quasi-stationary sources so that it evolves slowly during detector's observation mission, which can be expressed as
\begin{equation}
\label{averaged spectral power}
	\langle\tilde{h}(f) \tilde{h}^*(f)\rangle=\mathcal{R}(f)\left(|\tilde{h}_{+}(f)|^2+|\tilde{h}_{\times}(f)|^2\right).
\end{equation}
Here $\mathcal{R}(f)$ is sky and polarization averaged signal response function of the instrument. It has been numerically computed by \citet{Larson2000}, and then fitted by \citet{Robson2019} as 
%but due to the lack of closed expression then it is fit as \citep{robson2019construction}
\begin{equation}
\mathcal{R}(f)=\frac{3}{10} \frac{1}{(1+0.6\left(f / f_*\right)^2)}.
\end{equation}
The second term in right side of Equation (\ref{averaged spectral power}) gives the spectral powers of plus and cross GW,
\begin{equation}
\begin{aligned}
&\tilde{h}_{+}(f)=A(f) \frac{\left(1+\cos ^2 \iota\right)}{2} \mathrm{e}^{\mathrm{i} \Psi(f)} \\
&\tilde{h}_{\times}(f)=\mathrm{i} A(f) \cos \iota \mathrm{e}^{\mathrm{i} \Psi(f)},
\end{aligned}
\end{equation}
where $A(f)$ and $\Psi(f)$ represent the amplitude and phase of GW, respectively, $\iota$ describes the inclination of the orbital angular momentum of a binary with respect to the line of sight.

For sources with known positions, we should consider GW signal modulation to obtain more precise S/N. The orbital motion of LISA induces a power spreading effect, where the power spectrum of the GW signal exhibits sidebands with a bandwidth of $10^{-4} \,\mathrm{mHz}$ around the GW frequency $f$. As a result, the strain amplitude at the true frequency is diminished. This effect is more pronounced at higher frequencies due to the larger bandwidth. Consequently, WDWD binaries are particularly impacted due to their relatively higher detectable frequencies. We make the assumption that all sources detectable by LISA share the same sky location as M31, with the ecliptic coordinates ($\theta=33.35^{\circ}, \phi=27.85^{\circ}$). Additionally, we consider their polarization $\psi$ and inclination $\iota$ to be uniformly distributed within the allowed range. In this study, we utilize the orbit-averaged amplitude modulation, originally defined by \citet{Cornish2003} and subsequently revised by \citet{Wagg2022b}, as
\begin{equation}
A^2= \mathcal{A}^2 \cdot A_{\mathrm{mod}}^2,
\end{equation}
where $\mathcal{A}$ is the intrinsic amplitude of the source and $A$ is the amplitude recorded in the detector. The amplitude modulation is given as
\begin{equation}
    A_{\mathrm{mod}}^2=\frac{1}{4} (1+\cos ^{2} \iota )^{2} \langle F_{+}^{2} \rangle+\cos ^{2} \iota \langle F_{\times}^{2} \rangle,
\end{equation}
where the orbit-averaged detector responses $\langle F_{+}^{2} \rangle$ and $\langle F_{\times}^{2} \rangle$ are the functions of sky location and polarization of sources in the ecliptic coordinates.

For the noise, we apply LISA's sensitivity curve constructed by \citet{Robson2019}. The effective power spectral density of the noise $S_{\mathrm{n}}(f)$ is expressed as 
%has been discussed in detail,
\begin{equation}
    S_n(f)=\frac{P_n(f)}{\mathcal{R}(f)}+S_{c}(f),
\end{equation}
where $P_n(f)$ is the power spectral density of the LISA noise, including the single-link optical metrology noise $P_{\mathrm{OMS}}$ and the single test mass acceleration noise $P_{\mathrm{acc}}$,
\begin{equation}
    P_n(f)=\frac{P_{\mathrm{OMS}}}{L^2}+2\left(1+\cos ^2\left(f / f_*\right)\right) \frac{P_{\mathrm{acc}}}{(2 \pi f)^4 L^2},
\end{equation}
where $L=2.5 \mathrm{~Gm}$ is the arm-length of LISA and $f_*=c /(2 \pi L)=19.09 \mathrm{~mHz}$ is LISA transfer frequency.
Actually, $P_{\mathrm{acc}}$ dominates $P_n(f)$ when $f$ is below $\sim \mathrm{5mHz}$ otherwise $P_{\mathrm{OMS}}$ dominates. For convenience, we ignore the term $f^{4}$ in $P_{\mathrm{acc}}$ and the term $f^{-4}$ in $P_{\mathrm{OMS}}$ when calculating $P_n(f)$.
In addition, $S_{c}(f)$ is the effective noise source resulted from unresolved galactic binaries \citep{Cornish2017,Babak2021} and can be alleviated by the increase of observation time. In our calculations, we apply the $S_{c}(f)$ calculated by \citet{Cornish2017}. 

%The calculation of $S_{\mathrm{n}}(f)$ above is suitable for sky-averaged galactic binaries. Now we eliminate the inclination averaged factor $4/5$ \citep[Equation (16)]{robson2019construction} which is also added to the calculation of total power \citep[Equation (6)]{peters1963gravitational} and sky/polarization averaged factor $\mathcal{R}(f)$ , then $S_{n}$ will be redefined as 
%
%\begin{equation}
%    S_{n}=\frac{4}{5\mathrm{R}} (P_{n}+S_{c}\mathcal{R})
%\end{equation}

After the analyses of signal and noise, we can calculate the sky, inclination and polarization averaged S/N \citep{Wagg2022b} 
%which is abbreviated as $\rho$ below, 
\begin{equation}
\label{eq: S/N_all}
    \langle{\rm S/N}\rangle_{(\theta, \phi, \psi, \iota)}^2=\sum_{n=1}^{\infty}\left\langle\rm (S/N)_{n}\right\rangle^{2}=\sum_{n=1}^{\infty} \int_{0}^{\infty} d f_{n} \frac{h_{c, n}^{2}}{f_{n}^{2} S_{\mathrm{n}}\left(f_{n}\right)}.
\end{equation}
The equation above for the $\langle\rm S/N \rangle$ is suitable whether the binary is circular or eccentric and stationary or non-stationary.

It is noted that the amplitude modulation derived by \citet{Cornish2003} is just suitable for quasi-circular systems. For WDWD systems detected by LISA, it is proper to add the amplitude modulation when calculating S/N. While for eccentric binaries with BH and/or NS components, we take the averaged value of $A_{\mathrm{mod}}$. After calculation, we find that the ratio of S/N with amplitude modulation to S/N without amplitude modulation is about $0.6$ for the same circular source.

\begin{figure*}[htbp]
\centering
\includegraphics[width=15cm]{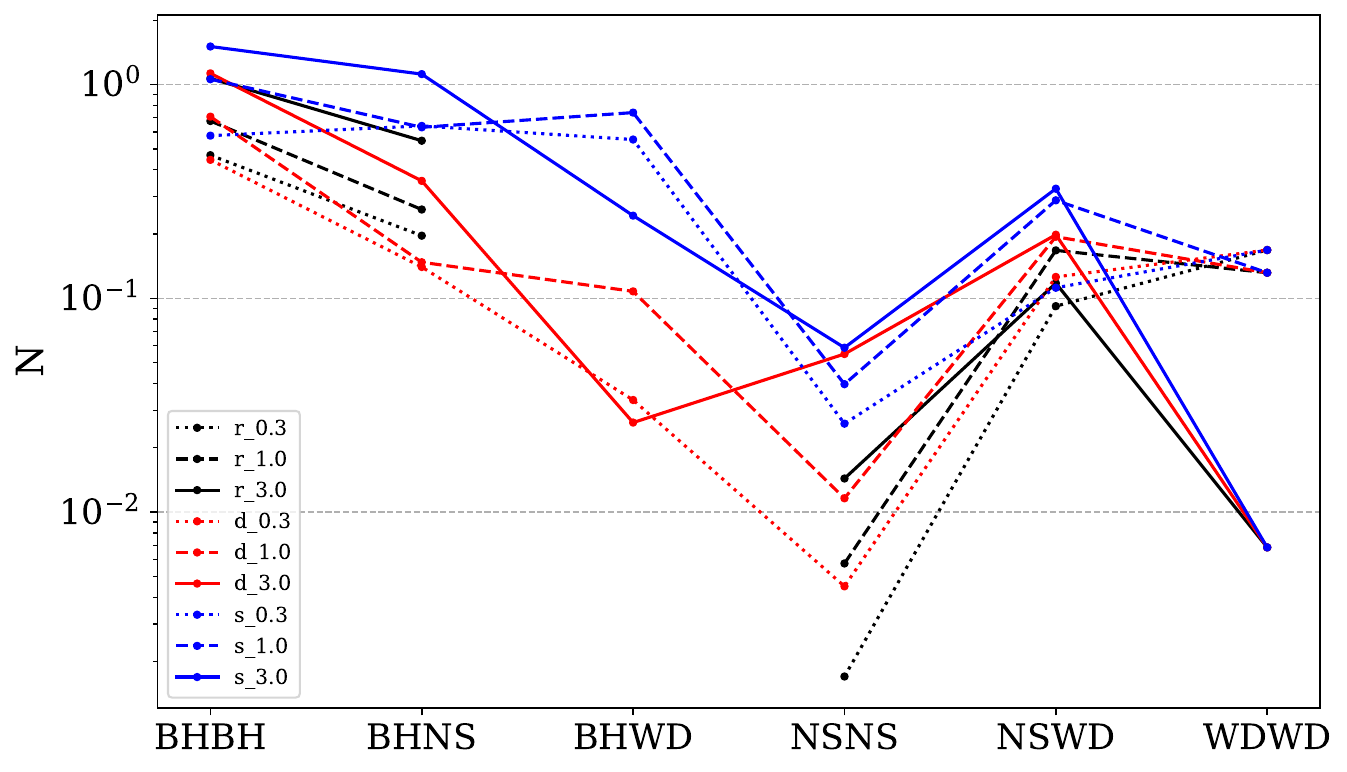}
\includegraphics[width=15cm]{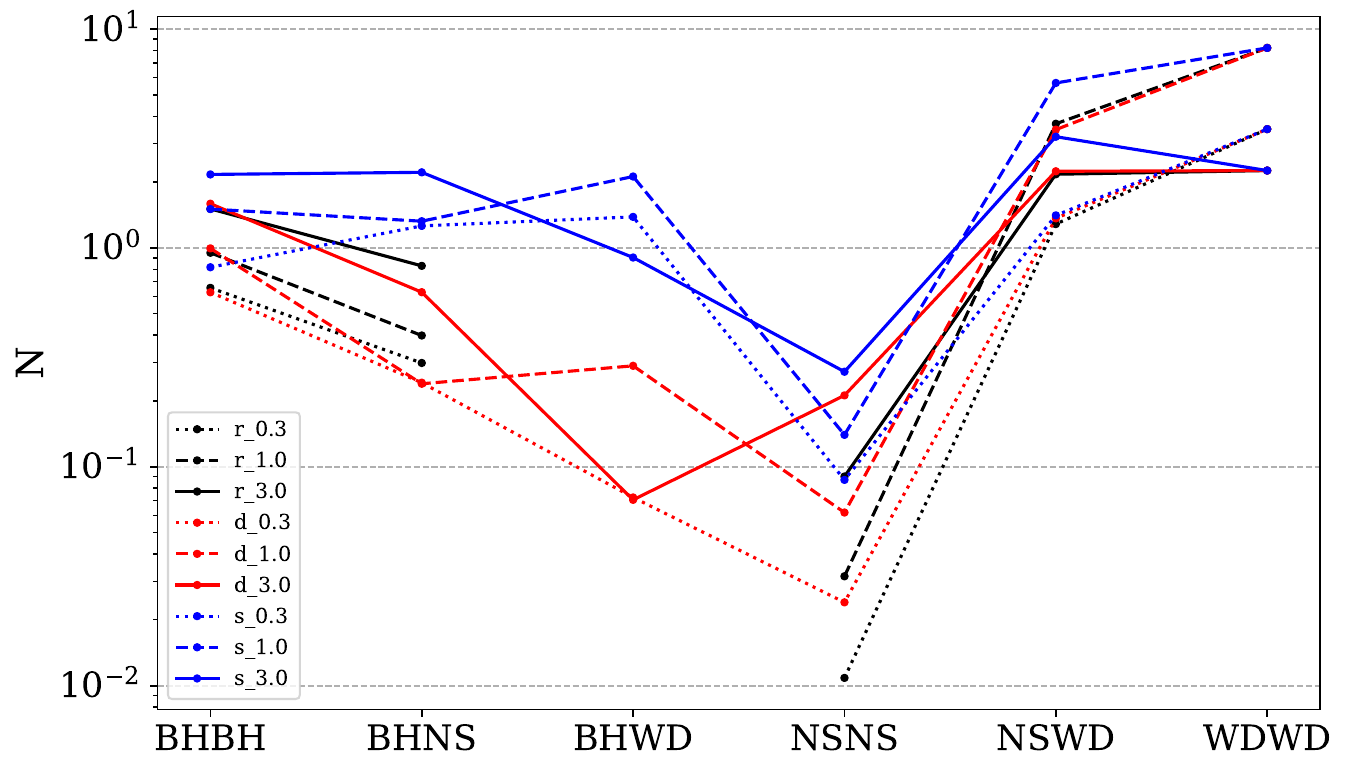}
\caption{Calculated numbers of various types of LISA binaries in M31 under different models. The top and bottom panels represent instrument's mission duration of 4 and 10 yr, respectively. In each panel, the black, red and blue curves correspond to the rapid, delay and stochastic supernova mechanisms, and the dotted, dashed and solid curves correspond to $\mathrm{\alpha_{CE}=0.3}$, 1.0 and 3.0, respectively.} 
\label{fig:num}
\end{figure*}

\section{Result} \label{sec:result}

\subsection{Overall Populations}

After analysing our obtained data, we find that the most common evolutionary path for the formation of LISA DCO binaries involves the canonical channel proposed by the reviews of \citet{Bhattacharya1991} and \citet{Tauris2023}. Evolved from a primordial binary, the primary star firstly evolves to fill its Roche lobe and stably transfers its envelope to the secondary star. After mass transfer, the primary star becomes a helium star and then evolves to be the first compact object. When the secondary star also evolves to fill its own Roche lobe, the binary enters the X-ray binary phase. Since the mass ratio between the secondary star and the compact star is relatively large, the mass transfer is dynamically unstable leading to the CE phase. During CE evolution, the secondary star fully loses its hydrogen envelope to become another helium star and finally evolves to be the second compact object. This CE phase leads to significant orbit decay of the binary so that the formed DCO is able to appear as a GW source. In some cases, the formation of tight DCO binaries with BH components does not involve this CE phase and the mass transfer is stable during the X-ray binary stage \citep[see also e.g.,][]{Pavlovskii2017,Shao2021}. According to our calculations, the formation of LISA BHBH and BHNS binaries can be dominated by the channel of stable Roche lobe overflow if one takes $\alpha_{\rm{CE}} = 0.3$ and $\alpha_{\rm{CE}} = 1.0$. In the $\alpha_{\rm{CE}} = 3.0$ case, all kinds of LISA DCO systems are predominantly formed through the channel of CE evolution.

Figure \ref{fig:num} presents estimated numbers of various types of DCOs in M31 that can be detected by LISA for a mission duration of  4 (top panel) or 10 yr (bottom panel), by assuming different models related to CE ejection efficiencies and supernova-explosion mechanisms. These numbers are also given in the Appendix. We can see that, for a 4 yr observation duration, at most $ \sim 3 $ DCO systems are likely to be identified by LISA. In this case, the systems with BH components dominate overall GW sources albeit in small numbers. For a 10 yr observation duration, the total number of LISA sources significantly increases to nearly a dozen. The numbers of the LISA binaries with BH components show a slight increase, while the numbers of the systems with WD components (i.e. WDWDs and NSWDs) increase by a factor of $\gtrsim 10$. 
%The reason is that the GW binaries with smaller chirp masses are more susceptible to the signal-to-noise ratios of sources. 
For the rapid supernova mechanism (indicated by black dots in the diagram), no BHWD binaries can form within the LISA band. As this mechanism is only able to produce BHs with masses  of $> 5 M_{\odot}$ and the masses of WD's progenitors are less than $\sim 8 M_{\odot}$, the mass ratios of the WD's progenitors to the BHs are so small that mass transfer occurs via stable Roche lobe overflow rather than CE evolution, eventually leading to the formation of long-period BHWD binaries \cite[see also][]{Shao2021}. Overall, the numbers of different types of LISA sources are affected by adopted CE ejection efficiencies and supernova-explosion mechanisms. There is a common feature that stochastic supernova mechanism tends to produce more LISA sources as relatively small natal kicks are imparted to NSs and BHs \citep{Mandel2020}.   

\begin{figure}[htbp]
\centering
\includegraphics[width=8.5cm]{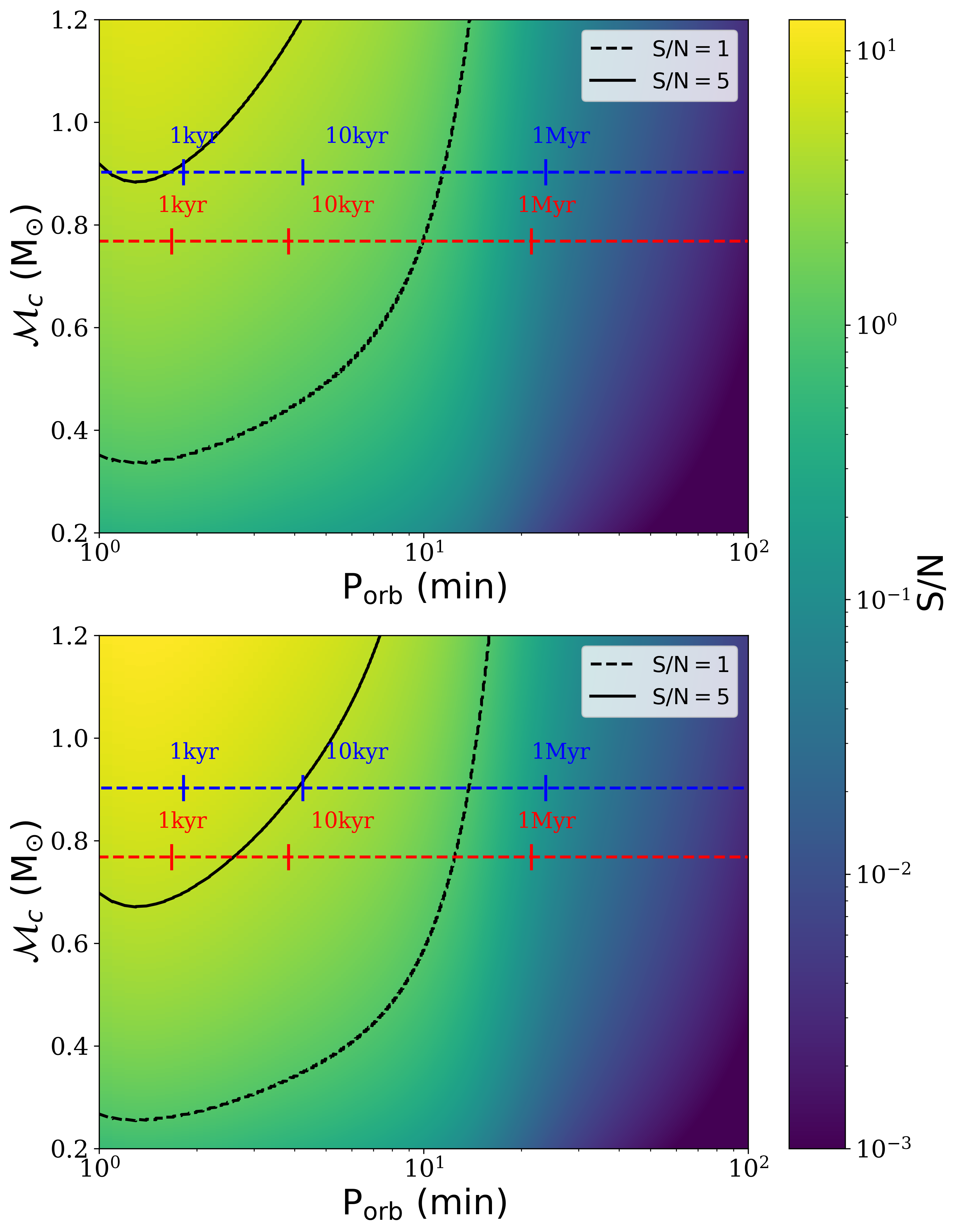}
\caption{S/N for circular binaries with different orbital periods and chirp masses when $T_{\rm obs}=4\rm\,yr$  (top panel) and $T_{\rm obs}=10\rm\,yr$ (bottom panel). The black solid and black dashed lines represent SNR = 5 and SNR = 1, respectively. The red dashed line denotes the evolutionary track of typical WDWD systems with different merger times marked. The blue dashed line shows a similar track but for typical NSWD systems.}
\label{fig:SNR}
\end{figure}

Figure \ref{fig:SNR} shows the $\rm S/N$ distribution of M31 LISA sources as a function of orbital periods and chirp masses for 4 (top panel) or 10 yr (bottom panel) observation duration. The black solid curve gives a boundary of $\rm S/N = 5$, indicating that only the DCO systems located
above this curve are considered detectable by LISA. The blue and red dashed lines correspond to the chirp masses of typical NSWD and WDWD systems, respectively. 
%Figure \ref{fig:SNR} show the S/N under different orbital periods and chirp masses of circular binaries for the 4-yr observation (top panel) and 10-yr observation (bottom panel), respectively. The black solid curve denotes S/N = 5 and DCOs in the area above the curve can be detected by LISA. The blue dotted line corresponds to the characteristic chirp mass of NSWD systems, while the red dotted line is for counterparts of WDWD binaries. 
In the case with $T_{\rm obs}=4\rm\,yr$, the red dashed line is below the black solid curve, indicating a lack of detectability of WDWD binaries. In addition, the blue dashed line is very close to the black solid curve. This means that the maximal $\rm S/N$ of merging NSWD binaries can just reach LISA's detection threshold. In the case with $T_{\rm obs}=10\rm\,yr$, we see that the regions where WDWD and NSWD systems can be detected are obviously enlarged. 
%Among whole LISA source population, WDWD systems dominate and their numbers can only be influenced by CE ejection efficiencies. 
By contrast, the detection of the systems with BH components is not sensitive to the options of instrument's observation lifetime since these systems have relatively large chirp masses. 

Considering that a number of GW sources are expected to be observed by LISA for a 10 yr mission lifetime, we will analyse their parameter distributions in the following.

\begin{figure*}[htbp]
\centering
\includegraphics[width=16cm]{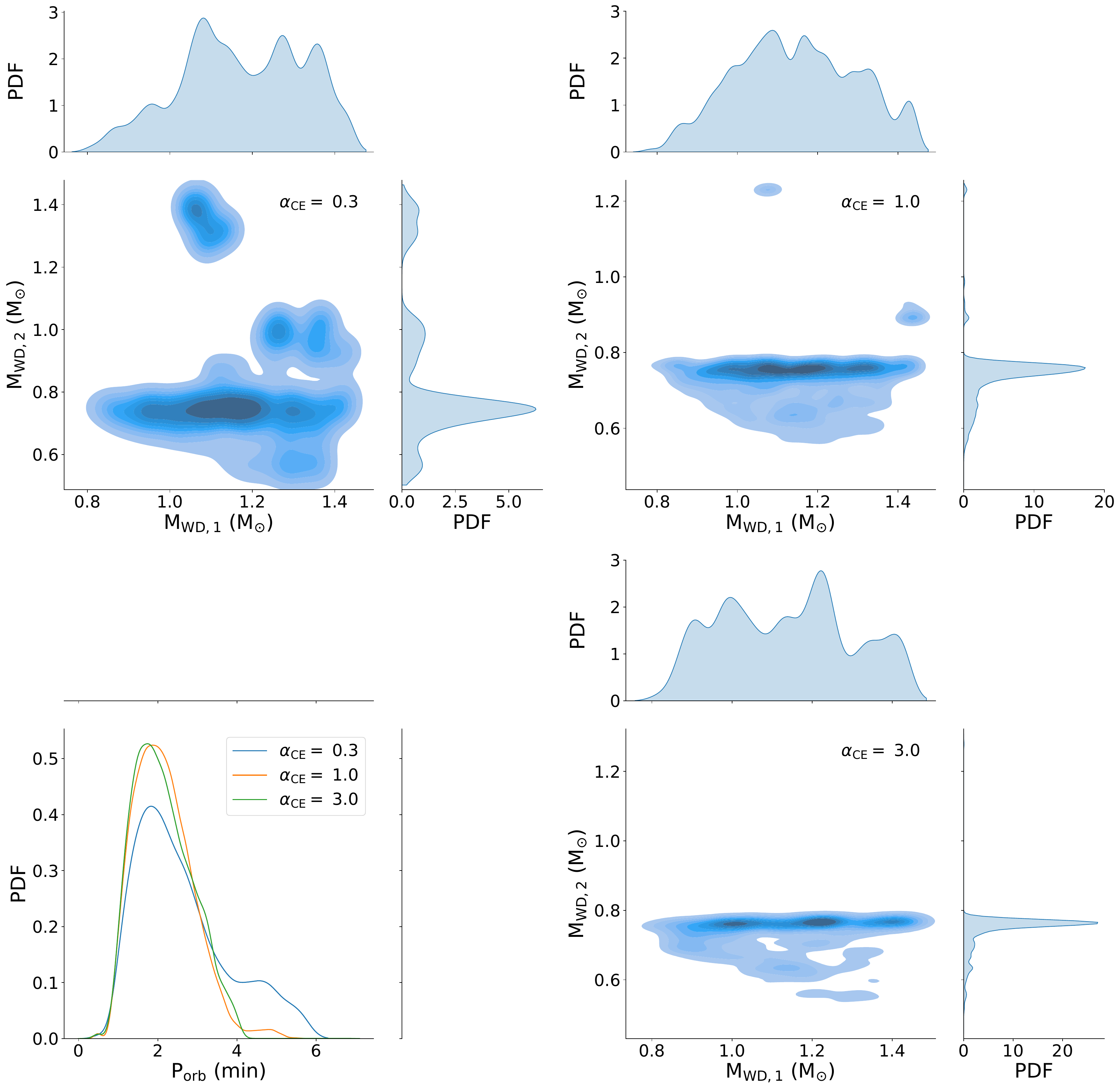}
\caption{Probability density functions (PDFs) of M31 LISA WDWDs as a function of component masses and orbital periods, by assuming different CE ejection efficiencies. The left upper, right upper and right lower panels correspond to $\alpha_{\rm{CE}} = 0.3$, $\alpha_{\rm{CE}} = 1.0$ and $\alpha_{\rm{CE}} = 3.0$, respectively.}
\label{fig:distribution of WDWD binaries}
\end{figure*}

\subsection{The case of WDWD binaries}

Figure \ref{fig:distribution of WDWD binaries} presents the probability density functions of M31 LISA WDWDs as a function of component masses and orbital periods. Since WDWD systems do not undergo supernova explosions during their formation, we discuss the calculated results with different CE ejection efficiencies.  Our results show that almost all of these LISA sources consist of CO WDs and/or ONe WDs. Note that the Galactic WDWD sample identified via electromagnetic observations mostly include He WDs \citep[][and references therein]{Korol2022}. These binaries, characterized by small chirp masses, would not
be detectable as M31 LISA sources. Under the assumption of different $\alpha_{\mathrm{CE}}$, the masses of the secondary WDs are distributed with a peak at $\sim 0.7-0.8M_{\odot}$ and the masses of  the primary WDs have a relatively flat distribution varying in the range of  $\sim 0.8-1.4 M_{\odot}$.  Importantly, these WDWD systems possess total masses exceeding the Chandrasekhar limit,  indicating they are promising progenitors of type Ia supernovae. 
%WDWD systems consisting of Helium (He) WDs is less than $1 \%$.
%Considering the independence of WD formation on the SN model, we just show the results under delayed model. The first three panels are the distribution of component masses corresponding to $\alpha_{\mathrm{CE}} = 0.3$, $\alpha_{\mathrm{CE}} = 1.0$ and $\alpha_{\mathrm{CE}} = 3.0$, respectively. The forth panel gives the distribution of orbital period for various $\alpha_{\mathrm{CE}}$.

%On the whole, the primary masses are mostly above $1.0 M_{\odot}$, while the secondary masses are distributed with a peak at $\sim 0.8M_{\odot}$ under different $\alpha_{\mathrm{CE}}$. Almost all WDWD binaries are consisting of carbon-oxygen (CO) WDs or oxygen-neon (ONe) WDs. Our result indicates that WDWD systems consisting of Helium (He) WDs is less than $1 \%$.

For $\alpha_{\mathrm{CE}} = 0.3$,  there is a remarkable subgroup of systems consisting of  a $\sim 1.0-1.2 M_{\odot}$ ONe WD and a $\sim 1.2-1.4 M_{\odot}$ CO WD. These systems are evolved from the primordial binaries with both component masses of $ \sim 6-8 M_{\odot}$ in $\sim 20$ d orbits. When $\alpha_{\mathrm{CE}} = 1.0$ or 3.0, the primordial binaries with similar parameters always evolve into wide WDWD systems that do not contribute to the population of LISA sources. 

\begin{figure*}[htbp]
\centering
\includegraphics[width=16cm]{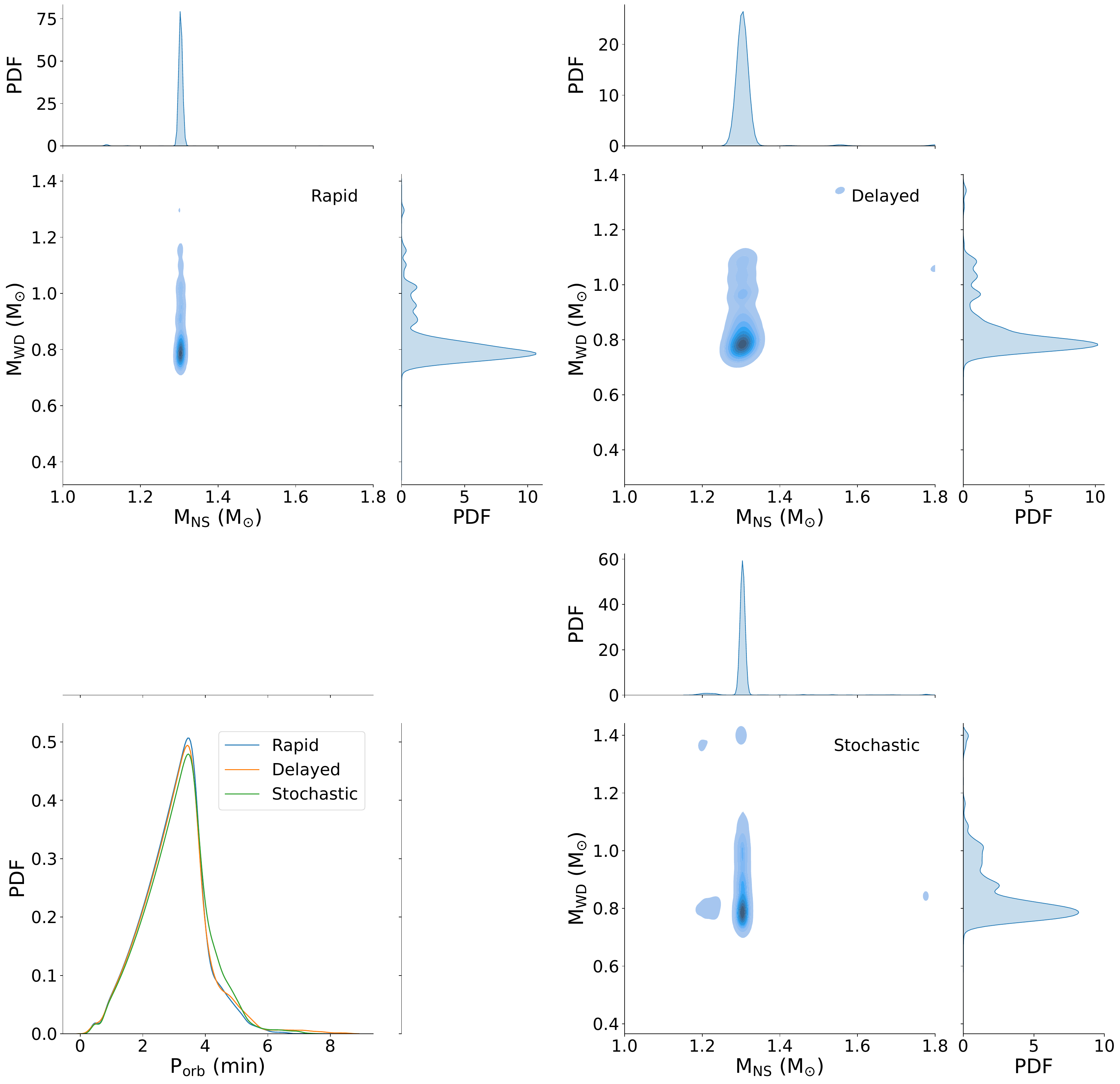}
\caption{Probability density functions (PDFs) of M31 LISA NSWDs as a function of component masses and orbital periods under the assumption of $\alpha_{\mathrm{CE}}=1.0$. The left upper, right upper and right lower panels correspond to the rapid, delayed and stochastic supernova mechanisms, respectively.}
\label{fig:distribution of NSWD binaries}
\end{figure*}

The orbital periods of WDWD binaries are always less than 6  min (see also Figure \ref{fig:SNR}) and distributed with a peak at 2 min. Note that the $\alpha_{\mathrm{CE}}=0.3$ scenario leads to a bulge at $\sim 4-6$  min in the curve of orbital period distribution, corresponding to the formation of very massive WDWD systems as mentioned above. The orbital period distributions of post-CE binaries are expected to be influenced by the CE ejection efficiencies. However,  it should be noted that very close systems detectable by LISA have undergone significant orbital shrinkage due to GW radiation. Therefore, different assumptions of $\alpha_{\mathrm{CE}}$  do not significantly alter the shape of the orbital period distributions. For other types of LISA DCO systems, we focus on the role of supernova-explosion mechanisms in their parameter distributions and take $\alpha_{\mathrm{CE}}=1.0$. 

% According the mass-radius relation of WD, lighter white dwarf would be the donor star if mass transfer take place. Binaries with the same accretor masses but larger donor masses have larger maximum mass transfer rates \citep{chen2022evolution}. It seems that the distribution of second mass should have been drop as mass increase, but as said before that higher chirp mass will also result more LISA systems. As modeled by \citet{chen2022evolution}, 

% The smaller $\alpha_{\mathrm{CE}}$, the higher chirp mass so the larger the longest orbital period. Orbital period will reduce in the wake of the radiation of gravitational wave, and then mass transfer would occur. Generally, mass transfer has a potential to widen the orbital when mass ratio is less than one. Under the influence of both, binary may leave a loop on the LISA sensitivity curve diagram. But we consider interacting system and low chirp mass binaries aren't our object so the further discussion about this can be put later. 

\begin{figure*}[htbp]
\centering
\includegraphics[width=16cm]{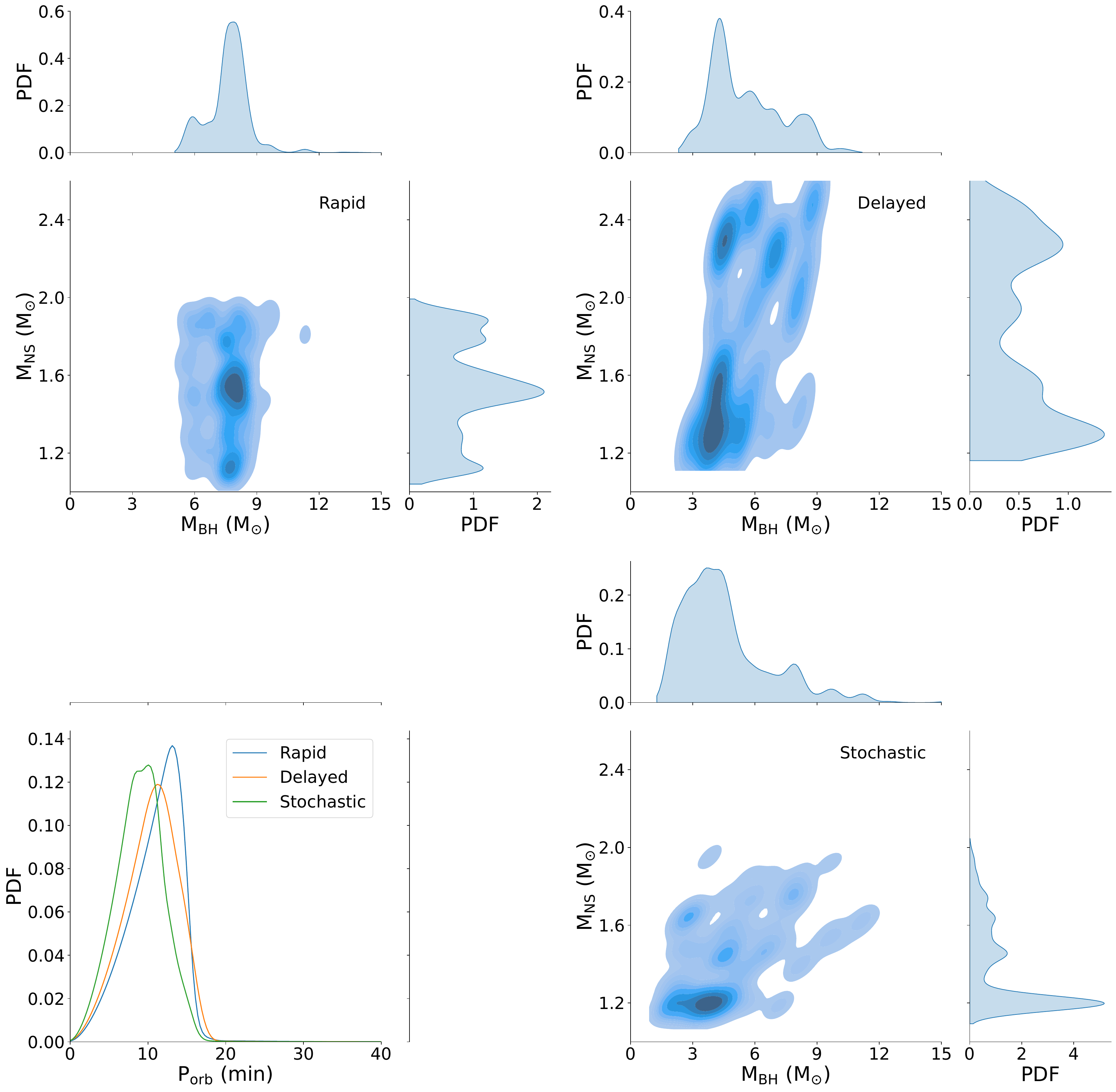}
\caption{Similar to Figure \ref{fig:distribution of NSWD binaries} but for LISA BHNS systems in M31. }
\label{fig:distribution of BHNS binaries}
\end{figure*}

\subsection{The case of NSWD binaries}
In Figure \ref{fig:distribution of NSWD binaries}, we show the probability density functions of M31 LISA NSWDs as a function of component masses and orbital periods under the assumption of $\alpha_{\mathrm{CE}}=1.0$. It can be clearly seen that  different supernova mechanisms we adopted tend to produce similar parameter distributions for LISA NSWD systems. The NS masses are distributed with a sharp peak at $\sim 1.3M_{\odot}$, and the WD masses are mainly distributed in the range of $\sim 0.7-1.2M_{\odot}$ with a peak at $\sim 0.8M_{\odot}$. The orbital periods of these binaries are narrowly distributed within the range of $\lesssim 6$ min with a peak at $\sim 4$ min, which are greatly lower than the minimal orbital period of $ \sim 0.1 $ d for Galactic WD binaries with a pulsar companion \citep{Manchester2005}.

For these LISA sources, almost all NSs  are formed via electron-capture rather than core-collapse supernovae.  For core-collapse NSs, the resulting NS binaries with  main-sequence companions of intermediate mass ($\sim 2-7M_\odot$) have orbital periods of $\lesssim 100$ d and most of them have orbital periods of less than a few days; while for electron-capture NSs,  the corresponding systems possess orbital periods of $\lesssim 1000$ d and the majority are relatively wide binaries \citep[see Figure~2 of][]{Shao2015}. It is possible that the former binaries with relatively close orbits undergo mergers in subsequent CE phases, avoiding the formation of tight NSWD binaries.

%\subsection{The case of BHNS binaries}
%Figure \ref{fig:distribution of BHNS binaries} presents the probability density distributions of M31 LISA BHNSs as a function of component masses and orbital periods. We discuss the calculated results with different supernova models and $\alpha_{\mathrm{CE}}=1$.

\subsection{The case of BHNS/BHBH binaries}

\begin{figure*}[htbp]
\centering
\includegraphics[width=16cm]{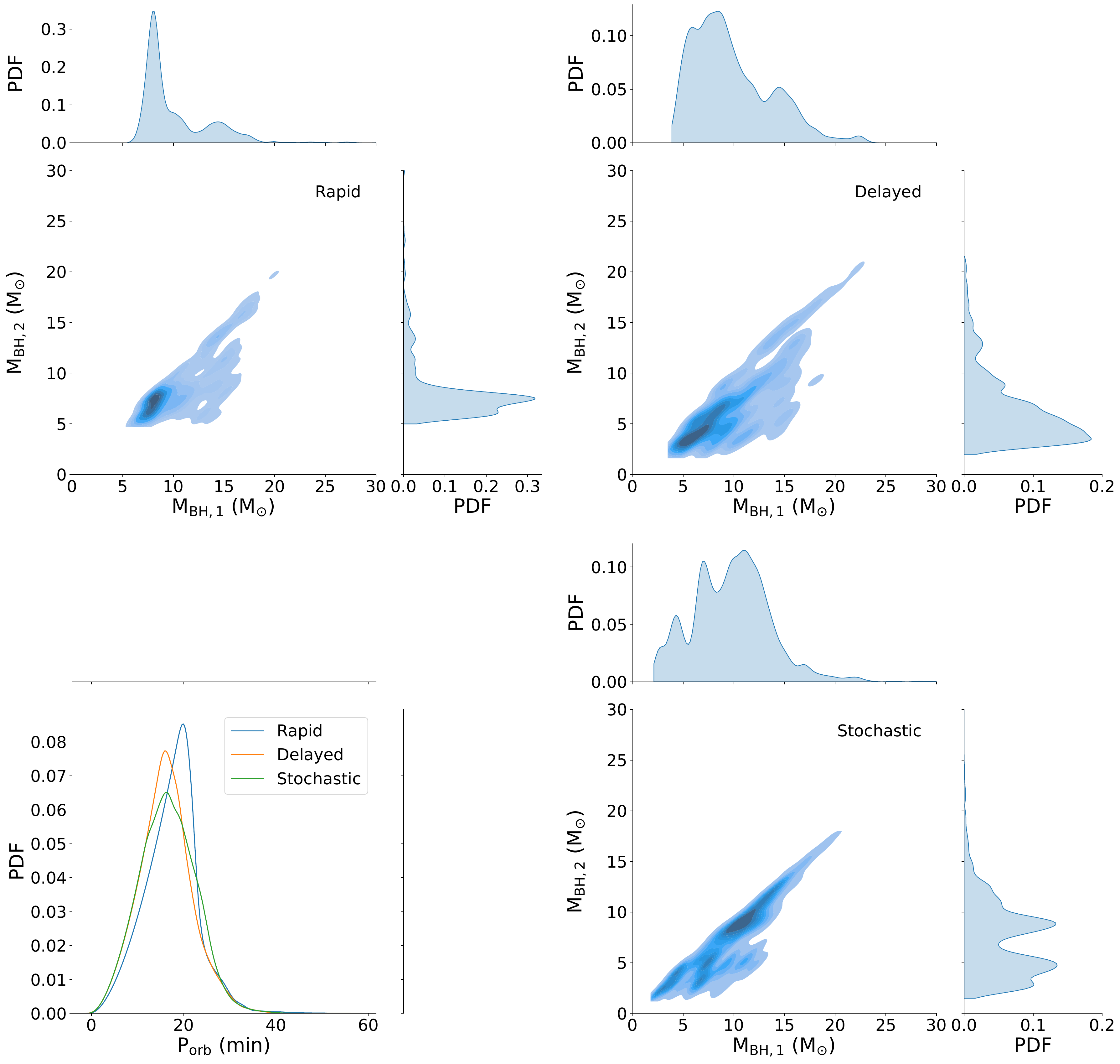}
\caption{Similar to Figure \ref{fig:distribution of NSWD binaries} but for LISA BHBH systems in M31.}
\label{fig:distribution of BHBH binaries}
\end{figure*}

Figures \ref{fig:distribution of BHNS binaries}  and \ref{fig:distribution of BHBH binaries} present the probability density functions of LISA BHNS and BHBH systems in M31, respectively,  as a function of component masses and orbital periods. Here the CE ejection efficiency is taken to be 1.0.
%To research on the effect of SN model, we analyze the result under $\alpha_{\mathrm{CE}} = 1.0$. The first three panels are the distribution of component masses corresponding to the rapid, delayed and stochastic model, respectively. The forth panel gives the distribution of orbital period for various SN model.
Since the rapid supernova mechanism allows the existence of  a $ 2-5$ mass gap between NSs and BHs, it predicts that the BH masses in both BHNS and BHBH systems are always larger than $5M_{\odot}$ and distributed with a peak at $\sim 7-8M_{\odot}$. On the contrary, the delayed and stochastic mechanisms are likely to form the systems with low-mass ($<5M_\odot$) BH components.
%As the rapid model predicted, no BHs with mass below $ 5M_{\odot}$ can form for this model. While results for the other two models do not suggest an existing mass gap. The mass of primary BH peaks at the $7-8 ~M_{\odot}$ both for the rapid and delayed model. While counterparts of stochastic models have a peak at $11-12 ~M_{\odot}$. Compared with WDWD binaries, BHBH systems have a long orbital period around twenty minutes for their lager chirp masses.
Compared with the LISA sources with WD components,  the BHNS (BHBH) systems  have wider orbital period distributions in the range of $\lesssim 20$ min ($\lesssim 40$ min) with peaks at $\sim 10$ min ($\sim 20$ min).

%\subsection{Distribution of GW characteristics}

%We also research the characteristic distribution of GW detected by LISA. The GW signal from BH-CSs are distributed around $\mathrm{mHz}$. While radiation of WDWD binaries tend to peak at the vicinity of $10 \mathrm{~mHz}$ which corresponds to orbital period of three minutes. Meanwhile, the characteristic strain amplitude of GW from WDWD systems are lower than BH-CSs. 

%Actually, some GW coming from BHBH binaries with period of thousands minutes also can be detected due to the extreme eccentricity of BHBH systems. But these system are quite few so we neglect them.

%Then we research on the characteristics of GW signal detected by LISA. Figure \ref{fig:distribution of GW} shows the distribution of the GW frequency $f$ and strain characteristic amplitude $h_{\mathrm{c}}$ for various binaries under the delayed SN model and $\alpha_{\mathrm{CE}} = 1$.
%
%\begin{figure*}[htbp]
%\centering
%\includegraphics[width=18cm]{distribution_GW.png}
%\caption{Distribution of GW}
%\label{fig:distribution of GW}
%\end{figure*}
%
%It is obvious in Figure \ref{fig:distribution of GW} that GW from BH-CSs are distributed around $\mathrm{mHz}$. While radiation of DWDs tend to peak at the vicinity of $10 \mathrm{~mHz}$ which corresponds to orbital period of $3 \mathrm{~minutes}$. Meanwhile, the characteristic strain amplitude of GW from DWDs are lower than BH-CSs. Moreover we find that the number of BNSs is the smallest and all these systems are born in high metallicity environments.

\section{Discussion} \label{sec:discussion}

For LISA WDWD systems in M31, \citet{Korol2018} proposed that 17 (60) such sources with $\rm S/N \geq 7$ can be detected for 4 (10) yr observation lifetime, which are significantly larger than our obtained. A main reason for this  discrepancy  may be attributed to the calculation of $\rm S/N$. It is known that theoretical predictions for the detection of LISA WDWD binaries in M31 are very sensitive to the magnitude of $\rm S/N$ (see Figure \ref{fig:SNR}). Since the adopted values of $P_{\mathrm{OMS}}$ (from \citet{AS2017} and \citet{Robson2019}, respectively) in the effective power spectral density of the noise $S_{\mathrm{n}}(f)$ differ from each other by a factor of 1.5, the S/N calculated by \citet{Korol2018} is significantly higher than ours. In addition, \citet{Korol2018} made some physical assumptions that differ from our model: e.g. they assumed the total stellar mass of M31 to be twice that of
the Milky Way and  extrapolated
the properties of  Galactic WDWD population to that of M31;  they adopted the Kroupa's IMF and set the binary fraction to be $50\%$;  for CE evolution, they used a different prescription proposed by \citet{Nelemans2001} that the orbital evolution of a binary system is assumed to be caused by angular momentum loss due to mass outflow. Checking the influence of these different assumptions requires a more in-depth study of LISA WDWD sources, which is beyond the scope of this paper. 

Among all LISA sources, we predict that the number of NSNS binaries is the smallest, indicating that it is difficult to detect NSNS systems in M31 even during a 10 yr observation mission. Using the inferred merger rate density of $ 1540\rm\, Gpc^{-3}\, yr^{-1}$ from the detection of GW170817 \citep{Abbott2017} and combining it with  a simple relation to estimate the merger rate of NSNS systems in M31, \citet{Seto2019} suggested that LISA might identify $\sim 5 $ NSNS binaries in M31 with $\rm S/N > 10$. There is a significant difference between their predictions and ours. On the one hand, the recently estimated merger rate density of NSNS systems is between $10\rm\, Gpc^{-3}\, yr^{-1}$ and $1700\rm\, Gpc^{-3}\, yr^{-1}$ \citep{LVK2023}, which is still subject to a big uncertainty with probably a very low value. On the other hand, the overall distribution of orbital parameters for Galactic NSNS binaries implies that the small natal kicks are imparted to the second-born NSs \citep[e.g.,][]{Tauris2017,Shao2018}, we do not include an additional assumption on the natal kicks of second-born NSs in our current work.

% The main reason is that \citet{Korol2018} applied the lower noise spectral density $S_{\mathrm{n}}(f)$ which may attribute to the lower $P_{\mathrm{OMS}}$ \citep{AS2017}. While we adopt the higher value for $P_{\mathrm{OMS}}$ \citep{LISA2018} and follow the convention of more recent paper \citep{Robson2019}. As a consequence, the S/N obtained from the former is $\sim 1.5$ times higher than that from the latter at around $\mathrm{10mHz}$ where most LISA WDWD systems are gathered. Although \citet{Korol2018} took the LISA detection threshold of 7, it is still easier to be detected in their work for the same WDWD system. So the number of LISA WDWDs we obtain is reduced by several times compared with \citet{Korol2018}.

%\citet{Korol2018} applied a rough value which is the proposed requirement in the LISA proposal for ESA L3 \citep{AS2017}, while we adopt the 2018 LISA mission-performance-requirement design parameters \citep{LISA2018}. Their specific formula for $S_{\mathrm{n}}(f)$ is unclear while we . 

%In addition, the number of NSWD systems source is second only to WDWD binaries when $\mathrm{T_{obs}=10yr}$. Roughly there are several NSWD systems to be observed by LISA through 10-yr observation. Generally, NSWD binaries has two types according the formation order \citep{Nelemans2001} . If NS forms first the orbital will circularize during mass transfer otherwise the system will be eccentric if NS forms later through accretion. Almost all NSWD binaries detected by LISA contain first-forming NS.

Our calculations show that $ \sim1-5 $ NSWD systems in M31 are expected to be observed by LISA for 10 yr instrument's lifetime. Among them, almost all NSs may  originate from electron-capture supernovae, with corresponding masses of $ \sim 1.3M_\odot $. 
Different from WDWD and NSWD systems, the LISA sources with BH components are more vulnerable to adopted supernova-explosion mechanisms. It is predicted that the stochastic mechanism can produce the largest population of these types of binaries, with $ \sim 2-3 $ likely being detected in M31 (see the Appendix). Furthermore, these GW sources are likely to host mass-gap ($\sim 2-5M_\odot$) BHs. 
%These features mentioned above can be used to test the mechanisms of supernova explosions.

Our estimations of the number of M31 LISA sources include some uncertainties and simplifications. For example, we allow the binaries with Hertzsprung gap donors to survive CE phases. However, it remains uncertain whether such binaries can survive CE phases or directly lead to mergers \citep[e.g.,][]{Dominik2012}, as Hertzsprung gap donors are not expected to have developed a steep density gradient between envelope and core \citep{Ivanova2004}. If assuming all binaries with Hertzsprung gap donors to merge during the CE evolution, we obtain that the total number of M31 DCO binaries detectable by LISA during its 10~yr observation mission will decrease by a factor of less than 2. In contrast to WDWD binaries, which have almost unchanged numbers, other types of binaries are obviously vulnerable to this assumption (especially in the $\alpha_{\mathrm{CE}} = 3.0$ case). Furthermore, we assume that all primordial binaries have circular orbits in our calculations. If we choose the orbital eccentricities from a thermal distribution between 0 and 1 \citep{Heggie1975}, the total number of M31 LISA sources will slightly increase (decrease) by a factor of  $ \sim 1.04-1.09$ ($\sim 0.85-0.98$).

For the treatment of CE evolution with the energy balance method, we assume that $\alpha_{\mathrm{CE}}$ has a fixed value for the calculation of a specific population. However, some authors \citep[e.g.,][]{Davis2012} have tried to connect $\alpha_{\mathrm{CE}}$ to the binary parameters such as component masses and orbital periods at the onset of CE evolution, and given some possible relations between the parameters of post-CE binaries. It is possible that these relations are extrapolated and incorporated in BPS codes to determine the $\alpha_{\mathrm{CE}}$ values precisely.

Recently, \cite{Fryer2022} introduced a new treatment of supernova mechanisms by incorporating a convection mixing parameter, $f_{\mathrm{mix}}$. Typically, $f_{\mathrm{mix}}$ ranges from $0.5$ to $4.0$, where $f_{\mathrm{mix}}=0.5$ and $f_{\mathrm{mix}}=4.0$ approximately correspond to the delayed and the rapid mechanisms, respectively. This new method aims to mitigate the mass gap. 
%That is to say the gap is no longer a step but a smooth rise although it may be a little steep. 
Future observation of LISA sources can help settle a possible supernova mechanism and give a proper parameter in the remnant mass prescription.

In our calculations, we only consider the formation of DCO systems through isolated binary evolution in M31. As expected, LISA may detect GW sources originating from globular clusters and these sources are more likely to form through dynamical interactions. 

\section{Conclusion} \label{sec:conclusion}

In this work, we have investigated the prospects of detecting DCO GW sources in M31 with a BPS method, by considering the star-formation history of the M31 disk and a range of evolutionary models. For a 4-yr mission duration, we expect that LISA can resolve $\sim 0.4-1.6 $ BHBHs, $\sim 0.1-1.2 $ BHNSs, $ <0.8 $ BHWDs, $ <0.1 $ NSNSs, $\sim 0.1-0.4 $ NSWDs, and $ <0.2 $ WDWDs. For a 10-yr mission duration, LISA may detect $\sim 0.6-2.2 $ BHBHs, $\sim 0.2-2.3 $ BHNSs, $ <2.2 $ BHWDs, $ <0.3 $ NSNSs, $\sim 1.2-5.7  $ NSWDs, and $ \sim 2.3-8.2 $ WDWDs. Since M31 has a distance of 780 kpc, only very tight DCOs can be observed by LISA. These binaries provide an important category of GW sources that are fast-merging.

We thank the anonymous referee for valuable suggestions that helped to improve our manuscript. This work was supported by the National Key Research and Development Program of China (2021YFA0718500), the Natural Science Foundation of China (Nos. 11973026, 12041301, 12121003), and the Project U1838201 supported by NSFC and CAS.

\appendix

\section{Numbers of M31 LISA binaries}

Table \ref{tab:num for 4yr} shows the expected numbers for various types of DCO binaries in M31 that can be resolved by LISA during its mission duration of 4 or 10 yr, by assuming different models related to supernova-explosion mechanisms and CE ejection efficiencies.

\begin{deluxetable*}{ccccccc}
\tablecaption{Expected numbers of  various types of LISA sources for a 4 (10) yr observation duration.}
\label{tab:num for 4yr}
\tablehead{
\colhead{Models} & 
\colhead{$N_{\rm BHBH}$} & \colhead{$N_{\rm BHNS}$} & 
\colhead{$N_{\rm BHWD}$} & \colhead{$N_{\rm NSNS}$} & 
\colhead{$N_{\rm NSWD}$} & \colhead{$N_{\rm WDWD}$}
}
\startdata
$\mathrm{r\_0.3}$ &0.47 (0.66)& 0.20 (0.30)& 0 (0) & 0 (0.01)& 0.09 (1.28)& 0.17 (3.49) \\
$\mathrm{r\_1.0}$ &0.68 (0.95) & 0.26 (0.40) & 0 (0)& 0.01 (0.03)& 0.17 (3.69)& 0.13 (8.20)\\
$\mathrm{r\_3.0}$ &1.07 (1.51)& 0.55 (0.83)& 0 (0)& 0.01 (0.09)& 0.12 (2.17)& 0.01 (2.26)\\
$\mathrm{d\_0.3}$ &0.44 (0.63)& 0.14 (0.24)& 0.03 (0.07)& 0 (0.02)& 0.13 (1.36)& 0.17 (3.49)\\
$\mathrm{d\_1.0}$ &0.71 (1.00)& 0.15 (0.24)& 0.11 (0.29)& 0.01 (0.06)& 0.19 (3.47)& 0.13 (8.20)\\
$\mathrm{d\_3.0}$ &1.13 (1.59)& 0.35 (0.63)& 0.03 (0.07)& 0.05 (0.21)& 0.20 (2.24)& 0.01 (2.26)\\
$\mathrm{s\_0.3}$ &0.58 (0.82)& 0.64 (1.26)& 0.55 (1.39)& 0.03 (0.09)& 0.11 (1.41)& 0.17 (3.49)\\
$\mathrm{s\_1.0}$ &1.06 (1.50)& 0.63 (1.32)& 0.74 (2.12)& 0.04 (0.14)& 0.29 (5.67)& 0.13 (8.20)\\
$\mathrm{s\_3.0}$ &1.51 (2.17)& 1.12 (2.21)& 0.24 (0.90)& 0.06 (0.27)& 0.33 (3.22)& 0.01 (2.26)\\
\enddata	
%\tablecomments{}
\end{deluxetable*}

\bibliography{article}{}

\begin{thebibliography}{}
\expandafter\ifx\csname natexlab\endcsname\relax\def\natexlab#1{#1}\fi
\providecommand{\url}[1]{\href{#1}{#1}}
\providecommand{\dodoi}[1]{doi:~\href{http://doi.org/#1}{\nolinkurl{#1}}}
\providecommand{\doeprint}[1]{\href{http://ascl.net/#1}{\nolinkurl{http://ascl.net/#1}}}
\providecommand{\doarXiv}[1]{\href{https://arxiv.org/abs/#1}{\nolinkurl{https://arxiv.org/abs/#1}}}

\bibitem[{{Abbott} {et~al.}(2016){Abbott}, {Abbott}, {Abbott}, {Abernathy},
  {Acernese}, {Ackley}, {Adams}, {Adams}, {Addesso}, {Adhikari}, {Adya},
  {Affeldt}, {Agathos}, {Agatsuma}, {Aggarwal}, {Aguiar}, {Aiello}, {Ain},
  {Ajith}, {Allen}, {Allocca}, {Altin}, {Anderson}, {Anderson}, {Arai},
  {Arain}, {Araya}, {Arceneaux}, {Areeda}, {Arnaud}, {Arun}, {Ascenzi},
  {Ashton}, {Ast}, {Aston}, {Astone}, {Aufmuth}, {Aulbert}, {Babak}, {Bacon},
  {Bader}, {Baker}, {Baldaccini}, {Ballardin}, {Ballmer}, {Barayoga},
  {Barclay}, {Barish}, {Barker}, {Barone}, {Barr}, {Barsotti}, {Barsuglia},
  {Barta}, {Bartlett}, {Barton}, {Bartos}, {Bassiri}, {Basti}, {Batch},
  {Baune}, {Bavigadda}, {Bazzan}, {Behnke}, {Bejger}, {Belczynski}, {Bell},
  {Bell}, {Berger}, {Bergman}, {Bergmann}, {Berry}, {Bersanetti}, {Bertolini},
  {Betzwieser}, {Bhagwat}, {Bhandare}, {Bilenko}, {Billingsley}, {Birch},
  {Birney}, {Birnholtz}, {Biscans}, {Bisht}, {Bitossi}, {Biwer}, {Bizouard},
  {Blackburn}, {Blair}, {Blair}, {Blair}, {Bloemen}, {Bock}, {Bodiya}, {Boer},
  {Bogaert}, {Bogan}, {Bohe}, {Bojtos}, {Bond}, {Bondu}, {Bonnand}, {Boom},
  {Bork}, {Boschi}, {Bose}, {Bouffanais}, {Bozzi}, {Bradaschia}, {Brady},
  {Braginsky}, {Branchesi}, {Brau}, {Briant}, {Brillet}, {Brinkmann},
  {Brisson}, {Brockill}, {Brooks}, {Brown}, {Brown}, {Brown}, {Buchanan},
  {Buikema}, {Bulik}, {Bulten}, {Buonanno}, {Buskulic}, {Buy}, {Byer},
  {Cabero}, {Cadonati}, {Cagnoli}, {Cahillane}, {Bustillo}, {Callister},
  {Calloni}, {Camp}, {Cannon}, {Cao}, {Capano}, {Capocasa}, {Carbognani},
  {Caride}, {Casanueva Diaz}, {Casentini}, {Caudill}, {Cavagli{\`a}},
  {Cavalier}, {Cavalieri}, {Cella}, {Cepeda}, {Baiardi}, {Cerretani},
  {Cesarini}, {Chakraborty}, {Chalermsongsak}, {Chamberlin}, {Chan}, {Chao},
  {Charlton}, {Chassande-Mottin}, {Chen}, {Chen}, {Cheng}, {Chincarini},
  {Chiummo}, {Cho}, {Cho}, {Chow}, {Christensen}, {Chu}, {Chua}, {Chung},
  {Ciani}, {Clara}, {Clark}, {Cleva}, {Coccia}, {Cohadon}, {Colla}, {Collette},
  {Cominsky}, {Constancio}, {Conte}, {Conti}, {Cook}, {Corbitt}, {Cornish},
  {Corsi}, {Cortese}, {Costa}, {Coughlin}, {Coughlin}, {Coulon}, {Countryman},
  {Couvares}, {Cowan}, {Coward}, {Cowart}, {Coyne}, {Coyne}, {Craig},
  {Creighton}, {Creighton}, {Cripe}, {Crowder}, {Cruise}, {Cumming},
  {Cunningham}, {Cuoco}, {Dal Canton}, {Danilishin}, {D'Antonio}, {Danzmann},
  {Darman}, {Da Silva Costa}, {Dattilo}, {Dave}, {Daveloza}, {Davier},
  {Davies}, {Daw}, {Day}, {De}, {DeBra}, {Debreczeni}, {Degallaix}, {De
  Laurentis}, {Del{\'e}glise}, {Del Pozzo}, {Denker}, {Dent}, {Dereli},
  {Dergachev}, {DeRosa}, {De Rosa}, {DeSalvo}, {Dhurandhar}, {D{\'\i}az}, {Di
  Fiore}, {Di Giovanni}, {Di Lieto}, {Di Pace}, {Di Palma}, {Di Virgilio},
  {Dojcinoski}, {Dolique}, {Donovan}, {Dooley}, {Doravari}, {Douglas},
  {Downes}, {Drago}, {Drever}, {Driggers}, {Du}, {Ducrot}, {Dwyer}, {Edo},
  {Edwards}, {Effler}, {Eggenstein}, {Ehrens}, {Eichholz}, {Eikenberry},
  {Engels}, {Essick}, {Etzel}, {Evans}, {Evans}, {Everett}, {Factourovich},
  {Fafone}, {Fair}, {Fairhurst}, {Fan}, {Fang}, {Farinon}, {Farr}, {Farr},
  {Favata}, {Fays}, {Fehrmann}, {Fejer}, {Feldbaum}, {Ferrante}, {Ferreira},
  {Ferrini}, {Fidecaro}, {Finn}, {Fiori}, {Fiorucci}, {Fisher}, {Flaminio},
  {Fletcher}, {Fong}, {Fournier}, {Franco}, {Frasca}, {Frasconi}, {Frede},
  {Frei}, {Freise}, {Frey}, {Frey}, {Fricke}, {Fritschel}, {Frolov}, {Fulda},
  {Fyffe}, {Gabbard}, {Gair}, {Gammaitoni}, {Gaonkar}, {Garufi}, {Gatto},
  {Gaur}, {Gehrels}, {Gemme}, {Gendre}, {Genin}, {Gennai}, {George}, {Gergely},
  {Germain}, {Ghosh}, {Ghosh}, {Ghosh}, {Giaime}, {Giardina}, {Giazotto},
  {Gill}, {Glaefke}, {Gleason}, {Goetz}, {Goetz}, {Gondan}, {Gonz{\'a}lez},
  {Castro}, {Gopakumar}, {Gordon}, {Gorodetsky}, {Gossan}, {Gosselin},
  {Gouaty}, {Graef}, {Graff}, {Granata}, {Grant}, {Gras}, {Gray}, {Greco},
  {Green}, {Greenhalgh}, {Groot}, {Grote}, {Grunewald}, {Guidi}, {Guo},
  {Gupta}, {Gupta}, {Gushwa}, {Gustafson}, {Gustafson}, {Hacker}, {Hall},
  {Hall}, {Hammond}, {Haney}, {Hanke}, {Hanks}, {Hanna}, {Hannam}, {Hanson},
  {Hardwick}, {Harms}, {Harry}, {Harry}, {Hart}, {Hartman}, {Haster},
  {Haughian}, {Healy}, {Heefner}, {Heidmann}, {Heintze}, {Heinzel}, {Heitmann},
  {Hello}, {Hemming}, {Hendry}, {Heng}, {Hennig}, {Heptonstall}, {Heurs},
  {Hild}, {Hoak}, {Hodge}, {Hofman}, {Hollitt}, {Holt}, {Holz}, {Hopkins},
  {Hosken}, {Hough}, {Houston}, {Howell}, {Hu}, {Huang}, {Huerta}, {Huet},
  {Hughey}, {Husa}, {Huttner}, {Huynh-Dinh}, {Idrisy}, {Indik}, {Ingram},
  {Inta}, {Isa}, {Isac}, {Isi}, {Islas}, {Isogai}, {Iyer}, {Izumi}, {Jacobson},
  {Jacqmin}, {Jang}, {Jani}, {Jaranowski}, {Jawahar}, {Jim{\'e}nez-Forteza},
  {Johnson}, {Johnson-McDaniel}, {Jones}, {Jones}, {Jonker}, {Ju}, {Haris},
  {Kalaghatgi}, {Kalogera}, {Kandhasamy}, {Kang}, {Kanner}, {Karki},
  {Kasprzack}, {Katsavounidis}, {Katzman}, {Kaufer}, {Kaur}, {Kawabe},
  {Kawazoe}, {K{\'e}f{\'e}lian}, {Kehl}, {Keitel}, {Kelley}, {Kells},
  {Kennedy}, {Keppel}, {Key}, {Khalaidovski}, {Khalili}, {Khan}, {Khan},
  {Khan}, {Khazanov}, {Kijbunchoo}, {Kim}, {Kim}, {Kim}, {Kim}, {Kim}, {Kim},
  {King}, {King}, {Kinzel}, {Kissel}, {Kleybolte}, {Klimenko}, {Koehlenbeck},
  {Kokeyama}, {Koley}, {Kondrashov}, {Kontos}, {Koranda}, {Korobko}, {Korth},
  {Kowalska}, {Kozak}, {Kringel}, {Krishnan}, {Kr{\'o}lak}, {Krueger}, {Kuehn},
  {Kumar}, {Kumar}, {Kuo}, {Kutynia}, {Kwee}, {Lackey}, {Landry}, {Lange},
  {Lantz}, {Lasky}, {Lazzarini}, {Lazzaro}, {Leaci}, {Leavey}, {Lebigot},
  {Lee}, {Lee}, {Lee}, {Lee}, {Lenon}, {Leonardi}, {Leong}, {Leroy},
  {Letendre}, {Levin}, {Levine}, {Li}, {Libson}, {Littenberg}, {Lockerbie},
  {Logue}, {Lombardi}, {London}, {Lord}, {Lorenzini}, {Loriette}, {Lormand},
  {Losurdo}, {Lough}, {Lousto}, {Lovelace}, {L{\"u}ck}, {Lundgren}, {Luo},
  {Lynch}, {Ma}, {MacDonald}, {Machenschalk}, {MacInnis}, {Macleod},
  {Maga{\~n}a-Sandoval}, {Magee}, {Mageswaran}, {Majorana}, {Maksimovic},
  {Malvezzi}, {Man}, {Mandel}, {Mandic}, {Mangano}, {Mansell}, {Manske},
  {Mantovani}, {Marchesoni}, {Marion}, {M{\'a}rka}, {M{\'a}rka}, {Markosyan},
  {Maros}, {Martelli}, {Martellini}, {Martin}, {Martin}, {Martynov}, {Marx},
  {Mason}, {Masserot}, {Massinger}, {Masso-Reid}, {Matichard}, {Matone},
  {Mavalvala}, {Mazumder}, {Mazzolo}, {McCarthy}, {McClelland}, {McCormick},
  {McGuire}, {McIntyre}, {McIver}, {McManus}, {McWilliams}, {Meacher},
  {Meadors}, {Meidam}, {Melatos}, {Mendell}, {Mendoza-Gandara}, {Mercer},
  {Merilh}, {Merzougui}, {Meshkov}, {Messenger}, {Messick}, {Meyers},
  {Mezzani}, {Miao}, {Michel}, {Middleton}, {Mikhailov}, {Milano}, {Miller},
  {Millhouse}, {Minenkov}, {Ming}, {Mirshekari}, {Mishra}, {Mitra},
  {Mitrofanov}, {Mitselmakher}, {Mittleman}, {Moggi}, {Mohan}, {Mohapatra},
  {Montani}, {Moore}, {Moore}, {Moraru}, {Moreno}, {Morriss}, {Mossavi},
  {Mours}, {Mow-Lowry}, {Mueller}, {Mueller}, {Muir}, {Mukherjee}, {Mukherjee},
  {Mukherjee}, {Mukund}, {Mullavey}, {Munch}, {Murphy}, {Murray}, {Mytidis},
  {Nardecchia}, {Naticchioni}, {Nayak}, {Necula}, {Nedkova}, {Nelemans},
  {Neri}, {Neunzert}, {Newton}, {Nguyen}, {Nielsen}, {Nissanke}, {Nitz},
  {Nocera}, {Nolting}, {Normandin}, {Nuttall}, {Oberling}, {Ochsner}, {O'Dell},
  {Oelker}, {Ogin}, {Oh}, {Oh}, {Ohme}, {Oliver}, {Oppermann}, {Oram},
  {O'Reilly}, {O'Shaughnessy}, {Ott}, {Ottaway}, {Ottens}, {Overmier}, {Owen},
  {Pai}, {Pai}, {Palamos}, {Palashov}, {Palomba}, {Pal-Singh}, {Pan}, {Pan},
  {Pankow}, {Pannarale}, {Pant}, {Paoletti}, {Paoli}, {Papa}, {Paris},
  {Parker}, {Pascucci}, {Pasqualetti}, {Passaquieti}, {Passuello},
  {Patricelli}, {Patrick}, {Pearlstone}, {Pedraza}, {Pedurand}, {Pekowsky},
  {Pele}, {Penn}, {Perreca}, {Pfeiffer}, {Phelps}, {Piccinni}, {Pichot},
  {Pickenpack}, {Piergiovanni}, {Pierro}, {Pillant}, {Pinard}, {Pinto},
  {Pitkin}, {Poeld}, {Poggiani}, {Popolizio}, {Post}, {Powell}, {Prasad},
  {Predoi}, {Premachandra}, {Prestegard}, {Price}, {Prijatelj}, {Principe},
  {Privitera}, {Prix}, {Prodi}, {Prokhorov}, {Puncken}, {Punturo}, {Puppo},
  {P{\"u}rrer}, {Qi}, {Qin}, {Quetschke}, {Quintero}, {Quitzow-James}, {Raab},
  {Rabeling}, {Radkins}, {Raffai}, {Raja}, {Rakhmanov}, {Ramet}, {Rapagnani},
  {Raymond}, {Razzano}, {Re}, {Read}, {Reed}, {Regimbau}, {Rei}, {Reid},
  {Reitze}, {Rew}, {Reyes}, {Ricci}, {Riles}, {Robertson}, {Robie}, {Robinet},
  {Rocchi}, {Rolland}, {Rollins}, {Roma}, {Romano}, {Romano}, {Romanov},
  {Romie}, {Rosi{\'n}ska}, {Rowan}, {R{\"u}diger}, {Ruggi}, {Ryan}, {Sachdev},
  {Sadecki}, {Sadeghian}, {Salconi}, {Saleem}, {Salemi}, {Samajdar}, {Sammut},
  {Sampson}, {Sanchez}, {Sandberg}, {Sandeen}, {Sanders}, {Sanders},
  {Sassolas}, {Sathyaprakash}, {Saulson}, {Sauter}, {Savage}, {Sawadsky},
  {Schale}, {Schilling}, {Schmidt}, {Schmidt}, {Schnabel}, {Schofield},
  {Sch{\"o}nbeck}, {Schreiber}, {Schuette}, {Schutz}, {Scott}, {Scott},
  {Sellers}, {Sengupta}, {Sentenac}, {Sequino}, {Sergeev}, {Serna},
  {Setyawati}, {Sevigny}, {Shaddock}, {Shaffer}, {Shah}, {Shahriar}, {Shaltev},
  {Shao}, {Shapiro}, {Shawhan}, {Sheperd}, {Shoemaker}, {Shoemaker}, {Siellez},
  {Siemens}, {Sigg}, {Silva}, {Simakov}, {Singer}, {Singer}, {Singh}, {Singh},
  {Singhal}, {Sintes}, {Slagmolen}, {Smith}, {Smith}, {Smith}, {Smith}, {Son},
  {Sorazu}, {Sorrentino}, {Souradeep}, {Srivastava}, {Staley}, {Steinke},
  {Steinlechner}, {Steinlechner}, {Steinmeyer}, {Stephens}, {Stevenson},
  {Stone}, {Strain}, {Straniero}, {Stratta}, {Strauss}, {Strigin}, {Sturani},
  {Stuver}, {Summerscales}, {Sun}, {Sutton}, {Swinkels}, {Szczepa{\'n}czyk},
  {Tacca}, {Talukder}, {Tanner}, {T{\'a}pai}, {Tarabrin}, {Taracchini},
  {Taylor}, {Theeg}, {Thirugnanasambandam}, {Thomas}, {Thomas}, {Thomas},
  {Thorne}, {Thorne}, {Thrane}, {Tiwari}, {Tiwari}, {Tokmakov}, {Tomlinson},
  {Tonelli}, {Torres}, {Torrie}, {T{\"o}yr{\"a}}, {Travasso}, {Traylor},
  {Trifir{\`o}}, {Tringali}, {Trozzo}, {Tse}, {Turconi}, {Tuyenbayev},
  {Ugolini}, {Unnikrishnan}, {Urban}, {Usman}, {Vahlbruch}, {Vajente},
  {Valdes}, {Vallisneri}, {van Bakel}, {van Beuzekom}, {van den Brand}, {Van
  Den Broeck}, {Vander-Hyde}, {van der Schaaf}, {van Heijningen}, {van Veggel},
  {Vardaro}, {Vass}, {Vas{\'u}th}, {Vaulin}, {Vecchio}, {Vedovato}, {Veitch},
  {Veitch}, {Venkateswara}, {Verkindt}, {Vetrano}, {Vicer{\'e}}, {Vinciguerra},
  {Vine}, {Vinet}, {Vitale}, {Vo}, {Vocca}, {Vorvick}, {Voss}, {Vousden},
  {Vyatchanin}, {Wade}, {Wade}, {Wade}, {Waldman}, {Walker}, {Wallace},
  {Walsh}, {Wang}, {Wang}, {Wang}, {Wang}, {Wang}, {Ward}, {Ward}, {Warner},
  {Was}, {Weaver}, {Wei}, {Weinert}, {Weinstein}, {Weiss}, {Welborn}, {Wen},
  {We{\ss}els}, {Westphal}, {Wette}, {Whelan}, {Whitcomb}, {White}, {Whiting},
  {Wiesner}, {Wilkinson}, {Willems}, {Williams}, {Williams}, {Williamson},
  {Willis}, {Willke}, {Wimmer}, {Winkelmann}, {Winkler}, {Wipf}, {Wiseman},
  {Wittel}, {Woan}, {Worden}, {Wright}, {Wu}, {Yablon}, {Yakushin}, {Yam},
  {Yamamoto}, {Yancey}, {Yap}, {Yu}, {Yvert}, {Zadro{\.Z}ny}, {Zangrando},
  {Zanolin}, {Zendri}, {Zevin}, {Zhang}, {Zhang}, {Zhang}, {Zhang}, {Zhao},
  {Zhou}, {Zhou}, {Zhu}, {Zucker}, {Zuraw}, {Zweizig}, {LIGO Scientific
  Collaboration}, \& {Virgo Collaboration}}]{Abbott2016}
{Abbott}, B.~P., {Abbott}, R., {Abbott}, T.~D., {et~al.} 2016, \prl, 116,
  061102, \dodoi{10.1103/PhysRevLett.116.061102}

\bibitem[{{Abbott} {et~al.}(2017){Abbott}, {Abbott}, {Abbott}, {Acernese},
  {Ackley}, {Adams}, {Adams}, {Addesso}, {Adhikari}, {Adya}, {Affeldt},
  {Afrough}, {Agarwal}, {Agathos}, {Agatsuma}, {Aggarwal}, {Aguiar}, {Aiello},
  {Ain}, {Ajith}, {Allen}, {Allen}, {Allocca}, {Altin}, {Amato}, {Ananyeva},
  {Anderson}, {Anderson}, {Angelova}, {Antier}, {Appert}, {Arai}, {Araya},
  {Areeda}, {Arnaud}, {Arun}, {Ascenzi}, {Ashton}, {Ast}, {Aston}, {Astone},
  {Atallah}, {Aufmuth}, {Aulbert}, {AultONeal}, {Austin}, {Avila-Alvarez},
  {Babak}, {Bacon}, {Bader}, {Bae}, {Bailes}, {Baker}, {Baldaccini},
  {Ballardin}, {Ballmer}, {Banagiri}, {Barayoga}, {Barclay}, {Barish},
  {Barker}, {Barkett}, {Barone}, {Barr}, {Barsotti}, {Barsuglia}, {Barta},
  {Barthelmy}, {Bartlett}, {Bartos}, {Bassiri}, {Basti}, {Batch}, {Bawaj},
  {Bayley}, {Bazzan}, {B{\'e}csy}, {Beer}, {Bejger}, {Belahcene}, {Bell},
  {Berger}, {Bergmann}, {Bernuzzi}, {Bero}, {Berry}, {Bersanetti}, {Bertolini},
  {Betzwieser}, {Bhagwat}, {Bhandare}, {Bilenko}, {Billingsley}, {Billman},
  {Birch}, {Birney}, {Birnholtz}, {Biscans}, {Biscoveanu}, {Bisht}, {Bitossi},
  {Biwer}, {Bizouard}, {Blackburn}, {Blackman}, {Blair}, {Blair}, {Blair},
  {Bloemen}, {Bock}, {Bode}, {Boer}, {Bogaert}, {Bohe}, {Bondu}, {Bonilla},
  {Bonnand}, {Boom}, {Bork}, {Boschi}, {Bose}, {Bossie}, {Bouffanais}, {Bozzi},
  {Bradaschia}, {Brady}, {Branchesi}, {Brau}, {Briant}, {Brillet}, {Brinkmann},
  {Brisson}, {Brockill}, {Broida}, {Brooks}, {Brown}, {Brown}, {Brunett},
  {Buchanan}, {Buikema}, {Bulik}, {Bulten}, {Buonanno}, {Buskulic}, {Buy},
  {Byer}, {Cabero}, {Cadonati}, {Cagnoli}, {Cahillane}, {Calder{\'o}n
  Bustillo}, {Callister}, {Calloni}, {Camp}, {Canepa}, {Canizares}, {Cannon},
  {Cao}, {Cao}, {Capano}, {Capocasa}, {Carbognani}, {Caride}, {Carney},
  {Carullo}, {Casanueva Diaz}, {Casentini}, {Caudill}, {Cavagli{\`a}},
  {Cavalier}, {Cavalieri}, {Cella}, {Cepeda}, {Cerd{\'a}-Dur{\'a}n},
  {Cerretani}, {Cesarini}, {Chamberlin}, {Chan}, {Chao}, {Charlton}, {Chase},
  {Chassande-Mottin}, {Chatterjee}, {Chatziioannou}, {Cheeseboro}, {Chen},
  {Chen}, {Chen}, {Cheng}, {Chia}, {Chincarini}, {Chiummo}, {Chmiel}, {Cho},
  {Cho}, {Chow}, {Christensen}, {Chu}, {Chua}, {Chua}, {Chung}, {Chung},
  {Ciani}, {Ciolfi}, {Cirelli}, {Cirone}, {Clara}, {Clark}, {Clearwater},
  {Cleva}, {Cocchieri}, {Coccia}, {Cohadon}, {Cohen}, {Colla}, {Collette},
  {Cominsky}, {Constancio}, {Conti}, {Cooper}, {Corban}, {Corbitt},
  {Cordero-Carri{\'o}n}, {Corley}, {Cornish}, {Corsi}, {Cortese}, {Costa},
  {Coughlin}, {Coughlin}, {Coulon}, {Countryman}, {Couvares}, {Covas}, {Cowan},
  {Coward}, {Cowart}, {Coyne}, {Coyne}, {Creighton}, {Creighton}, {Cripe},
  {Crowder}, {Cullen}, {Cumming}, {Cunningham}, {Cuoco}, {Dal Canton},
  {D{\'a}lya}, {Danilishin}, {D'Antonio}, {Danzmann}, {Dasgupta}, {Da Silva
  Costa}, {Dattilo}, {Dave}, {Davier}, {Davis}, {Daw}, {Day}, {De}, {DeBra},
  {Degallaix}, {De Laurentis}, {Del{\'e}glise}, {Del Pozzo}, {Demos}, {Denker},
  {Dent}, {De Pietri}, {Dergachev}, {De Rosa}, {DeRosa}, {De Rossi}, {DeSalvo},
  {de Varona}, {Devenson}, {Dhurandhar}, {D{\'\i}az}, {Dietrich}, {Di Fiore},
  {Di Giovanni}, {Di Girolamo}, {Di Lieto}, {Di Pace}, {Di Palma}, {Di Renzo},
  {Doctor}, {Dolique}, {Donovan}, {Dooley}, {Doravari}, {Dorrington},
  {Douglas}, {Dovale {\'A}lvarez}, {Downes}, {Drago}, {Dreissigacker},
  {Driggers}, {Du}, {Ducrot}, {Dudi}, {Dupej}, {Dwyer}, {Edo}, {Edwards},
  {Effler}, {Eggenstein}, {Ehrens}, {Eichholz}, {Eikenberry}, {Eisenstein},
  {Essick}, {Estevez}, {Etienne}, {Etzel}, {Evans}, {Evans}, {Factourovich},
  {Fafone}, {Fair}, {Fairhurst}, {Fan}, {Farinon}, {Farr}, {Farr},
  {Fauchon-Jones}, {Favata}, {Fays}, {Fee}, {Fehrmann}, {Feicht}, {Fejer},
  {Fernandez-Galiana}, {Ferrante}, {Ferreira}, {Ferrini}, {Fidecaro},
  {Finstad}, {Fiori}, {Fiorucci}, {Fishbach}, {Fisher}, {Fitz-Axen},
  {Flaminio}, {Fletcher}, {Fong}, {Font}, {Forsyth}, {Forsyth}, {Fournier},
  {Frasca}, {Frasconi}, {Frei}, {Freise}, {Frey}, {Frey}, {Fries}, {Fritschel},
  {Frolov}, {Fulda}, {Fyffe}, {Gabbard}, {Gadre}, {Gaebel}, {Gair},
  {Gammaitoni}, {Ganija}, {Gaonkar}, {Garcia-Quiros}, {Garufi}, {Gateley},
  {Gaudio}, {Gaur}, {Gayathri}, {Gehrels}, {Gemme}, {Genin}, {Gennai},
  {George}, {George}, {Gergely}, {Germain}, {Ghonge}, {Ghosh}, {Ghosh},
  {Ghosh}, {Giaime}, {Giardina}, {Giazotto}, {Gill}, {Glover}, {Goetz},
  {Goetz}, {Gomes}, {Goncharov}, {Gonz{\'a}lez}, {Gonzalez Castro},
  {Gopakumar}, {Gorodetsky}, {Gossan}, {Gosselin}, {Gouaty}, {Grado}, {Graef},
  {Granata}, {Grant}, {Gras}, {Gray}, {Greco}, {Green}, {Gretarsson}, {Groot},
  {Grote}, {Grunewald}, {Gruning}, {Guidi}, {Guo}, {Gupta}, {Gupta}, {Gushwa},
  {Gustafson}, {Gustafson}, {Halim}, {Hall}, {Hall}, {Hamilton}, {Hammond},
  {Haney}, {Hanke}, {Hanks}, {Hanna}, {Hannam}, {Hannuksela}, {Hanson},
  {Hardwick}, {Harms}, {Harry}, {Harry}, {Hart}, {Haster}, {Haughian}, {Healy},
  {Heidmann}, {Heintze}, {Heitmann}, {Hello}, {Hemming}, {Hendry}, {Heng},
  {Hennig}, {Heptonstall}, {Heurs}, {Hild}, {Hinderer}, {Ho}, {Hoak}, {Hofman},
  {Holt}, {Holz}, {Hopkins}, {Horst}, {Hough}, {Houston}, {Howell}, {Hreibi},
  {Hu}, {Huerta}, {Huet}, {Hughey}, {Husa}, {Huttner}, {Huynh-Dinh}, {Indik},
  {Inta}, {Intini}, {Isa}, {Isac}, {Isi}, {Iyer}, {Izumi}, {Jacqmin}, {Jani},
  {Jaranowski}, {Jawahar}, {Jim{\'e}nez-Forteza}, {Johnson},
  {Johnson-McDaniel}, {Jones}, {Jones}, {Jonker}, {Ju}, {Junker}, {Kalaghatgi},
  {Kalogera}, {Kamai}, {Kandhasamy}, {Kang}, {Kanner}, {Kapadia}, {Karki},
  {Karvinen}, {Kasprzack}, {Kastaun}, {Katolik}, {Katsavounidis}, {Katzman},
  {Kaufer}, {Kawabe}, {K{\'e}f{\'e}lian}, {Keitel}, {Kemball}, {Kennedy},
  {Kent}, {Key}, {Khalili}, {Khan}, {Khan}, {Khan}, {Khazanov}, {Kijbunchoo},
  {Kim}, {Kim}, {Kim}, {Kim}, {Kim}, {Kim}, {Kimbrell}, {King}, {King},
  {Kinley-Hanlon}, {Kirchhoff}, {Kissel}, {Kleybolte}, {Klimenko}, {Knowles},
  {Koch}, {Koehlenbeck}, {Koley}, {Kondrashov}, {Kontos}, {Korobko}, {Korth},
  {Kowalska}, {Kozak}, {Kr{\"a}mer}, {Kringel}, {Krishnan}, {Kr{\'o}lak},
  {Kuehn}, {Kumar}, {Kumar}, {Kumar}, {Kuo}, {Kutynia}, {Kwang}, {Lackey},
  {Lai}, {Landry}, {Lang}, {Lange}, {Lantz}, {Lanza}, {Larson},
  {Lartaux-Vollard}, {Lasky}, {Laxen}, {Lazzarini}, {Lazzaro}, {Leaci},
  {Leavey}, {Lee}, {Lee}, {Lee}, {Lee}, {Lee}, {Lehmann}, {Lenon}, {Leon},
  {Leonardi}, {Leroy}, {Letendre}, {Levin}, {Li}, {Linker}, {Littenberg},
  {Liu}, {Liu}, {Lo}, {Lockerbie}, {London}, {Lord}, {Lorenzini}, {Loriette},
  {Lormand}, {Losurdo}, {Lough}, {Lousto}, {Lovelace}, {L{\"u}ck}, {Lumaca},
  {Lundgren}, {Lynch}, {Ma}, {Macas}, {Macfoy}, {Machenschalk}, {MacInnis},
  {Macleod}, {Maga{\~n}a Hernandez}, {Maga{\~n}a-Sandoval}, {Maga{\~n}a
  Zertuche}, {Magee}, {Majorana}, {Maksimovic}, {Man}, {Mandic}, {Mangano},
  {Mansell}, {Manske}, {Mantovani}, {Marchesoni}, {Marion}, {M{\'a}rka},
  {M{\'a}rka}, {Markakis}, {Markosyan}, {Markowitz}, {Maros}, {Marquina},
  {Marsh}, {Martelli}, {Martellini}, {Martin}, {Martin}, {Martynov}, {Marx},
  {Mason}, {Massera}, {Masserot}, {Massinger}, {Masso-Reid}, {Mastrogiovanni},
  {Matas}, {Matichard}, {Matone}, {Mavalvala}, {Mazumder}, {McCarthy},
  {McClelland}, {McCormick}, {McCuller}, {McGuire}, {McIntyre}, {McIver},
  {McManus}, {McNeill}, {McRae}, {McWilliams}, {Meacher}, {Meadors}, {Mehmet},
  {Meidam}, {Mejuto-Villa}, {Melatos}, {Mendell}, {Mercer}, {Merilh},
  {Merzougui}, {Meshkov}, {Messenger}, {Messick}, {Metzdorff}, {Meyers},
  {Miao}, {Michel}, {Middleton}, {Mikhailov}, {Milano}, {Miller}, {Miller},
  {Miller}, {Millhouse}, {Milovich-Goff}, {Minazzoli}, {Minenkov}, {Ming},
  {Mishra}, {Mitra}, {Mitrofanov}, {Mitselmakher}, {Mittleman}, {Moffa},
  {Moggi}, {Mogushi}, {Mohan}, {Mohapatra}, {Molina}, {Montani}, {Moore},
  {Moraru}, {Moreno}, {Morisaki}, {Morriss}, {Mours}, {Mow-Lowry}, {Mueller},
  {Muir}, {Mukherjee}, {Mukherjee}, {Mukherjee}, {Mukund}, {Mullavey}, {Munch},
  {Mu{\~n}iz}, {Muratore}, {Murray}, {Nagar}, {Napier}, {Nardecchia},
  {Naticchioni}, {Nayak}, {Neilson}, {Nelemans}, {Nelson}, {Nery}, {Neunzert},
  {Nevin}, {Newport}, {Newton}, {Ng}, {Nguyen}, {Nguyen}, {Nichols}, {Nielsen},
  {Nissanke}, {Nitz}, {Noack}, {Nocera}, {Nolting}, {North}, {Nuttall},
  {Oberling}, {O'Dea}, {Ogin}, {Oh}, {Oh}, {Ohme}, {Okada}, {Oliver},
  {Oppermann}, {Oram}, {O'Reilly}, {Ormiston}, {Ortega}, {O'Shaughnessy},
  {Ossokine}, {Ottaway}, {Overmier}, {Owen}, {Pace}, {Page}, {Page}, {Pai},
  {Pai}, {Palamos}, {Palashov}, {Palomba}, {Pal-Singh}, {Pan}, {Pan}, {Pang},
  {Pang}, {Pankow}, {Pannarale}, {Pant}, {Paoletti}, {Paoli}, {Papa}, {Parida},
  {Parker}, {Pascucci}, {Pasqualetti}, {Passaquieti}, {Passuello}, {Patil},
  {Patricelli}, {Pearlstone}, {Pedraza}, {Pedurand}, {Pekowsky}, {Pele},
  {Penn}, {Perez}, {Perreca}, {Perri}, {Pfeiffer}, {Phelps}, {Piccinni},
  {Pichot}, {Piergiovanni}, {Pierro}, {Pillant}, {Pinard}, {Pinto}, {Pirello},
  {Pitkin}, {Poe}, {Poggiani}, {Popolizio}, {Porter}, {Post}, {Powell},
  {Prasad}, {Pratt}, {Pratten}, {Predoi}, {Prestegard}, {Prijatelj},
  {Principe}, {Privitera}, {Prix}, {Prodi}, {Prokhorov}, {Puncken}, {Punturo},
  {Puppo}, {P{\"u}rrer}, {Qi}, {Quetschke}, {Quintero}, {Quitzow-James},
  {Raab}, {Rabeling}, {Radkins}, {Raffai}, {Raja}, {Rajan}, {Rajbhandari},
  {Rakhmanov}, {Ramirez}, {Ramos-Buades}, {Rapagnani}, {Raymond}, {Razzano},
  {Read}, {Regimbau}, {Rei}, {Reid}, {Reitze}, {Ren}, {Reyes}, {Ricci},
  {Ricker}, {Rieger}, {Riles}, {Rizzo}, {Robertson}, {Robie}, {Robinet},
  {Rocchi}, {Rolland}, {Rollins}, {Roma}, {Romano}, {Romano}, {Romel}, {Romie},
  {Rosi{\'n}ska}, {Ross}, {Rowan}, {R{\"u}diger}, {Ruggi}, {Rutins}, {Ryan},
  {Sachdev}, {Sadecki}, {Sadeghian}, {Sakellariadou}, {Salconi}, {Saleem},
  {Salemi}, {Samajdar}, {Sammut}, {Sampson}, {Sanchez}, {Sanchez},
  {Sanchis-Gual}, {Sandberg}, {Sanders}, {Sassolas}, {Sathyaprakash},
  {Saulson}, {Sauter}, {Savage}, {Sawadsky}, {Schale}, {Scheel}, {Scheuer},
  {Schmidt}, {Schmidt}, {Schnabel}, {Schofield}, {Sch{\"o}nbeck}, {Schreiber},
  {Schuette}, {Schulte}, {Schutz}, {Schwalbe}, {Scott}, {Scott}, {Seidel},
  {Sellers}, {Sengupta}, {Sentenac}, {Sequino}, {Sergeev}, {Shaddock},
  {Shaffer}, {Shah}, {Shahriar}, {Shaner}, {Shao}, {Shapiro}, {Shawhan},
  {Sheperd}, {Shoemaker}, {Shoemaker}, {Siellez}, {Siemens}, {Sieniawska},
  {Sigg}, {Silva}, {Singer}, {Singh}, {Singhal}, {Sintes}, {Slagmolen},
  {Smith}, {Smith}, {Smith}, {Somala}, {Son}, {Sonnenberg}, {Sorazu},
  {Sorrentino}, {Souradeep}, {Spencer}, {Srivastava}, {Staats}, {Staley},
  {Steinke}, {Steinlechner}, {Steinlechner}, {Steinmeyer}, {Stevenson},
  {Stone}, {Stops}, {Strain}, {Stratta}, {Strigin}, {Strunk}, {Sturani},
  {Stuver}, {Summerscales}, {Sun}, {Sunil}, {Suresh}, {Sutton}, {Swinkels},
  {Szczepa{\'n}czyk}, {Tacca}, {Tait}, {Talbot}, {Talukder}, {Tanner},
  {T{\'a}pai}, {Taracchini}, {Tasson}, {Taylor}, {Taylor}, {Tewari}, {Theeg},
  {Thies}, {Thomas}, {Thomas}, {Thomas}, {Thorne}, {Thorne}, {Thrane},
  {Tiwari}, {Tiwari}, {Tokmakov}, {Toland}, {Tonelli}, {Tornasi},
  {Torres-Forn{\'e}}, {Torrie}, {T{\"o}yr{\"a}}, {Travasso}, {Traylor},
  {Trinastic}, {Tringali}, {Trozzo}, {Tsang}, {Tse}, {Tso}, {Tsukada}, {Tsuna},
  {Tuyenbayev}, {Ueno}, {Ugolini}, {Unnikrishnan}, {Urban}, {Usman},
  {Vahlbruch}, {Vajente}, {Valdes}, {Vallisneri}, {van Bakel}, {van Beuzekom},
  {van den Brand}, {Van Den Broeck}, {Vander-Hyde}, {van der Schaaf}, {van
  Heijningen}, {van Veggel}, {Vardaro}, {Varma}, {Vass}, {Vas{\'u}th},
  {Vecchio}, {Vedovato}, {Veitch}, {Veitch}, {Venkateswara}, {Venugopalan},
  {Verkindt}, {Vetrano}, {Vicer{\'e}}, {Viets}, {Vinciguerra}, {Vine}, {Vinet},
  {Vitale}, {Vo}, {Vocca}, {Vorvick}, {Vyatchanin}, {Wade}, {Wade}, {Wade},
  {Walet}, {Walker}, {Wallace}, {Walsh}, {Wang}, {Wang}, {Wang}, {Wang},
  {Wang}, {Ward}, {Warner}, {Was}, {Watchi}, {Weaver}, {Wei}, {Weinert},
  {Weinstein}, {Weiss}, {Wen}, {Wessel}, {We{\ss}els}, {Westerweck},
  {Westphal}, {Wette}, {Whelan}, {Whitcomb}, {Whiting}, {Whittle}, {Wilken},
  {Williams}, {Williams}, {Williamson}, {Willis}, {Willke}, {Wimmer},
  {Winkler}, {Wipf}, {Wittel}, {Woan}, {Woehler}, {Wofford}, {Wong}, {Worden},
  {Wright}, {Wu}, {Wysocki}, {Xiao}, {Yamamoto}, {Yancey}, {Yang}, {Yap},
  {Yazback}, {Yu}, {Yu}, {Yvert}, {Zadro{\.Z}ny}, {Zanolin}, {Zelenova},
  {Zendri}, {Zevin}, {Zhang}, {Zhang}, {Zhang}, {Zhang}, {Zhao}, {Zhou},
  {Zhou}, {Zhu}, {Zhu}, {Zimmerman}, {Zucker}, {Zweizig}, {LIGO Scientific
  Collaboration}, \& {Virgo Collaboration}}]{Abbott2017}
---. 2017, \prl, 119, 161101, \dodoi{10.1103/PhysRevLett.119.161101}

\bibitem[{{Abbott} {et~al.}(2020){Abbott}, {Abbott}, {Abraham}, {Acernese},
  {Ackley}, {Adams}, {Adhikari}, {Adya}, {Affeldt}, {Agathos}, \&
  et~al.}]{Abbott2020}
{Abbott}, R., {Abbott}, T.~D., {Abraham}, S., {et~al.} 2020, \apjl, 896, L44,
  \dodoi{10.3847/2041-8213/ab960f}

\bibitem[{{Abbott} {et~al.}(2023){Abbott}, {Abbott}, {Acernese}, {Ackley},
  {Adams}, {Adhikari}, {Adhikari}, {Adya}, {Affeldt}, {Agarwal}, {Agathos},
  {Agatsuma}, {Aggarwal}, {Aguiar}, {Aiello}, {Ain}, {Ajith}, {Akutsu}, {de
  Alarc{\'o}n}, {Akcay}, {Albanesi}, {Allocca}, {Altin}, {Amato}, {Anand},
  {Anand}, {Ananyeva}, {Anderson}, {Anderson}, {Ando}, {Andrade}, {Andres},
  {Andri{\'c}}, {Angelova}, {Ansoldi}, {Antelis}, {Antier}, {Antonini},
  {Appert}, {Arai}, {Arai}, {Arai}, {Araki}, {Araya}, {Araya}, {Areeda},
  {Ar{\`e}ne}, {Aritomi}, {Arnaud}, {Arogeti}, {Aronson}, {Arun}, {Asada},
  {Asali}, {Ashton}, {Aso}, {Assiduo}, {Aston}, {Astone}, {Aubin}, {Austin},
  {Babak}, {Badaracco}, {Bader}, {Badger}, {Bae}, {Bae}, {Baer}, {Bagnasco},
  {Bai}, {Baiotti}, {Baird}, {Bajpai}, {Ball}, {Ballardin}, {Ballmer},
  {Balsamo}, {Baltus}, {Banagiri}, {Bankar}, {Barayoga}, {Barbieri}, {Barish},
  {Barker}, {Barneo}, {Barone}, {Barr}, {Barsotti}, {Barsuglia}, {Barta},
  {Bartlett}, {Barton}, {Bartos}, {Bassiri}, {Basti}, {Bawaj}, {Bayley},
  {Baylor}, {Bazzan}, {B{\'e}csy}, {Bedakihale}, {Bejger}, {Belahcene},
  {Benedetto}, {Beniwal}, {Bennett}, {Bentley}, {Benyaala}, {Bergamin},
  {Berger}, {Bernuzzi}, {Berry}, {Bersanetti}, {Bertolini}, {Betzwieser},
  {Beveridge}, {Bhandare}, {Bhardwaj}, {Bhattacharjee}, {Bhaumik}, {Bilenko},
  {Billingsley}, {Bini}, {Birney}, {Birnholtz}, {Biscans}, {Bischi},
  {Biscoveanu}, {Bisht}, {Biswas}, {Bitossi}, {Bizouard}, {Blackburn}, {Blair},
  {Blair}, {Blair}, {Bobba}, {Bode}, {Boer}, {Bogaert}, {Boldrini}, {Bonavena},
  {Bondu}, {Bonilla}, {Bonnand}, {Booker}, {Boom}, {Bork}, {Boschi}, {Bose},
  {Bose}, {Bossilkov}, {Boudart}, {Bouffanais}, {Bozzi}, {Bradaschia}, {Brady},
  {Bramley}, {Branch}, {Branchesi}, {Brandt}, {Brau}, {Breschi}, {Briant},
  {Briggs}, {Brillet}, {Brinkmann}, {Brockill}, {Brooks}, {Brooks}, {Brown},
  {Brunett}, {Bruno}, {Bruntz}, {Bryant}, {Bulik}, {Bulten}, {Buonanno},
  {Buscicchio}, {Buskulic}, {Buy}, {Byer}, {Cadonati}, {Cagnoli}, {Cahillane},
  {Bustillo}, {Callaghan}, {Callister}, {Calloni}, {Cameron}, {Camp}, {Canepa},
  {Canevarolo}, {Cannavacciuolo}, {Cannon}, {Cao}, {Cao}, {Capocasa}, {Capote},
  {Carapella}, {Carbognani}, {Carlin}, {Carney}, {Carpinelli}, {Carrillo},
  {Carullo}, {Carver}, {Diaz}, {Casentini}, {Castaldi}, {Caudill},
  {Cavagli{\`a}}, {Cavalier}, {Cavalieri}, {Ceasar}, {Cella},
  {Cerd{\'a}-Dur{\'a}n}, {Cesarini}, {Chaibi}, {Chakravarti}, {Subrahmanya},
  {Champion}, {Chan}, {Chan}, {Chan}, {Chan}, {Chan}, {Chandra}, {Chanial},
  {Chao}, {Chapman-Bird}, {Charlton}, {Chase}, {Chassande-Mottin},
  {Chatterjee}, {Chatterjee}, {Chatterjee}, {Chaturvedi}, {Chaty},
  {Chatziioannou}, {Chen}, {Chen}, {Chen}, {Chen}, {Chen}, {Chen}, {Chen},
  {Chen}, {Cheng}, {Cheong}, {Cheung}, {Chia}, {Chiadini}, {Chiang},
  {Chiarini}, {Chierici}, {Chincarini}, {Chiofalo}, {Chiummo}, {Cho}, {Cho},
  {Choudhary}, {Choudhary}, {Christensen}, {Chu}, {Chu}, {Chu}, {Chua},
  {Chung}, {Ciani}, {Ciecielag}, {Cie{\'s}lar}, {Cifaldi}, {Ciobanu}, {Ciolfi},
  {Cipriano}, {Cirone}, {Clara}, {Clark}, {Clark}, {Clarke}, {Clearwater},
  {Clesse}, {Cleva}, {Coccia}, {Codazzo}, {Cohadon}, {Cohen}, {Cohen},
  {Colleoni}, {Collette}, {Colombo}, {Colpi}, {Compton}, {Constancio}, {Conti},
  {Cooper}, {Corban}, {Corbitt}, {Cordero-Carri{\'o}n}, {Corezzi}, {Corley},
  {Cornish}, {Corre}, {Corsi}, {Cortese}, {Costa}, {Cotesta}, {Coughlin},
  {Coulon}, {Countryman}, {Cousins}, {Couvares}, {Coward}, {Cowart}, {Coyne},
  {Coyne}, {Creighton}, {Creighton}, {Criswell}, {Croquette}, {Crowder},
  {Cudell}, {Cullen}, {Cumming}, {Cummings}, {Cunningham}, {Cuoco},
  {Cury{\l}o}, {Dabadie}, {Canton}, {Dall'Osso}, {D{\'a}lya}, {Dana},
  {Daneshgaranbajastani}, {D'Angelo}, {Danila}, {Danilishin}, {D'Antonio},
  {Danzmann}, {Darsow-Fromm}, {Dasgupta}, {Datrier}, {Datta}, {Dattilo},
  {Dave}, {Davier}, {Davies}, {Davis}, {Davis}, {Daw}, {Dean}, {Debra},
  {Deenadayalan}, {Degallaix}, {de Laurentis}, {Del{\'e}glise}, {Del Favero},
  {de Lillo}, {de Lillo}, {Del Pozzo}, {Demarchi}, {de Matteis}, {D'Emilio},
  {Demos}, {Dent}, {Depasse}, {de Pietri}, {De Rosa}, {de Rossi}, {Desalvo},
  {de Simone}, {Dhurandhar}, {D{\'\i}az}, {Diaz-Ortiz}, {Didio}, {Dietrich},
  {di Fiore}, {di Fronzo}, {di Giorgio}, {di Giovanni}, {di Giovanni}, {di
  Girolamo}, {di Lieto}, {Ding}, {di Pace}, {di Palma}, {di Renzo},
  {Divakarla}, {Dmitriev}, {Doctor}, {D'Onofrio}, {Donovan}, {Dooley},
  {Doravari}, {Dorrington}, {Drago}, {Driggers}, {Drori}, {Ducoin}, {Dupej},
  {Durante}, {D'Urso}, {Duverne}, {Dwyer}, {Eassa}, {Easter}, {Ebersold},
  {Eckhardt}, {Eddolls}, {Edelman}, {Edo}, {Edy}, {Effler}, {Eguchi},
  {Eichholz}, {Eikenberry}, {Eisenmann}, {Eisenstein}, {Ejlli}, {Engelby},
  {Enomoto}, {Errico}, {Essick}, {Estell{\'e}s}, {Estevez}, {Etienne}, {Etzel},
  {Evans}, {Evans}, {Ewing}, {Fafone}, {Fair}, {Fairhurst}, {Farah}, {Farinon},
  {Farr}, {Farr}, {Farrow}, {Fauchon-Jones}, {Favaro}, {Favata}, {Fays},
  {Fazio}, {Feicht}, {Fejer}, {Fenyvesi}, {Ferguson}, {Fernandez-Galiana},
  {Ferrante}, {Ferreira}, {Fidecaro}, {Figura}, {Fiori}, {Fishbach}, {Fisher},
  {Fittipaldi}, {Fiumara}, {Flaminio}, {Floden}, {Fong}, {Font}, {Fornal},
  {Forsyth}, {Franke}, {Frasca}, {Frasconi}, {Frederick}, {Freed}, {Frei},
  {Freise}, {Frey}, {Fritschel}, {Frolov}, {Fronz{\'e}}, {Fujii}, {Fujikawa},
  {Fukunaga}, {Fukushima}, {Fulda}, {Fyffe}, {Gabbard}, {Gadre}, {Gair},
  {Gais}, {Galaudage}, {Gamba}, {Ganapathy}, {Ganguly}, {Gao}, {Gaonkar},
  {Garaventa}, {Garc{\'\i}a}, {Garc{\'\i}a-N{\'u}{\~n}ez},
  {Garc{\'\i}a-Quir{\'o}s}, {Garufi}, {Gateley}, {Gaudio}, {Gayathri}, {Ge},
  {Gemme}, {Gennai}, {George}, {George}, {Gerberding}, {Gergely}, {Gewecke},
  {Ghonge}, {Ghosh}, {Ghosh}, {Ghosh}, {Ghosh}, {Giacomazzo}, {Giacoppo},
  {Giaime}, {Giardina}, {Gibson}, {Gier}, {Giesler}, {Giri}, {Gissi},
  {Glanzer}, {Gleckl}, {Godwin}, {Golomb}, {Goetz}, {Goetz}, {Gohlke},
  {Goncharov}, {Gonz{\'a}lez}, {Gopakumar}, {Gosselin}, {Gouaty}, {Gould},
  {Grace}, {Grado}, {Granata}, {Granata}, {Grant}, {Gras}, {Grassia}, {Gray},
  {Gray}, {Greco}, {Green}, {Green}, {Gretarsson}, {Gretarsson}, {Griffith},
  {Griffiths}, {Griggs}, {Grignani}, {Grimaldi}, {Grimm}, {Grote}, {Grunewald},
  {Gruning}, {Guerra}, {Guidi}, {Guimaraes}, {Guix{\'e}}, {Gulati}, {Guo},
  {Guo}, {Gupta}, {Gupta}, {Gupta}, {Gustafson}, {Gustafson}, {Guzman}, {Ha},
  {Haegel}, {Hagiwara}, {Haino}, {Halim}, {Hall}, {Hamilton}, {Hammond}, {Han},
  {Haney}, {Hanks}, {Hanna}, {Hannam}, {Hannuksela}, {Hansen}, {Hansen},
  {Hanson}, {Harder}, {Hardwick}, {Haris}, {Harms}, {Harry}, {Harry},
  {Hartwig}, {Hasegawa}, {Haskell}, {Hasskew}, {Haster}, {Hattori}, {Haughian},
  {Hayakawa}, {Hayama}, {Hayes}, {Healy}, {Heidmann}, {Heidt}, {Heintze},
  {Heinze}, {Heinzel}, {Heitmann}, {Hellman}, {Hello}, {Helmling-Cornell},
  {Hemming}, {Hendry}, {Heng}, {Hennes}, {Hennig}, {Hennig}, {Hernandez},
  {Vivanco}, {Heurs}, {Hild}, {Hill}, {Himemoto}, {Hines}, {Hiranuma},
  {Hirata}, {Hirose}, {Hochheim}, {Hofman}, {Hohmann}, {Holcomb}, {Holland},
  {Hollows}, {Holmes}, {Holt}, {Holz}, {Hong}, {Hopkins}, {Hough}, {Hourihane},
  {Howell}, {Hoy}, {Hoyland}, {Hreibi}, {Hsieh}, {Hsu}, {Huang}, {Huang},
  {Huang}, {Huang}, {Huang}, {Huang}, {H{\"u}bner}, {Huddart}, {Hughey}, {Hui},
  {Hui}, {Husa}, {Huttner}, {Huxford}, {Huynh-Dinh}, {Ide}, {Idzkowski},
  {Iess}, {Ikenoue}, {Imam}, {Inayoshi}, {Ingram}, {Inoue}, {Ioka}, {Isi},
  {Isleif}, {Ito}, {Itoh}, {Iyer}, {Izumi}, {Jaberianhamedan}, {Jacqmin},
  {Jadhav}, {Jadhav}, {James}, {Jan}, {Jani}, {Janquart}, {Janssens},
  {Janthalur}, {Jaranowski}, {Jariwala}, {Jaume}, {Jenkins}, {Jenner}, {Jeon},
  {Jeunon}, {Jia}, {Jin}, {Johns}, {Jones}, {Jones}, {Jones}, {Jones}, {Jones},
  {Jonker}, {Ju}, {Jung}, {Jung}, {Junker}, {Juste}, {Kaihotsu}, {Kajita},
  {Kakizaki}, {Kalaghatgi}, {Kalogera}, {Kamai}, {Kamiizumi}, {Kanda},
  {Kandhasamy}, {Kang}, {Kanner}, {Kao}, {Kapadia}, {Kapasi}, {Karat},
  {Karathanasis}, {Karki}, {Kashyap}, {Kasprzack}, {Kastaun}, {Katsanevas},
  {Katsavounidis}, {Katzman}, {Kaur}, {Kawabe}, {Kawaguchi}, {Kawai},
  {Kawasaki}, {K{\'e}f{\'e}lian}, {Keitel}, {Key}, {Khadka}, {Khalili}, {Khan},
  {Khazanov}, {Khetan}, {Khursheed}, {Kijbunchoo}, {Kim}, {Kim}, {Kim}, {Kim},
  {Kim}, {Kim}, {Kimball}, {Kimura}, {Kinley-Hanlon}, {Kirchhoff}, {Kissel},
  {Kita}, {Kitazawa}, {Kleybolte}, {Klimenko}, {Knee}, {Knowles}, {Knyazev},
  {Koch}, {Koekoek}, {Kojima}, {Kokeyama}, {Koley}, {Kolitsidou}, {Kolstein},
  {Komori}, {Kondrashov}, {Kong}, {Kontos}, {Koper}, {Korobko}, {Kotake},
  {Kovalam}, {Kozak}, {Kozakai}, {Kozu}, {Kringel}, {Krishnendu}, {Kr{\'o}lak},
  {Kuehn}, {Kuei}, {Kuijer}, {Kulkarni}, {Kumar}, {Kumar}, {Kumar}, {Kumar},
  {Kume}, {Kuns}, {Kuo}, {Kuo}, {Kuromiya}, {Kuroyanagi}, {Kusayanagi},
  {Kuwahara}, {Kwak}, {Lagabbe}, {Laghi}, {Lalande}, {Lam}, {Lamberts},
  {Landry}, {Landry}, {Lane}, {Lang}, {Lange}, {Lantz}, {La Rosa},
  {Lartaux-Vollard}, {Lasky}, {Laxen}, {Lazzarini}, {Lazzaro}, {Leaci},
  {Leavey}, {Lecoeuche}, {Lee}, {Lee}, {Lee}, {Lee}, {Lee}, {Lee}, {Lehmann},
  {Lema{\^\i}tre}, {Leonardi}, {Leroy}, {Letendre}, {Levesque}, {Levin},
  {Leviton}, {Leyde}, {Li}, {Li}, {Li}, {Li}, {Li}, {Li}, {Lin}, {Lin}, {Lin},
  {Lin}, {Lin}, {Linde}, {Linker}, {Linley}, {Littenberg}, {Liu}, {Liu}, {Liu},
  {Liu}, {Llamas}, {Llorens-Monteagudo}, {Lo}, {Lockwood}, {Loh}, {London},
  {Longo}, {Lopez}, {Portilla}, {Lorenzini}, {Loriette}, {Lormand}, {Losurdo},
  {Lott}, {Lough}, {Lousto}, {Lovelace}, {Lucaccioni}, {L{\"u}ck}, {Lumaca},
  {Lundgren}, {Luo}, {Lynam}, {Macas}, {Macinnis}, {MacLeod}, {MacMillan},
  {Macquet}, {Hernandez}, {Magazz{\`u}}, {Magee}, {Maggiore}, {Magnozzi},
  {Mahesh}, {Majorana}, {Makarem}, {Maksimovic}, {Maliakal}, {Malik}, {Man},
  {Mandic}, {Mangano}, {Mango}, {Mansell}, {Manske}, {Mantovani}, {Mapelli},
  {Marchesoni}, {Marchio}, {Marion}, {Mark}, {M{\'a}rka}, {M{\'a}rka},
  {Markakis}, {Markosyan}, {Markowitz}, {Maros}, {Marquina}, {Marsat},
  {Martelli}, {Martin}, {Martin}, {Martinez}, {Martinez}, {Martinez},
  {Martinovic}, {Martynov}, {Marx}, {Masalehdan}, {Mason}, {Massera},
  {Masserot}, {Massinger}, {Masso-Reid}, {Mastrogiovanni}, {Matas},
  {Mateu-Lucena}, {Matichard}, {Matiushechkina}, {Mavalvala}, {McCann},
  {McCarthy}, {McClelland}, {McClincy}, {McCormick}, {McCuller}, {McGhee},
  {McGuire}, {McIsaac}, {McIver}, {McRae}, {McWilliams}, {Meacher}, {Mehmet},
  {Mehta}, {Meijer}, {Melatos}, {Melchor}, {Mendell}, {Menendez-Vazquez},
  {Menoni}, {Mercer}, {Mereni}, {Merfeld}, {Merilh}, {Merritt}, {Merzougui},
  {Meshkov}, {Messenger}, {Messick}, {Meyers}, {Meylahn}, {Mhaske}, {Miani},
  {Miao}, {Michaloliakos}, {Michel}, {Michimura}, {Middleton}, {Milano},
  {Miller}, {Miller}, {Miller}, {Miller}, {Millhouse}, {Mills}, {Milotti},
  {Minazzoli}, {Minenkov}, {Mio}, {Mir}, {Miravet-Ten{\'e}s}, {Mishra},
  {Mishra}, {Mistry}, {Mitra}, {Mitrofanov}, {Mitselmakher}, {Mittleman},
  {Miyakawa}, {Miyamoto}, {Miyazaki}, {Miyo}, {Miyoki}, {Mo}, {Modafferi},
  {Moguel}, {Mogushi}, {Mohapatra}, {Mohite}, {Molina}, {Molina-Ruiz},
  {Mondin}, {Montani}, {Moore}, {Moraru}, {Morawski}, {More}, {Moreno},
  {Moreno}, {Mori}, {Morisaki}, {Moriwaki}, {Morr{\'a}s}, {Mours}, {Mow-Lowry},
  {Mozzon}, {Muciaccia}, {Mukherjee}, {Mukherjee}, {Mukherjee}, {Mukherjee},
  {Mukherjee}, {Mukund}, {Mullavey}, {Munch}, {Mu{\~n}iz}, {Murray},
  {Musenich}, {Muusse}, {Nadji}, {Nagano}, {Nagano}, {Nagar}, {Nakamura},
  {Nakano}, {Nakano}, {Nakashima}, {Nakayama}, {Napolano}, {Nardecchia},
  {Narikawa}, {Naticchioni}, {Nayak}, {Nayak}, {Negishi}, {Neil}, {Neilson},
  {Nelemans}, {Nelson}, {Nery}, {Neubauer}, {Neunzert}, {Ng}, {Ng}, {Nguyen},
  {Nguyen}, {Nguyen}, {Quynh}, {Ni}, {Nichols}, {Nishizawa}, {Nissanke},
  {Nitoglia}, {Nocera}, {Norman}, {North}, {Nozaki}, {Siles}, {Nuttall},
  {Oberling}, {O'Brien}, {Obuchi}, {O'Dell}, {Oelker}, {Ogaki}, {Oganesyan},
  {Oh}, {Oh}, {Oh}, {Ohashi}, {Ohishi}, {Ohkawa}, {Ohme}, {Ohta}, {Okada},
  {Okutani}, {Okutomi}, {Olivetto}, {Oohara}, {Ooi}, {Oram}, {O'Reilly},
  {Ormiston}, {Ormsby}, {Ortega}, {O'Shaughnessy}, {O'Shea}, {Oshino},
  {Ossokine}, {Osthelder}, {Otabe}, {Ottaway}, {Overmier}, {Pace}, {Pagano},
  {Page}, {Pagliaroli}, {Pai}, {Pai}, {Palamos}, {Palashov}, {Palomba}, {Pan},
  {Pan}, {Panda}, {Pang}, {Pang}, {Pankow}, {Pannarale}, {Pant}, {Panther},
  {Paoletti}, {Paoli}, {Paolone}, {Parisi}, {Park}, {Park}, {Parker},
  {Pascucci}, {Pasqualetti}, {Passaquieti}, {Passuello}, {Patel}, {Pathak},
  {Patricelli}, {Patron}, {Paul}, {Payne}, {Pedraza}, {Pegoraro}, {Pele},
  {Arellano}, {Penn}, {Perego}, {Pereira}, {Pereira}, {Perez}, {P{\'e}rigois},
  {Perkins}, {Perreca}, {Perri{\`e}s}, {Petermann}, {Petterson}, {Pfeiffer},
  {Pham}, {Phukon}, {Piccinni}, {Pichot}, {Piendibene}, {Piergiovanni},
  {Pierini}, {Pierro}, {Pillant}, {Pillas}, {Pilo}, {Pinard}, {Pinto}, {Pinto},
  {Piotrzkowski}, {Piotrzkowski}, {Pirello}, {Pitkin}, {Placidi}, {Planas},
  {Plastino}, {Pluchar}, {Poggiani}, {Polini}, {Pong}, {Ponrathnam},
  {Popolizio}, {Porter}, {Poulton}, {Powell}, {Pracchia}, {Pradier},
  {Prajapati}, {Prasai}, {Prasanna}, {Pratten}, {Principe}, {Prodi},
  {Prokhorov}, {Prosposito}, {Prudenzi}, {Puecher}, {Punturo}, {Puosi},
  {Puppo}, {P{\"u}rrer}, {Qi}, {Quetschke}, {Quitzow-James}, {Raab},
  {Raaijmakers}, {Radkins}, {Radulesco}, {Raffai}, {Rail}, {Raja}, {Rajan},
  {Ramirez}, {Ramirez}, {Ramos-Buades}, {Rana}, {Rapagnani}, {Rapol}, {Ray},
  {Raymond}, {Raza}, {Razzano}, {Read}, {Rees}, {Regimbau}, {Rei}, {Reid},
  {Reid}, {Reitze}, {Relton}, {Renzini}, {Rettegno}, {Reza}, {Rezac}, {Ricci},
  {Richards}, {Richardson}, {Richardson}, {Riemenschneider}, {Riles},
  {Rinaldi}, {Rink}, {Rizzo}, {Robertson}, {Robie}, {Robinet}, {Rocchi},
  {Rodriguez}, {Rolland}, {Rollins}, {Romanelli}, {Romano}, {Romel},
  {Romero-Rodr{\'\i}guez}, {Romero-Shaw}, {Romie}, {Ronchini}, {Rosa}, {Rose},
  {Rosi{\'n}ska}, {Ross}, {Rowan}, {Rowlinson}, {Roy}, {Roy}, {Roy}, {Rozza},
  {Ruggi}, {Ryan}, {Sachdev}, {Sadecki}, {Sadiq}, {Sago}, {Saito}, {Saito},
  {Sakai}, {Sakai}, {Sakellariadou}, {Sakuno}, {Salafia}, {Salconi}, {Saleem},
  {Salemi}, {Samajdar}, {Sanchez}, {Sanchez}, {Sanchez}, {Sanchis-Gual},
  {Sanders}, {Sanuy}, {Saravanan}, {Sarin}, {Sassolas}, {Satari},
  {Sathyaprakash}, {Sato}, {Sato}, {Sauter}, {Savage}, {Sawada}, {Sawant},
  {Sawant}, {Sayah}, {Schaetzl}, {Scheel}, {Scheuer}, {Schiworski}, {Schmidt},
  {Schmidt}, {Schnabel}, {Schneewind}, {Schofield}, {Sch{\"o}nbeck}, {Schulte},
  {Schutz}, {Schwartz}, {Scott}, {Scott}, {Seglar-Arroyo}, {Sekiguchi},
  {Sekiguchi}, {Sellers}, {Sengupta}, {Sentenac}, {Seo}, {Sequino}, {Sergeev},
  {Setyawati}, {Shaffer}, {Shahriar}, {Shams}, {Shao}, {Sharma}, {Sharma},
  {Shawhan}, {Shcheblanov}, {Shibagaki}, {Shikauchi}, {Shimizu}, {Shimoda},
  {Shimode}, {Shinkai}, {Shishido}, {Shoda}, {Shoemaker}, {Shoemaker},
  {Shyamsundar}, {Sieniawska}, {Sigg}, {Singer}, {Singh}, {Singh}, {Singha},
  {Sintes}, {Sipala}, {Skliris}, {Slagmolen}, {Slaven-Blair}, {Smetana},
  {Smith}, {Smith}, {Soldateschi}, {Somala}, {Somiya}, {Son}, {Soni}, {Soni},
  {Sordini}, {Sorrentino}, {Sorrentino}, {Sotani}, {Soulard}, {Souradeep},
  {Sowell}, {Spagnuolo}, {Spencer}, {Spera}, {Srinivasan}, {Srivastava},
  {Srivastava}, {Staats}, {Stachie}, {Steer}, {Steinhoff}, {Steinlechner},
  {Steinlechner}, {Stevenson}, {Stops}, {Stover}, {Strain}, {Strang},
  {Stratta}, {Strunk}, {Sturani}, {Stuver}, {Sudhagar}, {Sudhir}, {Sugimoto},
  {Suh}, {Sullivan}, {Summerscales}, {Sun}, {Sun}, {Sunil}, {Sur}, {Suresh},
  {Sutton}, {Suzuki}, {Suzuki}, {Swinkels}, {Szczepa{\'n}czyk}, {Szewczyk},
  {Tacca}, {Tagoshi}, {Tait}, {Takahashi}, {Takahashi}, {Takamori}, {Takano},
  {Takeda}, {Takeda}, {Talbot}, {Talbot}, {Tanaka}, {Tanaka}, {Tanaka},
  {Tanaka}, {Tanaka}, {Tanasijczuk}, {Tanioka}, {Tanner}, {Tao}, {Tao},
  {Mart{\'\i}n}, {Taranto}, {Tasson}, {Telada}, {Tenorio}, {Terhune},
  {Terkowski}, {Thirugnanasambandam}, {Thomas}, {Thomas}, {Thomas}, {Thompson},
  {Thondapu}, {Thorne}, {Thrane}, {Tiwari}, {Tiwari}, {Tiwari}, {Toivonen},
  {Toland}, {Tolley}, {Tomaru}, {Tomigami}, {Tomura}, {Tonelli},
  {Torres-Forn{\'e}}, {Torrie}, {E Melo}, {T{\"o}yr{\"a}}, {Trapananti},
  {Travasso}, {Traylor}, {Trevor}, {Tringali}, {Tripathee}, {Troiano},
  {Trovato}, {Trozzo}, {Trudeau}, {Tsai}, {Tsai}, {Tsang}, {Tsang}, {Tsao},
  {Tse}, {Tso}, {Tsubono}, {Tsuchida}, {Tsukada}, {Tsuna}, {Tsutsui},
  {Tsuzuki}, {Turbang}, {Turconi}, {Tuyenbayev}, {Ubhi}, {Uchikata},
  {Uchiyama}, {Udall}, {Ueda}, {Uehara}, {Ueno}, {Ueshima}, {Unnikrishnan},
  {Uraguchi}, {Urban}, {Ushiba}, {Utina}, {Vahlbruch}, {Vajente}, {Vajpeyi},
  {Valdes}, {Valentini}, {Valsan}, {van Bakel}, {van Beuzekom}, {van den
  Brand}, {van den Broeck}, {Vander-Hyde}, {van der Schaaf}, {van Heijningen},
  {Vanosky}, {van Putten}, {van Remortel}, {Vardaro}, {Vargas}, {Varma},
  {Vas{\'u}th}, {Vecchio}, {Vedovato}, {Veitch}, {Veitch}, {Venneberg},
  {Venugopalan}, {Verkindt}, {Verma}, {Verma}, {Veske}, {Vetrano},
  {Vicer{\'e}}, {Vidyant}, {Viets}, {Vijaykumar}, {Villa-Ortega}, {Vinet},
  {Virtuoso}, {Vitale}, {Vo}, {Vocca}, {von Reis}, {von Wrangel}, {Vorvick},
  {Vyatchanin}, {Wade}, {Wade}, {Wagner}, {Walet}, {Walker}, {Wallace},
  {Wallace}, {Walsh}, {Wang}, {Wang}, {Wang}, {Ward}, {Warner}, {Was},
  {Washimi}, {Washington}, {Watchi}, {Weaver}, {Webster}, {Weinert},
  {Weinstein}, {Weiss}, {Weller}, {Wellmann}, {Wen}, {We{\ss}els}, {Wette},
  {Whelan}, {White}, {Whiting}, {Whittle}, {Wilken}, {Williams}, {Williams},
  {Williamson}, {Willis}, {Willke}, {Wilson}, {Winkler}, {Wipf}, {Wlodarczyk},
  {Woan}, {Woehler}, {Wofford}, {Wong}, {Wu}, {Wu}, {Wu}, {Wu}, {Wysocki},
  {Xiao}, {Xu}, {Yamada}, {Yamamoto}, {Yamamoto}, {Yamamoto}, {Yamamoto},
  {Yamashita}, {Yamazaki}, {Yang}, {Yang}, {Yang}, {Yang}, {Yang}, {Yap},
  {Yeeles}, {Yelikar}, {Ying}, {Yokogawa}, {Yokoyama}, {Yokozawa}, {Yoo},
  {Yoshioka}, {Yu}, {Yu}, {Yuzurihara}, {Zadro{\.z}ny}, {Zanolin}, {Zeidler},
  {Zelenova}, {Zendri}, {Zevin}, {Zhan}, {Zhang}, {Zhang}, {Zhang}, {Zhang},
  {Zhang}, {Zhao}, {Zhao}, {Zhao}, {Zhao}, {Zheng}, {Zhou}, {Zhou}, {Zhu},
  {Zhu}, {Zimmerman}, {Zlochower}, {Zucker}, {Zweizig}, {LIGO Scientific
  Collaboration}, {VIRGO Collaboration}, \& {KAGRA Collaboration}}]{LVK2023}
{Abbott}, R., {Abbott}, T.~D., {Acernese}, F., {et~al.} 2023, Physical Review
  X, 13, 011048, \dodoi{10.1103/PhysRevX.13.011048}

\bibitem[{{Amaro-Seoane} {et~al.}(2017){Amaro-Seoane}, {Audley}, {Babak},
  {Baker}, {Barausse}, {Bender}, {Berti}, {Binetruy}, {Born}, {Bortoluzzi},
  {Camp}, {Caprini}, {Cardoso}, {Colpi}, {Conklin}, {Cornish}, {Cutler},
  {Danzmann}, {Dolesi}, {Ferraioli}, {Ferroni}, {Fitzsimons}, {Gair}, {Gesa
  Bote}, {Giardini}, {Gibert}, {Grimani}, {Halloin}, {Heinzel}, {Hertog},
  {Hewitson}, {Holley-Bockelmann}, {Hollington}, {Hueller}, {Inchauspe},
  {Jetzer}, {Karnesis}, {Killow}, {Klein}, {Klipstein}, {Korsakova}, {Larson},
  {Livas}, {Lloro}, {Man}, {Mance}, {Martino}, {Mateos}, {McKenzie},
  {McWilliams}, {Miller}, {Mueller}, {Nardini}, {Nelemans}, {Nofrarias},
  {Petiteau}, {Pivato}, {Plagnol}, {Porter}, {Reiche}, {Robertson},
  {Robertson}, {Rossi}, {Russano}, {Schutz}, {Sesana}, {Shoemaker}, {Slutsky},
  {Sopuerta}, {Sumner}, {Tamanini}, {Thorpe}, {Troebs}, {Vallisneri},
  {Vecchio}, {Vetrugno}, {Vitale}, {Volonteri}, {Wanner}, {Ward}, {Wass},
  {Weber}, {Ziemer}, \& {Zweifel}}]{AS2017}
{Amaro-Seoane}, P., {Audley}, H., {Babak}, S., {et~al.} 2017, arXiv e-prints,
  arXiv:1702.00786.
\newblock \doarXiv{1702.00786}

\bibitem[{{Amaro-Seoane} {et~al.}(2023){Amaro-Seoane}, {Andrews}, {Arca Sedda},
  {Askar}, {Baghi}, {Balasov}, {Bartos}, {Bavera}, {Bellovary}, {Berry},
  {Berti}, {Bianchi}, {Blecha}, {Blondin}, {Bogdanovi{\'c}}, {Boissier},
  {Bonetti}, {Bonoli}, {Bortolas}, {Breivik}, {Capelo}, {Caramete},
  {Cattorini}, {Charisi}, {Chaty}, {Chen}, {Chru{\'s}li{\'n}ska}, {Chua},
  {Church}, {Colpi}, {D'Orazio}, {Danielski}, {Davies}, {Dayal}, {De Rosa},
  {Derdzinski}, {Destounis}, {Dotti}, {Du{\r{A}}{\textsterling}an}, {Dvorkin},
  {Fabj}, {Foglizzo}, {Ford}, {Fouvry}, {Franchini}, {Fragos}, {Fryer},
  {Gaspari}, {Gerosa}, {Graziani}, {Groot}, {Habouzit}, {Haggard}, {Haiman},
  {Han}, {Istrate}, {Johansson}, {Khan}, {Kimpson}, {Kokkotas}, {Kong},
  {Korol}, {Kremer}, {Kupfer}, {Lamberts}, {Larson}, {Lau}, {Liu},
  {Lloyd-Ronning}, {Lodato}, {Lupi}, {Ma}, {Maccarone}, {Mandel}, {Mangiagli},
  {Mapelli}, {Mathis}, {Mayer}, {McGee}, {McKernan}, {Miller}, {Mota},
  {Mumpower}, {Nasim}, {Nelemans}, {Noble}, {Pacucci}, {Panessa},
  {Paschalidis}, {Pfister}, {Porquet}, {Quenby}, {Ricarte}, {R{\"o}pke},
  {Regan}, {Rosswog}, {Ruiter}, {Ruiz}, {Runnoe}, {Schneider}, {Schnittman},
  {Secunda}, {Sesana}, {Seto}, {Shao}, {Shapiro}, {Sopuerta}, {Stone},
  {Suvorov}, {Tamanini}, {Tamfal}, {Tauris}, {Temmink}, {Tomsick}, {Toonen},
  {Torres-Orjuela}, {Toscani}, {Tsokaros}, {Unal}, {V{\'a}zquez-Aceves},
  {Valiante}, {van Putten}, {van Roestel}, {Vignali}, {Volonteri}, {Wu},
  {Younsi}, {Yu}, {Zane}, {Zwick}, {Antonini}, {Baibhav}, {Barausse}, {Bonilla
  Rivera}, {Branchesi}, {Branduardi-Raymont}, {Burdge}, {Chakraborty},
  {Cuadra}, {Dage}, {Davis}, {de Mink}, {Decarli}, {Doneva}, {Escoffier},
  {Gandhi}, {Haardt}, {Lousto}, {Nissanke}, {Nordhaus}, {O'Shaughnessy},
  {Portegies Zwart}, {Pound}, {Schussler}, {Sergijenko}, {Spallicci},
  {Vernieri}, \& {Vigna-G{\'o}mez}}]{AS2023}
{Amaro-Seoane}, P., {Andrews}, J., {Arca Sedda}, M., {et~al.} 2023, Living
  Reviews in Relativity, 26, 2, \dodoi{10.1007/s41114-022-00041-y}

\bibitem[{{Andrews} {et~al.}(2020){Andrews}, {Breivik}, {Pankow}, {D'Orazio},
  \& {Safarzadeh}}]{Andrews2020}
{Andrews}, J.~J., {Breivik}, K., {Pankow}, C., {D'Orazio}, D.~J., \&
  {Safarzadeh}, M. 2020, \apjl, 892, L9, \dodoi{10.3847/2041-8213/ab5b9a}

\bibitem[{{Babak} {et~al.}(2021){Babak}, {Hewitson}, \& {Petiteau}}]{Babak2021}
{Babak}, S., {Hewitson}, M., \& {Petiteau}, A. 2021, arXiv e-prints,
  arXiv:2108.01167.
\newblock \doarXiv{2108.01167}

\bibitem[{{Barack} \& {Cutler}(2004)}]{Barack2004}
{Barack}, L., \& {Cutler}, C. 2004, \prd, 69, 082005,
  \dodoi{10.1103/PhysRevD.69.082005}

\bibitem[{{Belczynski} {et~al.}(2010{\natexlab{a}}){Belczynski}, {Benacquista},
  \& {Bulik}}]{Belczynski2010a}
{Belczynski}, K., {Benacquista}, M., \& {Bulik}, T. 2010{\natexlab{a}}, \apj,
  725, 816, \dodoi{10.1088/0004-637X/725/1/816}

\bibitem[{{Belczynski} {et~al.}(2010{\natexlab{b}}){Belczynski}, {Bulik},
  {Fryer}, {Ruiter}, {Valsecchi}, {Vink}, \& {Hurley}}]{Belczynski2010}
{Belczynski}, K., {Bulik}, T., {Fryer}, C.~L., {et~al.} 2010{\natexlab{b}},
  \apj, 714, 1217, \dodoi{10.1088/0004-637X/714/2/1217}

\bibitem[{{Bhattacharya} \& {van den Heuvel}(1991)}]{Bhattacharya1991}
{Bhattacharya}, D., \& {van den Heuvel}, E.~P.~J. 1991, \physrep, 203, 1,
  \dodoi{10.1016/0370-1573(91)90064-S}

\bibitem[{{Breivik} {et~al.}(2020){Breivik}, {Coughlin}, {Zevin}, {Rodriguez},
  {Kremer}, {Ye}, {Andrews}, {Kurkowski}, {Digman}, {Larson}, \&
  {Rasio}}]{Breivik2020}
{Breivik}, K., {Coughlin}, S., {Zevin}, M., {et~al.} 2020, \apj, 898, 71,
  \dodoi{10.3847/1538-4357/ab9d85}

\bibitem[{{Chen} {et~al.}(2021){Chen}, {Tauris}, {Han}, \& {Chen}}]{Chen2021}
{Chen}, H.-L., {Tauris}, T.~M., {Han}, Z., \& {Chen}, X. 2021, \mnras, 503,
  3540, \dodoi{10.1093/mnras/stab670}

\bibitem[{{Chen} {et~al.}(2020){Chen}, {Liu}, \& {Wang}}]{Chen2020}
{Chen}, W.-C., {Liu}, D.-D., \& {Wang}, B. 2020, \apjl, 900, L8,
  \dodoi{10.3847/2041-8213/abae66}

\bibitem[{{Cornish} \& {Robson}(2017)}]{Cornish2017}
{Cornish}, N., \& {Robson}, T. 2017, in Journal of Physics Conference Series,
  Vol. 840, Journal of Physics Conference Series, 012024,
  \dodoi{10.1088/1742-6596/840/1/012024}

\bibitem[{{Cornish} \& {Larson}(2003)}]{Cornish2003}
{Cornish}, N.~J., \& {Larson}, S.~L. 2003, \prd, 67, 103001,
  \dodoi{10.1103/PhysRevD.67.103001}

\bibitem[{{Davis} {et~al.}(2012){Davis}, {Kolb}, \& {Knigge}}]{Davis2012}
{Davis}, P.~J., {Kolb}, U., \& {Knigge}, C. 2012, \mnras, 419, 287,
  \dodoi{10.1111/j.1365-2966.2011.19690.x}

\bibitem[{{Davis} {et~al.}(2010){Davis}, {Kolb}, \& {Willems}}]{Davis2010}
{Davis}, P.~J., {Kolb}, U., \& {Willems}, B. 2010, \mnras, 403, 179,
  \dodoi{10.1111/j.1365-2966.2009.16138.x}

\bibitem[{{Dominik} {et~al.}(2012){Dominik}, {Belczynski}, {Fryer}, {Holz},
  {Berti}, {Bulik}, {Mandel}, \& {O'Shaughnessy}}]{Dominik2012}
{Dominik}, M., {Belczynski}, K., {Fryer}, C., {et~al.} 2012, \apj, 759, 52,
  \dodoi{10.1088/0004-637X/759/1/52}

\bibitem[{{Feng} {et~al.}(2023){Feng}, {Chen}, {Wang}, {Mohanty}, \&
  {Shao}}]{Feng2023}
{Feng}, W.-F., {Chen}, J.-W., {Wang}, Y., {Mohanty}, S.~D., \& {Shao}, Y. 2023,
  \prd, 107, 103035, \dodoi{10.1103/PhysRevD.107.103035}

\bibitem[{{Fragos} {et~al.}(2019){Fragos}, {Andrews}, {Ramirez-Ruiz}, {Meynet},
  {Kalogera}, {Taam}, \& {Zezas}}]{Fragos2019}
{Fragos}, T., {Andrews}, J.~J., {Ramirez-Ruiz}, E., {et~al.} 2019, \apjl, 883,
  L45, \dodoi{10.3847/2041-8213/ab40d1}

\bibitem[{{Fryer} {et~al.}(2012){Fryer}, {Belczynski}, {Wiktorowicz},
  {Dominik}, {Kalogera}, \& {Holz}}]{Fryer2012}
{Fryer}, C.~L., {Belczynski}, K., {Wiktorowicz}, G., {et~al.} 2012, \apj, 749,
  91, \dodoi{10.1088/0004-637X/749/1/91}

\bibitem[{{Fryer} {et~al.}(2022){Fryer}, {Olejak}, \& {Belczynski}}]{Fryer2022}
{Fryer}, C.~L., {Olejak}, A., \& {Belczynski}, K. 2022, \apj, 931, 94,
  \dodoi{10.3847/1538-4357/ac6ac9}

\bibitem[{{Fryer} {et~al.}(1999){Fryer}, {Woosley}, {Herant}, \&
  {Davies}}]{Fryer1999}
{Fryer}, C.~L., {Woosley}, S.~E., {Herant}, M., \& {Davies}, M.~B. 1999, \apj,
  520, 650, \dodoi{10.1086/307467}

\bibitem[{{Gallegos-Garcia} {et~al.}(2022){Gallegos-Garcia}, {Berry}, \&
  {Kalogera}}]{Gallegos-Garcia2022}
{Gallegos-Garcia}, M., {Berry}, C. P.~L., \& {Kalogera}, V. 2022, arXiv
  e-prints, arXiv:2211.15693, \dodoi{10.48550/arXiv.2211.15693}

\bibitem[{{Gao} {et~al.}(2022){Gao}, {Li}, \& {Shao}}]{Gao2022}
{Gao}, S.-J., {Li}, X.-D., \& {Shao}, Y. 2022, \mnras, 514, 1054,
  \dodoi{10.1093/mnras/stac1426}

\bibitem[{{Gompertz} {et~al.}(2020){Gompertz}, {Levan}, \&
  {Tanvir}}]{Gompertz2020}
{Gompertz}, B.~P., {Levan}, A.~J., \& {Tanvir}, N.~R. 2020, \apj, 895, 58,
  \dodoi{10.3847/1538-4357/ab8d24}

\bibitem[{{Hamann} {et~al.}(1995){Hamann}, {Koesterke}, \&
  {Wessolowski}}]{Hamann1995}
{Hamann}, W.~R., {Koesterke}, L., \& {Wessolowski}, U. 1995, \aap, 299, 151

\bibitem[{{Heggie}(1975)}]{Heggie1975}
{Heggie}, D.~C. 1975, \mnras, 173, 729, \dodoi{10.1093/mnras/173.3.729}

\bibitem[{{Hobbs} {et~al.}(2005){Hobbs}, {Lorimer}, {Lyne}, \&
  {Kramer}}]{Hobbs2005}
{Hobbs}, G., {Lorimer}, D.~R., {Lyne}, A.~G., \& {Kramer}, M. 2005, \mnras,
  360, 974, \dodoi{10.1111/j.1365-2966.2005.09087.x}

\bibitem[{{Hurley} {et~al.}(2002){Hurley}, {Tout}, \& {Pols}}]{hurley2002}
{Hurley}, J.~R., {Tout}, C.~A., \& {Pols}, O.~R. 2002, \mnras, 329, 897,
  \dodoi{10.1046/j.1365-8711.2002.05038.x}

\bibitem[{{Iben} \& {Tutukov}(1984)}]{Iben1984}
{Iben}, I., J., \& {Tutukov}, A.~V. 1984, \apjs, 54, 335,
  \dodoi{10.1086/190932}

\bibitem[{{Ivanova} \& {Taam}(2004)}]{Ivanova2004}
{Ivanova}, N., \& {Taam}, R.~E. 2004, \apj, 601, 1058, \dodoi{10.1086/380561}

\bibitem[{{Ivanova} {et~al.}(2013){Ivanova}, {Justham}, {Chen}, {De Marco},
  {Fryer}, {Gaburov}, {Ge}, {Glebbeek}, {Han}, {Li}, {Lu}, {Marsh},
  {Podsiadlowski}, {Potter}, {Soker}, {Taam}, {Tauris}, {van den Heuvel}, \&
  {Webbink}}]{Ivanova2013}
{Ivanova}, N., {Justham}, S., {Chen}, X., {et~al.} 2013, \aapr, 21, 59,
  \dodoi{10.1007/s00159-013-0059-2}

\bibitem[{{Kapil} {et~al.}(2023){Kapil}, {Mandel}, {Berti}, \&
  {M{\"u}ller}}]{Kapil2022}
{Kapil}, V., {Mandel}, I., {Berti}, E., \& {M{\"u}ller}, B. 2023, \mnras, 519,
  5893, \dodoi{10.1093/mnras/stad019}

\bibitem[{{Kemp} {et~al.}(2022){Kemp}, {Karakas}, {Casey}, {Kobayashi}, \&
  {Izzard}}]{Kemp2022}
{Kemp}, A.~J., {Karakas}, A.~I., {Casey}, A.~R., {Kobayashi}, C., \& {Izzard},
  R.~G. 2022, \mnras, 509, 1175, \dodoi{10.1093/mnras/stab3103}

\bibitem[{{Kiel} \& {Hurley}(2006)}]{Kiel2006}
{Kiel}, P.~D., \& {Hurley}, J.~R. 2006, \mnras, 369, 1152,
  \dodoi{10.1111/j.1365-2966.2006.10400.x}

\bibitem[{{Klencki} {et~al.}(2021){Klencki}, {Nelemans}, {Istrate}, \&
  {Chruslinska}}]{Klencki2021}
{Klencki}, J., {Nelemans}, G., {Istrate}, A.~G., \& {Chruslinska}, M. 2021,
  \aap, 645, A54, \dodoi{10.1051/0004-6361/202038707}

\bibitem[{{Kobulnicky} \& {Fryer}(2007)}]{Kobulnicky2007}
{Kobulnicky}, H.~A., \& {Fryer}, C.~L. 2007, \apj, 670, 747,
  \dodoi{10.1086/522073}

\bibitem[{{Korol} {et~al.}(2022){Korol}, {Hallakoun}, {Toonen}, \&
  {Karnesis}}]{Korol2022}
{Korol}, V., {Hallakoun}, N., {Toonen}, S., \& {Karnesis}, N. 2022, \mnras,
  511, 5936, \dodoi{10.1093/mnras/stac415}

\bibitem[{{Korol} {et~al.}(2018){Korol}, {Koop}, \& {Rossi}}]{Korol2018}
{Korol}, V., {Koop}, O., \& {Rossi}, E.~M. 2018, \apjl, 866, L20,
  \dodoi{10.3847/2041-8213/aae587}

\bibitem[{{Korol} {et~al.}(2017){Korol}, {Rossi}, {Groot}, {Nelemans},
  {Toonen}, \& {Brown}}]{Korol2017}
{Korol}, V., {Rossi}, E.~M., {Groot}, P.~J., {et~al.} 2017, \mnras, 470, 1894,
  \dodoi{10.1093/mnras/stx1285}

\bibitem[{{Kroupa} {et~al.}(1993){Kroupa}, {Tout}, \& {Gilmore}}]{Kroupa1993}
{Kroupa}, P., {Tout}, C.~A., \& {Gilmore}, G. 1993, \mnras, 262, 545,
  \dodoi{10.1093/mnras/262.3.545}

\bibitem[{{Lamberts} {et~al.}(2019){Lamberts}, {Blunt}, {Littenberg},
  {Garrison-Kimmel}, {Kupfer}, \& {Sanderson}}]{Lamberts2019}
{Lamberts}, A., {Blunt}, S., {Littenberg}, T.~B., {et~al.} 2019, \mnras, 490,
  5888, \dodoi{10.1093/mnras/stz2834}

\bibitem[{{Lamberts} {et~al.}(2018){Lamberts}, {Garrison-Kimmel}, {Hopkins},
  {Quataert}, {Bullock}, {Faucher-Gigu{\`e}re}, {Wetzel}, {Kere{\v{s}}},
  {Drango}, \& {Sanderson}}]{Lamberts2018}
{Lamberts}, A., {Garrison-Kimmel}, S., {Hopkins}, P.~F., {et~al.} 2018, \mnras,
  480, 2704, \dodoi{10.1093/mnras/sty2035}

\bibitem[{{Larson} {et~al.}(2000){Larson}, {Hiscock}, \&
  {Hellings}}]{Larson2000}
{Larson}, S.~L., {Hiscock}, W.~A., \& {Hellings}, R.~W. 2000, \prd, 62, 062001,
  \dodoi{10.1103/PhysRevD.62.062001}

\bibitem[{{Lau} {et~al.}(2020){Lau}, {Mandel}, {Vigna-G{\'o}mez}, {Neijssel},
  {Stevenson}, \& {Sesana}}]{Lau2020}
{Lau}, M. Y.~M., {Mandel}, I., {Vigna-G{\'o}mez}, A., {et~al.} 2020, \mnras,
  492, 3061, \dodoi{10.1093/mnras/staa002}

\bibitem[{{Li} \& {Paczy{\'n}ski}(1998)}]{Li1998}
{Li}, L.-X., \& {Paczy{\'n}ski}, B. 1998, \apjl, 507, L59,
  \dodoi{10.1086/311680}

\bibitem[{{Liu} {et~al.}(2010){Liu}, {Han}, {Zhang}, \& {Zhang}}]{Liu2010}
{Liu}, J., {Han}, Z., {Zhang}, F., \& {Zhang}, Y. 2010, \apj, 719, 1546,
  \dodoi{10.1088/0004-637X/719/2/1546}

\bibitem[{{Liu} \& {Zhang}(2014)}]{Liu2014}
{Liu}, J., \& {Zhang}, Y. 2014, \pasp, 126, 211, \dodoi{10.1086/675721}

\bibitem[{{Luo} {et~al.}(2016){Luo}, {Chen}, {Duan}, {Gong}, {Hu}, {Ji}, {Liu},
  {Mei}, {Milyukov}, {Sazhin}, {Shao}, {Toth}, {Tu}, {Wang}, {Wang}, {Yeh},
  {Zhan}, {Zhang}, {Zharov}, \& {Zhou}}]{Luo2016}
{Luo}, J., {Chen}, L.-S., {Duan}, H.-Z., {et~al.} 2016, Classical and Quantum
  Gravity, 33, 035010, \dodoi{10.1088/0264-9381/33/3/035010}

\bibitem[{{Manchester} {et~al.}(2005){Manchester}, {Hobbs}, {Teoh}, \&
  {Hobbs}}]{Manchester2005}
{Manchester}, R.~N., {Hobbs}, G.~B., {Teoh}, A., \& {Hobbs}, M. 2005, \aj, 129,
  1993, \dodoi{10.1086/428488}

\bibitem[{{Mandel} \& {M{\"u}ller}(2020)}]{Mandel2020}
{Mandel}, I., \& {M{\"u}ller}, B. 2020, \mnras, 499, 3214,
  \dodoi{10.1093/mnras/staa3043}

\bibitem[{{Marchant} {et~al.}(2021){Marchant}, {Pappas}, {Gallegos-Garcia},
  {Berry}, {Taam}, {Kalogera}, \& {Podsiadlowski}}]{Marchant2021}
{Marchant}, P., {Pappas}, K. M.~W., {Gallegos-Garcia}, M., {et~al.} 2021, \aap,
  650, A107, \dodoi{10.1051/0004-6361/202039992}

\bibitem[{{Neijssel} {et~al.}(2019){Neijssel}, {Vigna-G{\'o}mez}, {Stevenson},
  {Barrett}, {Gaebel}, {Broekgaarden}, {de Mink}, {Sz{\'e}csi}, {Vinciguerra},
  \& {Mandel}}]{Neijssel2019}
{Neijssel}, C.~J., {Vigna-G{\'o}mez}, A., {Stevenson}, S., {et~al.} 2019,
  \mnras, 490, 3740, \dodoi{10.1093/mnras/stz2840}

\bibitem[{{Nelemans} {et~al.}(2001){Nelemans}, {Yungelson}, \& {Portegies
  Zwart}}]{Nelemans2001}
{Nelemans}, G., {Yungelson}, L.~R., \& {Portegies Zwart}, S.~F. 2001, \aap,
  375, 890, \dodoi{10.1051/0004-6361:20010683}

\bibitem[{{Nelson} \& {Eggleton}(2001)}]{Nelson2001}
{Nelson}, C.~A., \& {Eggleton}, P.~P. 2001, \apj, 552, 664,
  \dodoi{10.1086/320560}

\bibitem[{{Nitz} {et~al.}(2021){Nitz}, {Kumar}, {Wang}, {Kastha}, {Wu},
  {Sch{\"a}fer}, {Dhurkunde}, \& {Capano}}]{Nitz2021}
{Nitz}, A.~H., {Kumar}, S., {Wang}, Y.-F., {et~al.} 2021, arXiv e-prints,
  arXiv:2112.06878.
\newblock \doarXiv{2112.06878}

\bibitem[{{Olejak} {et~al.}(2021){Olejak}, {Belczynski}, \&
  {Ivanova}}]{Olejak2021}
{Olejak}, A., {Belczynski}, K., \& {Ivanova}, N. 2021, \aap, 651, A100,
  \dodoi{10.1051/0004-6361/202140520}

\bibitem[{{Packet}(1981)}]{Packet1981}
{Packet}, W. 1981, \aap, 102, 17

\bibitem[{{Paczynski}(1986)}]{Paczynski1986}
{Paczynski}, B. 1986, \apjl, 308, L43, \dodoi{10.1086/184740}

\bibitem[{{Pavlovskii} {et~al.}(2017){Pavlovskii}, {Ivanova}, {Belczynski}, \&
  {Van}}]{Pavlovskii2017}
{Pavlovskii}, K., {Ivanova}, N., {Belczynski}, K., \& {Van}, K.~X. 2017,
  \mnras, 465, 2092, \dodoi{10.1093/mnras/stw2786}

\bibitem[{{Peters} \& {Mathews}(1963)}]{Peters1963}
{Peters}, P.~C., \& {Mathews}, J. 1963, Physical Review, 131, 435,
  \dodoi{10.1103/PhysRev.131.435}

\bibitem[{{Qin} {et~al.}(2023){Qin}, {Jiang}, \& {Chen}}]{Qin2023}
{Qin}, K., {Jiang}, L., \& {Chen}, W.-C. 2023, \apj, 944, 83,
  \dodoi{10.3847/1538-4357/acb340}

\bibitem[{{Ribas} {et~al.}(2005){Ribas}, {Jordi}, {Vilardell}, {Fitzpatrick},
  {Hilditch}, \& {Guinan}}]{Ribas2005}
{Ribas}, I., {Jordi}, C., {Vilardell}, F., {et~al.} 2005, \apjl, 635, L37,
  \dodoi{10.1086/499161}

\bibitem[{{Robson} {et~al.}(2019){Robson}, {Cornish}, \& {Liu}}]{Robson2019}
{Robson}, T., {Cornish}, N.~J., \& {Liu}, C. 2019, Classical and Quantum
  Gravity, 36, 105011, \dodoi{10.1088/1361-6382/ab1101}

\bibitem[{{Ruan} {et~al.}(2020){Ruan}, {Guo}, {Cai}, \& {Zhang}}]{Ruan2020}
{Ruan}, W.-H., {Guo}, Z.-K., {Cai}, R.-G., \& {Zhang}, Y.-Z. 2020,
  International Journal of Modern Physics A, 35, 2050075,
  \dodoi{10.1142/S0217751X2050075X}

\bibitem[{{Ruiter} {et~al.}(2010){Ruiter}, {Belczynski}, {Benacquista},
  {Larson}, \& {Williams}}]{Ruiter2010}
{Ruiter}, A.~J., {Belczynski}, K., {Benacquista}, M., {Larson}, S.~L., \&
  {Williams}, G. 2010, \apj, 717, 1006, \dodoi{10.1088/0004-637X/717/2/1006}

\bibitem[{{Scherbak} \& {Fuller}(2023)}]{Scherbak2023}
{Scherbak}, P., \& {Fuller}, J. 2023, \mnras, 518, 3966,
  \dodoi{10.1093/mnras/stac3313}

\bibitem[{{Sesana} {et~al.}(2020){Sesana}, {Lamberts}, \&
  {Petiteau}}]{Sesana2020}
{Sesana}, A., {Lamberts}, A., \& {Petiteau}, A. 2020, \mnras, 494, L75,
  \dodoi{10.1093/mnrasl/slaa039}

\bibitem[{{Seto}(2019)}]{Seto2019}
{Seto}, N. 2019, \mnras, 489, 4513, \dodoi{10.1093/mnras/stz2439}

\bibitem[{{Shao}(2022)}]{Shao2022}
{Shao}, Y. 2022, Research in Astronomy and Astrophysics, 22, 122002,
  \dodoi{10.1088/1674-4527/ac995e}

\bibitem[{{Shao} \& {Li}(2014)}]{Shao2014}
{Shao}, Y., \& {Li}, X.-D. 2014, \apj, 796, 37,
  \dodoi{10.1088/0004-637X/796/1/37}

\bibitem[{{Shao} \& {Li}(2015)}]{Shao2015}
---. 2015, \apj, 809, 99, \dodoi{10.1088/0004-637X/809/1/99}

\bibitem[{{Shao} \& {Li}(2018)}]{Shao2018}
---. 2018, \apj, 867, 124, \dodoi{10.3847/1538-4357/aae648}

\bibitem[{{Shao} \& {Li}(2021)}]{Shao2021}
---. 2021, \apj, 920, 81, \dodoi{10.3847/1538-4357/ac173e}

\bibitem[{{Soberman} {et~al.}(1997){Soberman}, {Phinney}, \& {van den
  Heuvel}}]{Soberman1997}
{Soberman}, G.~E., {Phinney}, E.~S., \& {van den Heuvel}, E.~P.~J. 1997, \aap,
  327, 620, \dodoi{10.48550/arXiv.astro-ph/9703016}

\bibitem[{{Tauris}(2018)}]{Tauris2018}
{Tauris}, T.~M. 2018, \prl, 121, 131105, \dodoi{10.1103/PhysRevLett.121.131105}

\bibitem[{{Tauris} \& {van den Heuvel}(2023)}]{Tauris2023}
{Tauris}, T.~M., \& {van den Heuvel}, E. P.~J. 2023, {Physics of Binary Star
  Evolution. From Stars to X-ray Binaries and Gravitational Wave Sources}

\bibitem[{{Tauris} {et~al.}(2017){Tauris}, {Kramer}, {Freire}, {Wex}, {Janka},
  {Langer}, {Podsiadlowski}, {Bozzo}, {Chaty}, {Kruckow}, {van den Heuvel},
  {Antoniadis}, {Breton}, \& {Champion}}]{Tauris2017}
{Tauris}, T.~M., {Kramer}, M., {Freire}, P.~C.~C., {et~al.} 2017, \apj, 846,
  170, \dodoi{10.3847/1538-4357/aa7e89}

\bibitem[{{The LIGO Scientific Collaboration} {et~al.}(2021){The LIGO
  Scientific Collaboration}, {the Virgo Collaboration}, {the KAGRA
  Collaboration}, {Abbott}, {Abbott}, {Acernese}, {Ackley}, {Adams},
  {Adhikari}, {Adhikari}, {Adya}, {Affeldt}, {Agarwal}, {Agathos}, {Agatsuma},
  {Aggarwal}, {Aguiar}, {Aiello}, {Ain}, {Ajith}, {Akcay}, {Akutsu},
  {Albanesi}, {Allocca}, {Altin}, {Amato}, {Anand}, {Anand}, {Ananyeva},
  {Anderson}, {Anderson}, {Ando}, {Andrade}, {Andres}, {Andri{\'c}},
  {Angelova}, {Ansoldi}, {Antelis}, {Antier}, {Appert}, {Arai}, {Arai}, {Arai},
  {Araki}, {Araya}, {Araya}, {Areeda}, {Ar{\`e}ne}, {Aritomi}, {Arnaud},
  {Arogeti}, {Aronson}, {Arun}, {Asada}, {Asali}, {Ashton}, {Aso}, {Assiduo},
  {Aston}, {Astone}, {Aubin}, {Austin}, {Babak}, {Badaracco}, {Bader},
  {Badger}, {Bae}, {Bae}, {Baer}, {Bagnasco}, {Bai}, {Baiotti}, {Baird},
  {Bajpai}, {Ball}, {Ballardin}, {Ballmer}, {Balsamo}, {Baltus}, {Banagiri},
  {Bankar}, {Barayoga}, {Barbieri}, {Barish}, {Barker}, {Barneo}, {Barone},
  {Barr}, {Barsotti}, {Barsuglia}, {Barta}, {Bartlett}, {Barton}, {Bartos},
  {Bassiri}, {Basti}, {Bawaj}, {Bayley}, {Baylor}, {Bazzan}, {B{\'e}csy},
  {Bedakihale}, {Bejger}, {Belahcene}, {Benedetto}, {Beniwal}, {Bennett},
  {Bentley}, {BenYaala}, {Bergamin}, {Berger}, {Bernuzzi}, {Berry},
  {Bersanetti}, {Bertolini}, {Betzwieser}, {Beveridge}, {Bhandare}, {Bhardwaj},
  {Bhattacharjee}, {Bhaumik}, {Bilenko}, {Billingsley}, {Bini}, {Birney},
  {Birnholtz}, {Biscans}, {Bischi}, {Biscoveanu}, {Bisht}, {Biswas}, {Bitossi},
  {Bizouard}, {Blackburn}, {Blair}, {Blair}, {Blair}, {Bobba}, {Bode}, {Boer},
  {Bogaert}, {Boldrini}, {Bonavena}, {Bondu}, {Bonilla}, {Bonnand}, {Booker},
  {Boom}, {Bork}, {Boschi}, {Bose}, {Bose}, {Bossilkov}, {Boudart},
  {Bouffanais}, {Bozzi}, {Bradaschia}, {Brady}, {Bramley}, {Branch},
  {Branchesi}, {Brandt}, {Brau}, {Breschi}, {Briant}, {Briggs}, {Brillet},
  {Brinkmann}, {Brockill}, {Brooks}, {Brooks}, {Brown}, {Brunett}, {Bruno},
  {Bruntz}, {Bryant}, {Bulik}, {Bulten}, {Buonanno}, {Buscicchio}, {Buskulic},
  {Buy}, {Byer}, {Cabourn Davies}, {Cadonati}, {Cagnoli}, {Cahillane},
  {Calder{\'o}n Bustillo}, {Callaghan}, {Callister}, {Calloni}, {Cameron},
  {Camp}, {Canepa}, {Canevarolo}, {Cannavacciuolo}, {Cannon}, {Cao}, {Cao},
  {Capocasa}, {Capote}, {Carapella}, {Carbognani}, {Carlin}, {Carney},
  {Carpinelli}, {Carrillo}, {Carullo}, {Carver}, {Casanueva Diaz}, {Casentini},
  {Castaldi}, {Caudill}, {Cavagli{\`a}}, {Cavalier}, {Cavalieri}, {Ceasar},
  {Cella}, {Cerd{\'a}-Dur{\'a}n}, {Cesarini}, {Chaibi}, {Chakravarti},
  {Chalathadka Subrahmanya}, {Champion}, {Chan}, {Chan}, {Chan}, {Chan},
  {Chan}, {Chandra}, {Chanial}, {Chao}, {Chapman-Bird}, {Charlton}, {Chase},
  {Chassande-Mottin}, {Chatterjee}, {Chatterjee}, {Chatterjee}, {Chaturvedi},
  {Chaty}, {Chatziioannou}, {Chen}, {Chen}, {Chen}, {Chen}, {Chen}, {Chen},
  {Chen}, {Chen}, {Cheng}, {Cheong}, {Cheung}, {Chia}, {Chiadini}, {Chiang},
  {Chiarini}, {Chierici}, {Chincarini}, {Chiofalo}, {Chiummo}, {Cho}, {Cho},
  {Choudhary}, {Choudhary}, {Christensen}, {Chu}, {Chu}, {Chu}, {Chua},
  {Chung}, {Ciani}, {Ciecielag}, {Cie{\'s}lar}, {Cifaldi}, {Ciobanu}, {Ciolfi},
  {Cipriano}, {Cirone}, {Clara}, {Clark}, {Clark}, {Clarke}, {Clearwater},
  {Clesse}, {Cleva}, {Coccia}, {Codazzo}, {Cohadon}, {Cohen}, {Cohen},
  {Colleoni}, {Collette}, {Colombo}, {Colpi}, {Compton}, {Constancio}, {Conti},
  {Cooper}, {Corban}, {Corbitt}, {Cordero-Carri{\'o}n}, {Corezzi}, {Corley},
  {Cornish}, {Corre}, {Corsi}, {Cortese}, {Costa}, {Cotesta}, {Coughlin},
  {Coulon}, {Countryman}, {Cousins}, {Couvares}, {Coward}, {Cowart}, {Coyne},
  {Coyne}, {Creighton}, {Creighton}, {Criswell}, {Croquette}, {Crowder},
  {Cudell}, {Cullen}, {Cumming}, {Cummings}, {Cunningham}, {Cuoco},
  {Cury{\l}o}, {Dabadie}, {Dal Canton}, {Dall'Osso}, {D{\'a}lya}, {Dana},
  {DaneshgaranBajastani}, {D'Angelo}, {Danila}, {Danilishin}, {D'Antonio},
  {Danzmann}, {Darsow-Fromm}, {Dasgupta}, {Datrier}, {Datta}, {Dattilo},
  {Dave}, {Davier}, {Davis}, {Davis}, {Daw}, {de Alarc{\'o}n}, {Dean}, {DeBra},
  {Deenadayalan}, {Degallaix}, {De Laurentis}, {Del{\'e}glise}, {Del Favero},
  {De Lillo}, {De Lillo}, {Del Pozzo}, {DeMarchi}, {De Matteis}, {D'Emilio},
  {Demos}, {Dent}, {Depasse}, {De Pietri}, {De Rosa}, {De Rossi}, {DeSalvo},
  {De Simone}, {Dhurandhar}, {D{\'\i}az}, {Diaz-Ortiz}, {Didio}, {Dietrich},
  {Di Fiore}, {Di Fronzo}, {Di Giorgio}, {Di Giovanni}, {Di Giovanni}, {Di
  Girolamo}, {Di Lieto}, {Ding}, {Di Pace}, {Di Palma}, {Di Renzo},
  {Divakarla}, {Dmitriev}, {Doctor}, {D'Onofrio}, {Donovan}, {Dooley},
  {Doravari}, {Dorrington}, {Drago}, {Driggers}, {Drori}, {Ducoin}, {Dupej},
  {Durante}, {D'Urso}, {Duverne}, {Dwyer}, {Eassa}, {Easter}, {Ebersold},
  {Eckhardt}, {Eddolls}, {Edelman}, {Edo}, {Edy}, {Effler}, {Eguchi},
  {Eichholz}, {Eikenberry}, {Eisenmann}, {Eisenstein}, {Ejlli}, {Engelby},
  {Enomoto}, {Errico}, {Essick}, {Estell{\'e}s}, {Estevez}, {Etienne}, {Etzel},
  {Evans}, {Evans}, {Ewing}, {Fafone}, {Fair}, {Fairhurst}, {Farah}, {Farinon},
  {Farr}, {Farr}, {Farrow}, {Fauchon-Jones}, {Favaro}, {Favata}, {Fays},
  {Fazio}, {Feicht}, {Fejer}, {Fenyvesi}, {Ferguson}, {Fernandez-Galiana},
  {Ferrante}, {Ferreira}, {Fidecaro}, {Figura}, {Fiori}, {Fishbach}, {Fisher},
  {Fittipaldi}, {Fiumara}, {Flaminio}, {Floden}, {Fong}, {Font}, {Fornal},
  {Forsyth}, {Franke}, {Frasca}, {Frasconi}, {Frederick}, {Freed}, {Frei},
  {Freise}, {Frey}, {Fritschel}, {Frolov}, {Fronz{\'e}}, {Fujii}, {Fujikawa},
  {Fukunaga}, {Fukushima}, {Fulda}, {Fyffe}, {Gabbard}, {Gabella}, {Gadre},
  {Gair}, {Gais}, {Galaudage}, {Gamba}, {Ganapathy}, {Ganguly}, {Gao},
  {Gaonkar}, {Garaventa}, {Garc{\'\i}a}, {Garc{\'\i}a-N{\'u}{\~n}ez},
  {Garc{\'\i}a-Quir{\'o}s}, {Garufi}, {Gateley}, {Gaudio}, {Gayathri}, {Ge},
  {Gemme}, {Gennai}, {George}, {George}, {Gerberding}, {Gergely}, {Gewecke},
  {Ghonge}, {Ghosh}, {Ghosh}, {Ghosh}, {Ghosh}, {Giacomazzo}, {Giacoppo},
  {Giaime}, {Giardina}, {Gibson}, {Gier}, {Giesler}, {Giri}, {Gissi},
  {Glanzer}, {Gleckl}, {Godwin}, {Goetz}, {Goetz}, {Gohlke}, {Golomb},
  {Goncharov}, {Gonz{\'a}lez}, {Gopakumar}, {Gosselin}, {Gouaty}, {Gould},
  {Grace}, {Grado}, {Granata}, {Granata}, {Grant}, {Gras}, {Grassia}, {Gray},
  {Gray}, {Greco}, {Green}, {Green}, {Gretarsson}, {Gretarsson}, {Griffith},
  {Griffiths}, {Griggs}, {Grignani}, {Grimaldi}, {Grimm}, {Grote}, {Grunewald},
  {Gruning}, {Guerra}, {Guidi}, {Guimaraes}, {Guix{\'e}}, {Gulati}, {Guo},
  {Guo}, {Gupta}, {Gupta}, {Gupta}, {Gustafson}, {Gustafson}, {Guzman}, {Ha},
  {Haegel}, {Hagiwara}, {Haino}, {Halim}, {Hall}, {Hamilton}, {Hammond}, {Han},
  {Haney}, {Hanks}, {Hanna}, {Hannam}, {Hannuksela}, {Hansen}, {Hansen},
  {Hanson}, {Harder}, {Hardwick}, {Haris}, {Harms}, {Harry}, {Harry},
  {Hartwig}, {Hasegawa}, {Haskell}, {Hasskew}, {Haster}, {Hattori}, {Haughian},
  {Hayakawa}, {Hayama}, {Hayes}, {Healy}, {Heidmann}, {Heidt}, {Heintze},
  {Heinze}, {Heinzel}, {Heitmann}, {Hellman}, {Hello}, {Helmling-Cornell},
  {Hemming}, {Hendry}, {Heng}, {Hennes}, {Hennig}, {Hennig}, {Hernandez},
  {Hernandez Vivanco}, {Heurs}, {Hild}, {Hill}, {Himemoto}, {Hines},
  {Hiranuma}, {Hirata}, {Hirose}, {Hochheim}, {Hofman}, {Hohmann}, {Holcomb},
  {Holland}, {Holley-Bockelmann}, {Hollows}, {Holmes}, {Holt}, {Holz}, {Hong},
  {Hopkins}, {Hough}, {Hourihane}, {Howell}, {Hoy}, {Hoyland}, {Hreibi},
  {Hsieh}, {Hsu}, {Huang}, {Huang}, {Huang}, {Huang}, {Huang}, {Huang},
  {H{\"u}bner}, {Huddart}, {Hughey}, {Hui}, {Hui}, {Husa}, {Huttner},
  {Huxford}, {Huynh-Dinh}, {Ide}, {Idzkowski}, {Iess}, {Ikenoue}, {Imam},
  {Inayoshi}, {Ingram}, {Inoue}, {Ioka}, {Isi}, {Isleif}, {Ito}, {Itoh},
  {Iyer}, {Izumi}, {JaberianHamedan}, {Jacqmin}, {Jadhav}, {Jadhav}, {James},
  {Jan}, {Jani}, {Janquart}, {Janssens}, {Janthalur}, {Jaranowski}, {Jariwala},
  {Jaume}, {Jenkins}, {Jenner}, {Jeon}, {Jeunon}, {Jia}, {Jin}, {Johns},
  {Johnson-McDaniel}, {Jones}, {Jones}, {Jones}, {Jones}, {Jones}, {Jonker},
  {Ju}, {Jung}, {Jung}, {Junker}, {Juste}, {Kaihotsu}, {Kajita}, {Kakizaki},
  {Kalaghatgi}, {Kalogera}, {Kamai}, {Kamiizumi}, {Kanda}, {Kandhasamy},
  {Kang}, {Kanner}, {Kao}, {Kapadia}, {Kapasi}, {Karat}, {Karathanasis},
  {Karki}, {Kashyap}, {Kasprzack}, {Kastaun}, {Katsanevas}, {Katsavounidis},
  {Katzman}, {Kaur}, {Kawabe}, {Kawaguchi}, {Kawai}, {Kawasaki},
  {K{\'e}f{\'e}lian}, {Keitel}, {Key}, {Khadka}, {Khalili}, {Khan}, {Khazanov},
  {Khetan}, {Khursheed}, {Kijbunchoo}, {Kim}, {Kim}, {Kim}, {Kim}, {Kim},
  {Kim}, {Kimball}, {Kimura}, {Kinley-Hanlon}, {Kirchhoff}, {Kissel}, {Kita},
  {Kitazawa}, {Kleybolte}, {Klimenko}, {Knee}, {Knowles}, {Knyazev}, {Koch},
  {Koekoek}, {Kojima}, {Kokeyama}, {Koley}, {Kolitsidou}, {Kolstein}, {Komori},
  {Kondrashov}, {Kong}, {Kontos}, {Koper}, {Korobko}, {Kotake}, {Kovalam},
  {Kozak}, {Kozakai}, {Kozu}, {Kringel}, {Krishnendu}, {Kr{\'o}lak}, {Kuehn},
  {Kuei}, {Kuijer}, {Kulkarni}, {Kumar}, {Kumar}, {Kumar}, {Kumar}, {Kume},
  {Kuns}, {Kuo}, {Kuo}, {Kuromiya}, {Kuroyanagi}, {Kusayanagi}, {Kuwahara},
  {Kwak}, {Lagabbe}, {Laghi}, {Lalande}, {Lam}, {Lamberts}, {Landry}, {Lane},
  {Lang}, {Lange}, {Lantz}, {La Rosa}, {Lartaux-Vollard}, {Lasky}, {Laxen},
  {Lazzarini}, {Lazzaro}, {Leaci}, {Leavey}, {Lecoeuche}, {Lee}, {Lee}, {Lee},
  {Lee}, {Lee}, {Lee}, {Lehmann}, {Lema{\^\i}tre}, {Leonardi}, {Leroy},
  {Letendre}, {Levesque}, {Levin}, {Leviton}, {Leyde}, {Li}, {Li}, {Li}, {Li},
  {Li}, {Li}, {Lin}, {Lin}, {Lin}, {Lin}, {Lin}, {Linde}, {Linker}, {Linley},
  {Littenberg}, {Liu}, {Liu}, {Liu}, {Liu}, {Llamas}, {Llorens-Monteagudo},
  {Lo}, {Lockwood}, {Loh}, {London}, {Longo}, {Lopez}, {Lopez Portilla},
  {Lorenzini}, {Loriette}, {Lormand}, {Losurdo}, {Lott}, {Lough}, {Lousto},
  {Lovelace}, {Lucaccioni}, {L{\"u}ck}, {Lumaca}, {Lundgren}, {Luo}, {Lynam},
  {Macas}, {MacInnis}, {Macleod}, {MacMillan}, {Macquet}, {Maga{\~n}a
  Hernandez}, {Magazz{\`u}}, {Magee}, {Maggiore}, {Magnozzi}, {Mahesh},
  {Majorana}, {Makarem}, {Maksimovic}, {Maliakal}, {Malik}, {Man}, {Mandic},
  {Mangano}, {Mango}, {Mansell}, {Manske}, {Mantovani}, {Mapelli},
  {Marchesoni}, {Marchio}, {Marion}, {Mark}, {M{\'a}rka}, {M{\'a}rka},
  {Markakis}, {Markosyan}, {Markowitz}, {Maros}, {Marquina}, {Marsat},
  {Martelli}, {Martin}, {Martin}, {Martinez}, {Martinez}, {Martinez},
  {Martinovic}, {Martynov}, {Marx}, {Masalehdan}, {Mason}, {Massera},
  {Masserot}, {Massinger}, {Masso-Reid}, {Mastrogiovanni}, {Matas},
  {Mateu-Lucena}, {Matichard}, {Matiushechkina}, {Mavalvala}, {McCann},
  {McCarthy}, {McClelland}, {McClincy}, {McCormick}, {McCuller}, {McGhee},
  {McGuire}, {McIsaac}, {McIver}, {McRae}, {McWilliams}, {Meacher}, {Mehmet},
  {Mehta}, {Meijer}, {Melatos}, {Melchor}, {Mendell}, {Menendez-Vazquez},
  {Menoni}, {Mercer}, {Mereni}, {Merfeld}, {Merilh}, {Merritt}, {Merzougui},
  {Meshkov}, {Messenger}, {Messick}, {Meyers}, {Meylahn}, {Mhaske}, {Miani},
  {Miao}, {Michaloliakos}, {Michel}, {Michimura}, {Middleton}, {Milano},
  {Miller}, {Miller}, {Miller}, {Millhouse}, {Mills}, {Milotti}, {Minazzoli},
  {Minenkov}, {Mio}, {Mir}, {Miravet-Ten{\'e}s}, {Mishra}, {Mishra}, {Mistry},
  {Mitra}, {Mitrofanov}, {Mitselmakher}, {Mittleman}, {Miyakawa}, {Miyamoto},
  {Miyazaki}, {Miyo}, {Miyoki}, {Mo}, {Modafferi}, {Moguel}, {Mogushi},
  {Mohapatra}, {Mohite}, {Molina}, {Molina-Ruiz}, {Mondin}, {Montani}, {Moore},
  {Moraru}, {Morawski}, {More}, {Moreno}, {Moreno}, {Mori}, {Morisaki},
  {Moriwaki}, {Morr{\'a}s}, {Mours}, {Mow-Lowry}, {Mozzon}, {Muciaccia},
  {Mukherjee}, {Mukherjee}, {Mukherjee}, {Mukherjee}, {Mukherjee}, {Mukund},
  {Mullavey}, {Munch}, {Mu{\~n}iz}, {Murray}, {Musenich}, {Muusse}, {Nadji},
  {Nagano}, {Nagano}, {Nagar}, {Nakamura}, {Nakano}, {Nakano}, {Nakashima},
  {Nakayama}, {Napolano}, {Nardecchia}, {Narikawa}, {Naticchioni}, {Nayak},
  {Nayak}, {Negishi}, {Neil}, {Neilson}, {Nelemans}, {Nelson}, {Nery},
  {Neubauer}, {Neunzert}, {Ng}, {Ng}, {Nguyen}, {Nguyen}, {Nguyen}, {Nguyen
  Quynh}, {Ni}, {Nichols}, {Nishizawa}, {Nissanke}, {Nitoglia}, {Nocera},
  {Norman}, {North}, {Nozaki}, {Nu{\~n}o Siles}, {Nuttall}, {Oberling},
  {O'Brien}, {Obuchi}, {O'Dell}, {Oelker}, {Ogaki}, {Oganesyan}, {Oh}, {Oh},
  {Oh}, {Ohashi}, {Ohishi}, {Ohkawa}, {Ohme}, {Ohta}, {Okada}, {Okutani},
  {Okutomi}, {Olivetto}, {Oohara}, {Ooi}, {Oram}, {O'Reilly}, {Ormiston},
  {Ormsby}, {Ortega}, {O'Shaughnessy}, {O'Shea}, {Oshino}, {Ossokine},
  {Osthelder}, {Otabe}, {Ottaway}, {Overmier}, {Pace}, {Pagano}, {Page},
  {Pagliaroli}, {Pai}, {Pai}, {Palamos}, {Palashov}, {Palomba}, {Pan}, {Pan},
  {Panda}, {Pang}, {Pang}, {Pankow}, {Pannarale}, {Pant}, {Panther},
  {Paoletti}, {Paoli}, {Paolone}, {Parisi}, {Park}, {Park}, {Parker},
  {Pascucci}, {Pasqualetti}, {Passaquieti}, {Passuello}, {Patel}, {Pathak},
  {Patricelli}, {Patron}, {Paul}, {Payne}, {Pedraza}, {Pegoraro}, {Pele},
  {Pe{\~n}a Arellano}, {Penn}, {Perego}, {Pereira}, {Pereira}, {Perez},
  {P{\'e}rigois}, {Perkins}, {Perreca}, {Perri{\`e}s}, {Petermann},
  {Petterson}, {Pfeiffer}, {Pham}, {Phukon}, {Piccinni}, {Pichot},
  {Piendibene}, {Piergiovanni}, {Pierini}, {Pierro}, {Pillant}, {Pillas},
  {Pilo}, {Pinard}, {Pinto}, {Pinto}, {Piotrzkowski}, {Piotrzkowski},
  {Pirello}, {Pitkin}, {Placidi}, {Planas}, {Plastino}, {Pluchar}, {Poggiani},
  {Polini}, {Pong}, {Ponrathnam}, {Popolizio}, {Porter}, {Poulton}, {Powell},
  {Pracchia}, {Pradier}, {Prajapati}, {Prasai}, {Prasanna}, {Pratten},
  {Principe}, {Prodi}, {Prokhorov}, {Prosposito}, {Prudenzi}, {Puecher},
  {Punturo}, {Puosi}, {Puppo}, {P{\"u}rrer}, {Qi}, {Quetschke},
  {Quitzow-James}, {Qutob}, {Raab}, {Raaijmakers}, {Radkins}, {Radulesco},
  {Raffai}, {Rail}, {Raja}, {Rajan}, {Ramirez}, {Ramirez}, {Ramos-Buades},
  {Rana}, {Rapagnani}, {Rapol}, {Ray}, {Raymond}, {Raza}, {Razzano}, {Read},
  {Rees}, {Regimbau}, {Rei}, {Reid}, {Reid}, {Reitze}, {Relton}, {Renzini},
  {Rettegno}, {Reza}, {Rezac}, {Ricci}, {Richards}, {Richardson}, {Richardson},
  {Riemenschneider}, {Riles}, {Rinaldi}, {Rink}, {Rizzo}, {Robertson}, {Robie},
  {Robinet}, {Rocchi}, {Rodriguez}, {Rolland}, {Rollins}, {Romanelli},
  {Romano}, {Romel}, {Romero-Rodr{\'\i}guez}, {Romero-Shaw}, {Romie},
  {Ronchini}, {Rosa}, {Rose}, {Rosi{\'n}ska}, {Ross}, {Rowan}, {Rowlinson},
  {Roy}, {Roy}, {Roy}, {Rozza}, {Ruggi}, {Ruiz-Rocha}, {Ryan}, {Sachdev},
  {Sadecki}, {Sadiq}, {Sago}, {Saito}, {Saito}, {Sakai}, {Sakai},
  {Sakellariadou}, {Sakuno}, {Salafia}, {Salconi}, {Saleem}, {Salemi},
  {Samajdar}, {Sanchez}, {Sanchez}, {Sanchez}, {Sanchis-Gual}, {Sanders},
  {Sanuy}, {Saravanan}, {Sarin}, {Sassolas}, {Satari}, {Sathyaprakash}, {Sato},
  {Sato}, {Sauter}, {Savage}, {Sawada}, {Sawant}, {Sawant}, {Sayah},
  {Schaetzl}, {Scheel}, {Scheuer}, {Schiworski}, {Schmidt}, {Schmidt},
  {Schnabel}, {Schneewind}, {Schofield}, {Sch{\"o}nbeck}, {Schulte}, {Schutz},
  {Schwartz}, {Scott}, {Scott}, {Seglar-Arroyo}, {Sekiguchi}, {Sekiguchi},
  {Sellers}, {Sengupta}, {Sentenac}, {Seo}, {Sequino}, {Sergeev}, {Setyawati},
  {Shaffer}, {Shahriar}, {Shams}, {Shao}, {Sharma}, {Sharma}, {Shawhan},
  {Shcheblanov}, {Shibagaki}, {Shikauchi}, {Shimizu}, {Shimoda}, {Shimode},
  {Shinkai}, {Shishido}, {Shoda}, {Shoemaker}, {Shoemaker}, {ShyamSundar},
  {Sieniawska}, {Sigg}, {Singer}, {Singh}, {Singh}, {Singha}, {Sintes},
  {Sipala}, {Skliris}, {Slagmolen}, {Slaven-Blair}, {Smetana}, {Smith},
  {Smith}, {Soldateschi}, {Somala}, {Somiya}, {Son}, {Soni}, {Soni}, {Sordini},
  {Sorrentino}, {Sorrentino}, {Sotani}, {Soulard}, {Souradeep}, {Sowell},
  {Spagnuolo}, {Spencer}, {Spera}, {Srinivasan}, {Srivastava}, {Srivastava},
  {Staats}, {Stachie}, {Steer}, {Steinhoff}, {Steinlechner}, {Steinlechner},
  {Stevenson}, {Stops}, {Stover}, {Strain}, {Strang}, {Stratta}, {Strunk},
  {Sturani}, {Stuver}, {Sudhagar}, {Sudhir}, {Sugimoto}, {Suh}, {Sullivan},
  {Sullivan}, {Summerscales}, {Sun}, {Sun}, {Sunil}, {Sur}, {Suresh}, {Sutton},
  {Suzuki}, {Suzuki}, {Swinkels}, {Szczepa{\'n}czyk}, {Szewczyk}, {Tacca},
  {Tagoshi}, {Tait}, {Takahashi}, {Takahashi}, {Takamori}, {Takano}, {Takeda},
  {Takeda}, {Talbot}, {Talbot}, {Tanaka}, {Tanaka}, {Tanaka}, {Tanaka},
  {Tanaka}, {Tanasijczuk}, {Tanioka}, {Tanner}, {Tao}, {Tao}, {Tapia San
  Mart{\'\i}n}, {Taranto}, {Tasson}, {Telada}, {Tenorio}, {Terhune},
  {Terkowski}, {Thirugnanasambandam}, {Thomas}, {Thomas}, {Thomas}, {Thompson},
  {Thondapu}, {Thorne}, {Thrane}, {Tiwari}, {Tiwari}, {Tiwari}, {Toivonen},
  {Toland}, {Tolley}, {Tomaru}, {Tomigami}, {Tomura}, {Tonelli},
  {Torres-Forn{\'e}}, {Torrie}, {Tosta e Melo}, {T{\"o}yr{\"a}}, {Trapananti},
  {Travasso}, {Traylor}, {Trevor}, {Tringali}, {Tripathee}, {Troiano},
  {Trovato}, {Trozzo}, {Trudeau}, {Tsai}, {Tsai}, {Tsang}, {Tsang}, {Tsao},
  {Tse}, {Tso}, {Tsubono}, {Tsuchida}, {Tsukada}, {Tsuna}, {Tsutsui},
  {Tsuzuki}, {Turbang}, {Turconi}, {Tuyenbayev}, {Ubhi}, {Uchikata},
  {Uchiyama}, {Udall}, {Ueda}, {Uehara}, {Ueno}, {Ueshima}, {Unnikrishnan},
  {Uraguchi}, {Urban}, {Ushiba}, {Utina}, {Vahlbruch}, {Vajente}, {Vajpeyi},
  {Valdes}, {Valentini}, {Valsan}, {van Bakel}, {van Beuzekom}, {van den
  Brand}, {Van Den Broeck}, {Vander-Hyde}, {van der Schaaf}, {van Heijningen},
  {Vanosky}, {van Putten}, {van Remortel}, {Vardaro}, {Vargas}, {Varma},
  {Vas{\'u}th}, {Vecchio}, {Vedovato}, {Veitch}, {Veitch}, {Venneberg},
  {Venugopalan}, {Verkindt}, {Verma}, {Verma}, {Veske}, {Vetrano},
  {Vicer{\'e}}, {Vidyant}, {Viets}, {Vijaykumar}, {Villa-Ortega}, {Vinet},
  {Virtuoso}, {Vitale}, {Vo}, {Vocca}, {von Reis}, {von Wrangel}, {Vorvick},
  {Vyatchanin}, {Wade}, {Wade}, {Wagner}, {Walet}, {Walker}, {Wallace},
  {Wallace}, {Walsh}, {Wang}, {Wang}, {Wang}, {Ward}, {Warner}, {Was},
  {Washimi}, {Washington}, {Watchi}, {Weaver}, {Webster}, {Weinert},
  {Weinstein}, {Weiss}, {Weller}, {Weller}, {Wellmann}, {Wen}, {We{\ss}els},
  {Wette}, {Whelan}, {White}, {Whiting}, {Whittle}, {Wilken}, {Williams},
  {Williams}, {Williams}, {Williamson}, {Willis}, {Willke}, {Wilson},
  {Winkler}, {Wipf}, {Wlodarczyk}, {Woan}, {Woehler}, {Wofford}, {Wong}, {Wu},
  {Wu}, {Wu}, {Wu}, {Wysocki}, {Xiao}, {Xu}, {Yamada}, {Yamamoto}, {Yamamoto},
  {Yamamoto}, {Yamamoto}, {Yamashita}, {Yamazaki}, {Yang}, {Yang}, {Yang},
  {Yang}, {Yang}, {Yap}, {Yeeles}, {Yelikar}, {Ying}, {Yokogawa}, {Yokoyama},
  {Yokozawa}, {Yoo}, {Yoshioka}, {Yu}, {Yu}, {Yuzurihara}, {Zadro{\.z}ny},
  {Zanolin}, {Zeidler}, {Zelenova}, {Zendri}, {Zevin}, {Zhan}, {Zhang},
  {Zhang}, {Zhang}, {Zhang}, {Zhang}, {Zhao}, {Zhao}, {Zhao}, {Zhao}, {Zheng},
  {Zhou}, {Zhou}, {Zhu}, {Zhu}, {Zimmerman}, {Zlochower}, {Zucker}, \&
  {Zweizig}}]{LVK2021}
{The LIGO Scientific Collaboration}, {the Virgo Collaboration}, {the KAGRA
  Collaboration}, {et~al.} 2021, arXiv e-prints, arXiv:2111.03606,
  \dodoi{10.48550/arXiv.2111.03606}

\bibitem[{{van den Heuvel} {et~al.}(2017){van den Heuvel}, {Portegies Zwart},
  \& {de Mink}}]{vdH2017}
{van den Heuvel}, E.~P.~J., {Portegies Zwart}, S.~F., \& {de Mink}, S.~E. 2017,
  \mnras, 471, 4256, \dodoi{10.1093/mnras/stx1430}

\bibitem[{{Wagg} {et~al.}(2022{\natexlab{a}}){Wagg}, {Breivik}, \& {de
  Mink}}]{Wagg2022b}
{Wagg}, T., {Breivik}, K., \& {de Mink}, S.~E. 2022{\natexlab{a}}, \apjs, 260,
  52, \dodoi{10.3847/1538-4365/ac5c52}

\bibitem[{{Wagg} {et~al.}(2022{\natexlab{b}}){Wagg}, {Broekgaarden}, {de Mink},
  {Frankel}, {van Son}, \& {Justham}}]{Wagg2022a}
{Wagg}, T., {Broekgaarden}, F.~S., {de Mink}, S.~E., {et~al.}
  2022{\natexlab{b}}, \apj, 937, 118, \dodoi{10.3847/1538-4357/ac8675}

\bibitem[{{Wang} {et~al.}(2021){Wang}, {Chen}, {Liu}, {Chen}, {Wu}, {Tang},
  {Guo}, \& {Han}}]{Wang2021}
{Wang}, B., {Chen}, W.-C., {Liu}, D.-D., {et~al.} 2021, \mnras, 506, 4654,
  \dodoi{10.1093/mnras/stab2032}

\bibitem[{{Webbink}(1984)}]{Webbink1984}
{Webbink}, R.~F. 1984, \apj, 277, 355, \dodoi{10.1086/161701}

\bibitem[{{Weisz} {et~al.}(2015){Weisz}, {Johnson}, {Foreman-Mackey},
  {Dolphin}, {Beerman}, {Williams}, {Dalcanton}, {Rix}, {Hogg}, {Fouesneau},
  {Johnson}, {Bell}, {Boyer}, {Gouliermis}, {Guhathakurta}, {Kalirai}, {Lewis},
  {Seth}, \& {Skillman}}]{Weisz2015}
{Weisz}, D.~R., {Johnson}, L.~C., {Foreman-Mackey}, D., {et~al.} 2015, \apj,
  806, 198, \dodoi{10.1088/0004-637X/806/2/198}

\bibitem[{{Williams} {et~al.}(2017){Williams}, {Dolphin}, {Dalcanton}, {Weisz},
  {Bell}, {Lewis}, {Rosenfield}, {Choi}, {Skillman}, \&
  {Monachesi}}]{Williams2017}
{Williams}, B.~F., {Dolphin}, A.~E., {Dalcanton}, J.~J., {et~al.} 2017, \apj,
  846, 145, \dodoi{10.3847/1538-4357/aa862a}

\bibitem[{{Xu} \& {Li}(2010)}]{Xu2010}
{Xu}, X.-J., \& {Li}, X.-D. 2010, \apj, 716, 114,
  \dodoi{10.1088/0004-637X/716/1/114}

\bibitem[{{Yang} {et~al.}(2022){Yang}, {Ai}, {Zhang}, {Zhang}, {Liu}, {Wang},
  {Yang}, {Yin}, {Li}, \& {L{\"u}}}]{Yang2022}
{Yang}, J., {Ai}, S., {Zhang}, B.-B., {et~al.} 2022, \nat, 612, 232,
  \dodoi{10.1038/s41586-022-05403-8}

\bibitem[{{Yarza} {et~al.}(2022){Yarza}, {Everson}, \&
  {Ramirez-Ruiz}}]{Yarza2022}
{Yarza}, R., {Everson}, R.~W., \& {Ramirez-Ruiz}, E. 2022, arXiv e-prints,
  arXiv:2210.00010.
\newblock \doarXiv{2210.00010}

\bibitem[{{Zhu} {et~al.}(2021){Zhu}, {Wu}, {Yang}, {Zhang}, {Gao}, {Yu}, {Li},
  {Cao}, {Liu}, {Huang}, \& {Zhang}}]{Zhu2021}
{Zhu}, J.-P., {Wu}, S., {Yang}, Y.-P., {et~al.} 2021, \apj, 917, 24,
  \dodoi{10.3847/1538-4357/abfe5e}

\bibitem[{{Zorotovic} {et~al.}(2010){Zorotovic}, {Schreiber}, {G{\"a}nsicke},
  \& {Nebot G{\'o}mez-Mor{\'a}n}}]{Zorotovic2010}
{Zorotovic}, M., {Schreiber}, M.~R., {G{\"a}nsicke}, B.~T., \& {Nebot
  G{\'o}mez-Mor{\'a}n}, A. 2010, \aap, 520, A86,
  \dodoi{10.1051/0004-6361/200913658}

\bibitem[{{Zuo} \& {Li}(2014)}]{Zuo2014}
{Zuo}, Z.-Y., \& {Li}, X.-D. 2014, \mnras, 442, 1980,
  \dodoi{10.1093/mnras/stu993}

\end{thebibliography}
\bibliographystyle{aasjournal}



\end{document}